\documentclass{aa}
\usepackage{graphicx}
\usepackage[authoryear]{natbib}
\bibpunct{(}{)}{;}{a}{}{,}
\usepackage{psfig}

\begin{document}

\title{New Light on the Formation and Evolution of M31 and its Globular 
Cluster System}

\author{Thomas H. Puzia \inst{1,}\thanks{ESA Fellow, Space Telescope Division of ESA}, Kathy M. Perrett\inst{2,}\thanks{NSERC Post-doctoral Fellow}, \& Terry J. Bridges\inst{3}}
      
   \offprints{Thomas H. Puzia, \email{tpuzia@stsci.edu}}

   \institute{Space Telescope Science Institute,
        3700 San Martin Drive, Baltimore, MD 21218, USA, \\
        \email{tpuzia@stsci.edu}
      \and
	Department of Astronomy \& Astrophysics, University of Toronto, Toronto, ON, M5S 3H8, 
	Canada, \\
	\email{perrett@astro.utoronto.ca}
      \and
          Department of Physics, Queen's University, Kingston, ON, K7L 3N6, Canada, \\
	\email{tjb@astro.queensu.ca}
        }

   \authorrunning{Puzia et al.}  
   \titlerunning{New Light on the Formation and Evolution of M31 and its Globular 
Cluster System} 
      
   \date{Accepted January 2005 }
   
   \abstract{We present spectroscopic ages, metallicities, and [$\alpha$/Fe] ratios for
70 globular clusters in M31 that were derived from Lick line-index
measurements. A new interpolation technique of age-metallicity and
$\alpha$/Fe-diagnostic grids is used to account for changes in index
strength as a response to abundance-ratio variations, in particular for
all of the Balmer-line Lick indices. In addition to a population of old
($>10$ Gyr) globular clusters with a wide range of metallicities, from
about $-2.0$ dex to solar values, we find evidence for a population of
intermediate-age globular clusters with ages between $\sim5$ and 8 Gyr and
a mean metallicity [Z/H]~$\approx-0.6$. We also confirm the presence of
young M31 globular clusters that were recently identified by
\cite{beasley04}, which have ages $\la1$ Gyr and relatively high
metallicities around $-0.4$ dex. The M31 globular cluster system has a
clearly super-solar mean [$\alpha$/Fe]~$=0.14\pm0.04$ dex.
Intermediate-age and young objects show roughly solar abundance ratios. We
find evidence for an age-[$\alpha$/Fe] relation in the sense that younger
clusters have smaller mean [$\alpha$/Fe] ratios. From a comparison of
indices, mostly sensitive to carbon and/or nitrogen abundance, with SSP
model predictions for nitrogen-enhanced stellar populations, we find a
dichotomy in nitrogen enhancement between young and old M31 globular
clusters. The indices of objects older than 5 Gyr are consistent with a
factor of three or higher in nitrogen enhancement compared to their
younger counterparts. Using kinematical data from \cite{morrison04} we
find that the globular cluster sub-population with halo kinematics is old
($\ga9$ Gyr), has a bimodal metallicity distribution, and super-solar
[$\alpha$/Fe]. Disk globular clusters have a wider range of ages, are on
average more metal-rich, and have a slightly smaller mean [$\alpha$/Fe]
ratio. A cross-correlation of structural parameters for M31 globular
clusters with spectroscopically derived ages, metallicities, and
[$\alpha$/Fe] ratios shows a correlation between half-light/tidal radius
and metallicity, which is most likely due to the correlation of
half-light/tidal radius and galactocentric distance. We compare our
results for M31 globular clusters with those obtained with the same
technique for globular clusters in the Milky Way, Large Magellanic Cloud,
M81, and other spiral galaxies in the Sculptor group. Finally, we compare
the globular cluster systems of the two Local Group spirals, M31 and Milky
Way, with their integrated bulge light.}
   
   \maketitle

\keywords{Galaxies: star clusters, abundances, formation, evolution, Local Group --
Galaxy: globular clusters -- Galaxies: individual: M31, Milky Way, Large Magellanic Cloud, M81, NGC 55, NGC 247, NGC 253, NGC 300} 

\section{Introduction}
\label{ln:intro}
Because they can be resolved into single stars, the stellar populations in
the Local Group are milestones for understanding star formation and
chemical evolution histories of distant galaxies. Globular clusters host
simple stellar populations that are characterized by their small
dispersion in age and chemical composition, and are, compared to the mix
of stellar populations in galaxies, relatively easy to understand. All
theoretical predictions for observables of unresolved stellar systems are
calibrated on resolved stellar populations in Milky Way globular clusters.
Since the formation of globular clusters is temporally linked with major
star-formation episodes, globular cluster systems are fossil records of
the past star formation and chemical evolution history of their host
galaxies \citep{ashman98, harris01}. It is therefore of paramount
importance to ensure that the predictions drawn from a comparison of
Galactic and extragalactic globular clusters, and stellar populations in
general, are not biased by any peculiarity of our neighborhood.

The richest globular cluster systems in the Local Group are hosted by the
two major spiral galaxies Andromeda (M31; NGC 224) harboring $460\pm70$
\citep{barmby01} and the Milky Way with $160\pm20$ globular clusters
\citep{harris01}. Photometry has shown that the globular cluster systems of
M31 and Milky Way have a similar range in metallicity and that both
distributions are bimodal with peaks around [Fe/H]~$=-1.4$ and $-0.6$ dex
\citep[e.g.][]{barmby00, barmby01}. Recently, \cite{perrett02}
spectroscopically confirmed this metallicity bimodality. However, M31
hosts a larger number of metal-rich clusters, which is reflected in the
higher mean metallicity of the entire globular cluster system
\citep[e.g.][]{huchra91}. Although the globular cluster luminosity
functions are remarkably similar in both galaxies \citep{harris01}, they
may differ as a function of metallicity and galactocentric radius
\citep{barmby01b}. There is also a difference in spatial distribution
(scale size), where the M31 cluster system appears to be more extended
than the Galactic globular cluster system \citep{harris01}. Thus, despite
the overall similarities of the M31 and Milky Way globular cluster
systems, there are some striking differences in the details. Recently, a
revised optical/near-IR photometric catalog of 337 confirmed and 688
candidate globular clusters in M31, that combines and homogenizes previous
studies, was published by the Bologna group and will allow for more
accurate future analyses \citep[][ and references therein]{galleti04}.

Integrated-light spectroscopy of M31 globular clusters was pioneered by
\cite{vdB69} who found that the M31 globular cluster system extends to
much higher metallicities than the Milky Way globular cluster system.
Subsequent studies confirmed these results \citep[e.g.][]{spinrad72,
rabin81}. A thorough study, based on spectroscopic Lick line indices of 19
M31 and 17 Galactic globular clusters, by \cite{burstein84} confirmed this
and showed that M31 clusters have significantly stronger H$\beta$ and CN
absorption indices at a given metallicity. After carefully checking the
influence of hot horizontal-branch stars and/or blue stragglers, and
younger ages on these indices, the authors concluded that age is the most
favorable, but not sufficient, explanation for the systematic differences.
They speculated that their sample was dominated by relatively young disk
globular clusters, which could explain the offset in H$\beta$, but not the
strong CN indices. The possibility of an enhanced carbon and/or nitrogen
abundance was considered unlikely. \cite{burstein87} later showed that K
giants contributing to the integrated light of M31 clusters must have at
least as strong CN absorption bands as the most extreme Galactic giant
stars. In a subsequent analysis of the Ca{\sc ii} H+K absorption feature
of nine metal-rich M31 globular clusters, \cite{tripicco89} rejected the
hypothesis that hot horizontal-branch stars and/or blue stragglers boost
the H$\beta$ index of the integrated light. Modeling the spectra of M31
clusters he also found that the strong CN index is consistent with roughly
a factor of 10 in CN excess in both dwarf and giant stars. \cite{davidge90}
obtained near-infrared spectra of four luminous M31 globular clusters and
compared them with Galactic counterparts. He found a clear enhancement in
the 2 $\mu$m CN band and concluded that the increased CN abundance in M31
globular clusters is likely to be of primordial origin. From a comparison
of CO absorption bands in M31 clusters and the integrated light of M32,
which is known to host an intermediate-age population
\citep[e.g.][]{davidge90b}, he reasoned that, if both stellar populations
have similar metallicities, luminous M31 globular clusters may contain an
intermediate-age component.

In a comprehensive spectroscopic investigation of 149 globular clusters in
M31, using specifically defined line indices, \cite{brodie90} confirmed a
CN offset relative to Milky Way globular clusters, but failed to identify
a stronger mean H$\beta$ index, although good agreement was found between
the measurements of matched objects. This fact suggests that sample
selection in the M31 cluster system might be playing an important role and
implies the presence of multiple globular cluster sub-populations with
different mean H$\beta$ indices and potentially different chemical
compositions. \citeauthor{brodie90} reported also an enhancement in their
Ca{\sc ii} H+K index for M31 globular clusters, at a given metallicity.
Assuming that the CN enhancement is due to an increased nitrogen abundance
they identify aluminum satellite lines as responsible for an increased H+K
index, because of the correlation of nitrogen with aluminum abundance, at
least in metal-rich stars \citep{norris85}. 

An alternative explanation for the observed CN offset is lower mean
surface gravities, as these elevate both CN and Ca{\sc ii} H+K
\citep{oconnell73}. The surface gravity can be lowered by decreasing the
age; however, other indices such as Mg$_{2}$ (a measure of the Mg$b$
triplet and the magnesium hydride band, MgH) {\it increase significantly}
as age is decreased, and such an increase is not observed. However, the
Mg$_{2}$ index can be kept constant if the intrinsic magnesium abundance
{\it and} age are {\it both} decreased appropriately. The former could be
a result of a low [$\alpha$/Fe] ratio at constant total metallicity. This
is expected for stellar populations which formed late from material
enriched by type II and Ia supernovae \citep[e.g.][]{matteucci94,
matteucci01}. Later work using near-UV spectra analyzed the strength of
the NH band at 3360 \AA\ (closely linked to the nitrogen abundance) for
several M31 globular clusters \citep{ponder98, li03, burstein04}. These
studies confirmed that M31 globular clusters are indeed enhanced in
nitrogen. 

Furthermore, previous studies \citep{barmby00, beasley04, burstein04}
showed the presence of young M31 globular clusters. These young objects
have spectra which resemble those of young globular clusters in the Large
Magellanic Cloud with ages around 1 Gyr. Some of these M31 clusters appear
to belong to a distinct sub-population of thin disk globular clusters
discovered by \cite{morrison04}.

It is obvious that a study of ages and chemical compositions for a large
sample of M31 globular clusters is urgently needed to resolve some of the
above issues. Previous studies have shown indications of young ages for
some M31 globular clusters and clear signatures of chemical differences
between the globular cluster system in M31 and other galaxies. 

In this paper, we focus on the analysis of spectroscopic features of M31
globular clusters using Lick line indices from which we derive ages,
metallicities, and abundance ratios. Newly available simple stellar
population (SSP) models for well-defined chemical compositions
\citep{tmb03,tmk04} make these parameters accessible with unprecedented
systematic accuracy. We use high-quality Lick index measurements and a new
iterative prescription to derive these parameters that maximally reduces
systematic uncertainties in the internal calibration of Lick indices.

The paper is organized as follows: \S\ref{ln:data} describes the dataset
and its calibration to the Lick system. In the following Section
\S\ref{ln:analysis} we describe our method and perform a consistency check
for ages and metallicities derived from different diagnostic grids.
Results on M31 globular cluster ages, metallicities, and abundance ratios
are presented in Section \S\ref{ln:results}, along with a comparison with
globular cluster systems in other galaxies. Our findings are discussed in
\S\ref{ln:discussion} and summarized in \S\ref{ln:summary}.

\begin{table}[!t]
\centering
\caption{Radial coverage of background sectors. $r_{\rm min}$, $r_{\rm max}$,
and $r_{\rm median}$ are the radii of the fiber pointings with the minimal
and maximal galactocentric distance, and the median distance of the entire
sub-sample.}
\label{tab:bkgrad}
\begin{tabular}{lccc}
\hline\hline
\noalign{\smallskip}
sector & $r_{\rm min}$ [kpc] & $r_{\rm max}$ [kpc] & $r_{\rm median}$ [kpc]\\
\noalign{\smallskip}
\hline
\noalign{\smallskip}
z0  & 0.55 & 0.85 & 0.60 \\
z1  & 1.56 & 4.96 & 2.48 \\
\noalign{\smallskip}
\hline
\end{tabular}
\end{table}

\section{Data}
\label{ln:data}
\subsection{Globular Cluster Spectra}
\begin{figure*}[!ht]
\begin{center}
\includegraphics[width=15.8cm]{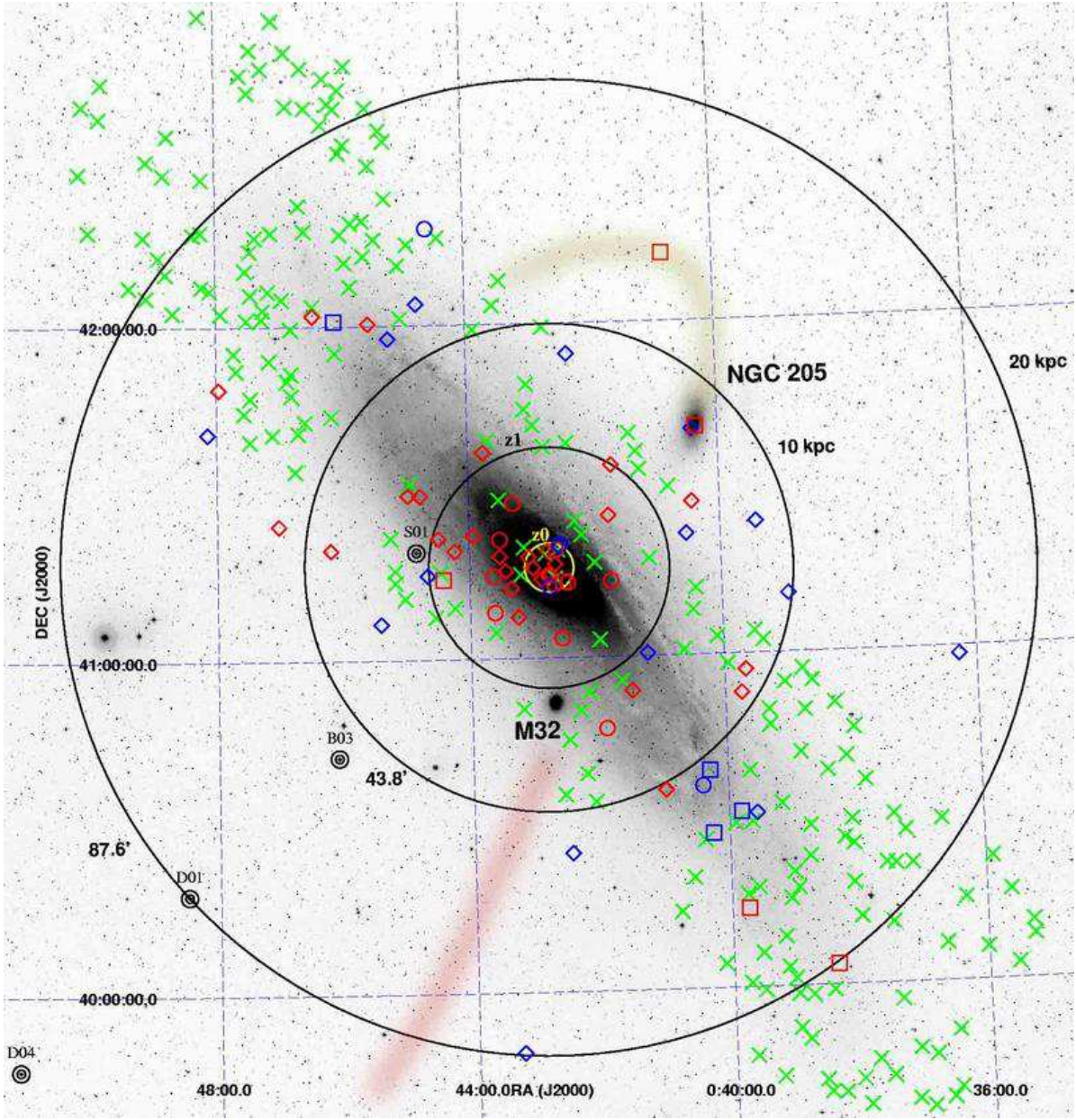}
\caption{Positions on the sky of studied globular clusters (open symbols) and
background fields (crosses). Globular clusters are parameterized by their
age and metallicity. Squares indicate globular clusters with ages below 5
Gyr, circles show intermediate-age globular clusters with ages between 5
and 8 Gyr, while diamonds mark the position of old globular clusters
($t>8$ Gyr). Old objects populate the entire studied field.
Intermediate-age globular clusters seem to prefer the region within the
central $\sim5$ kpc, while young clusters are preferentially located in
the outer disk. The metallicity is indicated by color-coded symbols. Red
symbols show metal-rich globular clusters with [Z/H]~$>-0.8$, while blue
symbols show their metal-poor counterparts. Note the clear concentration
of metal-rich objects around the nucleus of M31. Background spectra were
grouped in sectors z0 and z1 according to their galactocentric distance
indicated by the two inner circles. A few globular clusters (e.g. MII=G1)
which are located at large galactocentric distances fall off the image,
similar to some background pointings. The underlying image was taken from
DSS2/red. Shaded trails indicate stellar streams discovered by the
INT/CFHT collaboration \citep[see][]{ibata01,ferguson02,mcconnachie04}.
While the southern stream is the most massive ($\sim10^{8}-10^{9}$
M$_\odot$), the northern spur carries about an order of magnitude less
mass and appears to be kinematically associated with NGC 205
\citep{mcconnachie04}. Reticles mark pointings for which deep photometry
is available. Going from small to larger galactocentric distances the
fields are labeled as S01 \citep{sarajedini01}, B03 \citep{brown03}, D01
\citep{durrell01}, and D04 \citep{durrell04}.}
\label{ps:m31big}
\end{center}
\end{figure*}

We use spectroscopic observations obtained with the Wide Field Fibre Optic
Spectrograph \citep[WYFFOS,][]{bridges98} at the 4m William Herschel
Telescope on La Palma, Canary Islands. The dataset is described in detail
by \cite{perrett02} and we refer the reader to this paper for a discussion
of technical details. Figure~\ref{ps:m31big} shows the field of view with
all fiber positions. Open symbols indicate globular clusters; crosses mark
the position of background fibers. Sky corrections for globular cluster
spectra were performed by subtracting a mean sky spectrum that was
constructed from several nearby fibers placed in close vicinity to each
individual globular cluster. Throughout the following analysis we use
these sky-subtracted spectra.

\subsection{Spectra of the Diffuse Light}
In addition to our globular cluster measurements (see circles in
Fig.~\ref{ps:m31big}), we also obtained 207 integrated spectra of the
diffuse stellar population at different galactocentric radii. Each
individual spectrum has too low a signal-to-noise (S/N) to allow reliable
Lick index measurements. In order to maximize the S/N ratio, we select
spectra from the inner Bulge field (within 5 kpc projected galactocentric
distance) and subdivide this background sub-sample into an inner and outer
sector, z0 and z1, respectively. The radial sampling of the combined two
sectors is described in Table~\ref{tab:bkgrad}. Individual spectra in each
sector are then co-added. Before summing the spectra, a careful smoothing
to the Lick resolution of each single spectrum was performed, accounting
for the varying velocity dispersion at each individual fiber position and
the changing spectral resolution as a function of chip position. Velocity
dispersions as a function of galactocentric distance were calculated with
the disk-bulge-halo models of \cite{widrow03}. The average S/N ratio of
both combined background spectra exceeds $\sim25$ per \AA.

\subsection{Calibration of Lick Indices}
\label{ln:calib}
Prior to the measurement of Lick line indices, our spectra were broadened
to the characteristic Lick resolution. The steps undertaken during this
procedure were adopted from \cite{worthey97, kuntschner98, kuntschner00,
vazdekis99, puzia02, puzia04}, and \cite{puzia03phd}. A particular
characteristic of the WYFFOS instrument is its resolution as a function of
chip position. To account for this, we computed 2-dim. resolution maps
using arc exposures that were taken close in time with the science frames.
The resolution maps were then used to smooth our spectra to the
differential Lick resolution of $\ga8$ \AA\ \citep[see][ for
details]{worthey97}. We tested this transformation by matching our
globular cluster list with the M31 globular cluster sample of
\cite{worthey94} and \cite{kuntschner02}. The comparison gives six matches
(B012-G64, B015, B193-G244, B218-G272, B225-G280, B311-G33) that cover a
range in [Fe/H] from $-2.0$ to $-0.3$ dex. Small linear correction terms
were found for all indices, that in all cases had a significance of
$\la1~\sigma$ (see Table~\ref{tab:corr}). These systematic offsets were
applied to all subsequent Lick index measurements\footnote{Note that bad
pixel masking leaves two out of the six spectra useful for calibration of
the H$\gamma_{\rm A}$ index. The two remaining good-quality spectra are in
very good agreement with the Lick standard measurements. However, we warn
that the rms of this calibration cannot be trusted given the small
sample.}.

\begin{table}[!t]
\centering
\caption{Correction terms of the transformation to match the Lick/IDS instrumental
resolution. The corrections were used to transform our index measurements
to the Lick/IDS system \cite{worthey94}, in the sense $I_{\rm Lick} =
I_{\rm measured}+c$.}
\label{tab:corr}
\begin{tabular}{lrc}
\hline\hline
\noalign{\smallskip}
Index & $c$ & rms \\
\noalign{\smallskip}
\hline
\noalign{\smallskip}
H$\delta_{\rm A}$&$-0.127$ & 2.041 \\
H$\delta_{\rm F}$&$-0.437$ & 0.697 \\
CN$_{1}$ & $0.001$ & 0.096 \\
CN$_{2}$ & $0.051$ & 0.084 \\
Ca4227   &$-0.407$ & 0.956 \\
G4300    & $-0.315$ & 3.059 \\
H$\gamma_{\rm A}$&$-0.488$ & 0.514 \\
H$\gamma_{\rm F}$&$-0.094$ & 1.106 \\
Fe4383   & 0.764 & 2.480 \\ 
Ca4455  & 0.541 & 0.796 \\ 
Fe4531   & 1.086 & 1.445 \\  
Fe4668   & 0.007 & 2.193 \\ 
H$\beta$& 0.148 & 0.175 \\ 
Fe5015   &$-0.300$& 1.547 \\ 
Mg$_{1}$&0.004 & 0.025 \\
Mg$_{2}$&0.025 & 0.041 \\ 
Mg$b$   &$-0.123$ & 0.695 \\
Fe5270  & 0.730 & 0.589 \\
Fe5335  & 0.243 & 0.576 \\
Fe5406  &$-0.064$& 0.549 \\ 
Fe5709  &0.355& 0.489 \\
\noalign{\smallskip}
\hline
\end{tabular}
\end{table}

Blue Lick indices (from CN$_{1}$ to Fe4383) were measured on spectra taken
with the blue arm of WYFFOS, which covers the spectral range
$\sim3700-4500$ \AA. The remaining Lick indices (Ca4455 to Fe5709) were
measured on spectra taken with the red-arm with a spectral coverage from
$\sim4400$ to 5800 \AA. Spectra with S/N of $\sim25$ per \AA\ or less were
rejected from the subsequent analysis. As we propagate the full error
budget of the spectra into the line index measurement, some indices show
unusually large error bars due to masked bad pixels and/or cosmics. These
contaminated index measurements were also rejected from the following
analysis steps. With these data quality requirements 29 globular clusters 
remain in our sample.

\subsection{Data from other Sources}
\label{ln:inputdata}
To maximize the statistical significance of the globular cluster sample,
we include five M31 globular clusters from the Lick group
\citep{worthey94, worthey97} and 36 globular clusters from the
\cite{beasley04} sample. We note that all index measurements are performed
in the same passband system as the SSP models, in particular the
fitting functions that are used to compute theoretical predictions.

\begin{figure*}[!t]
\begin{center}
\includegraphics[width=11cm, bb=20 15 390 389]{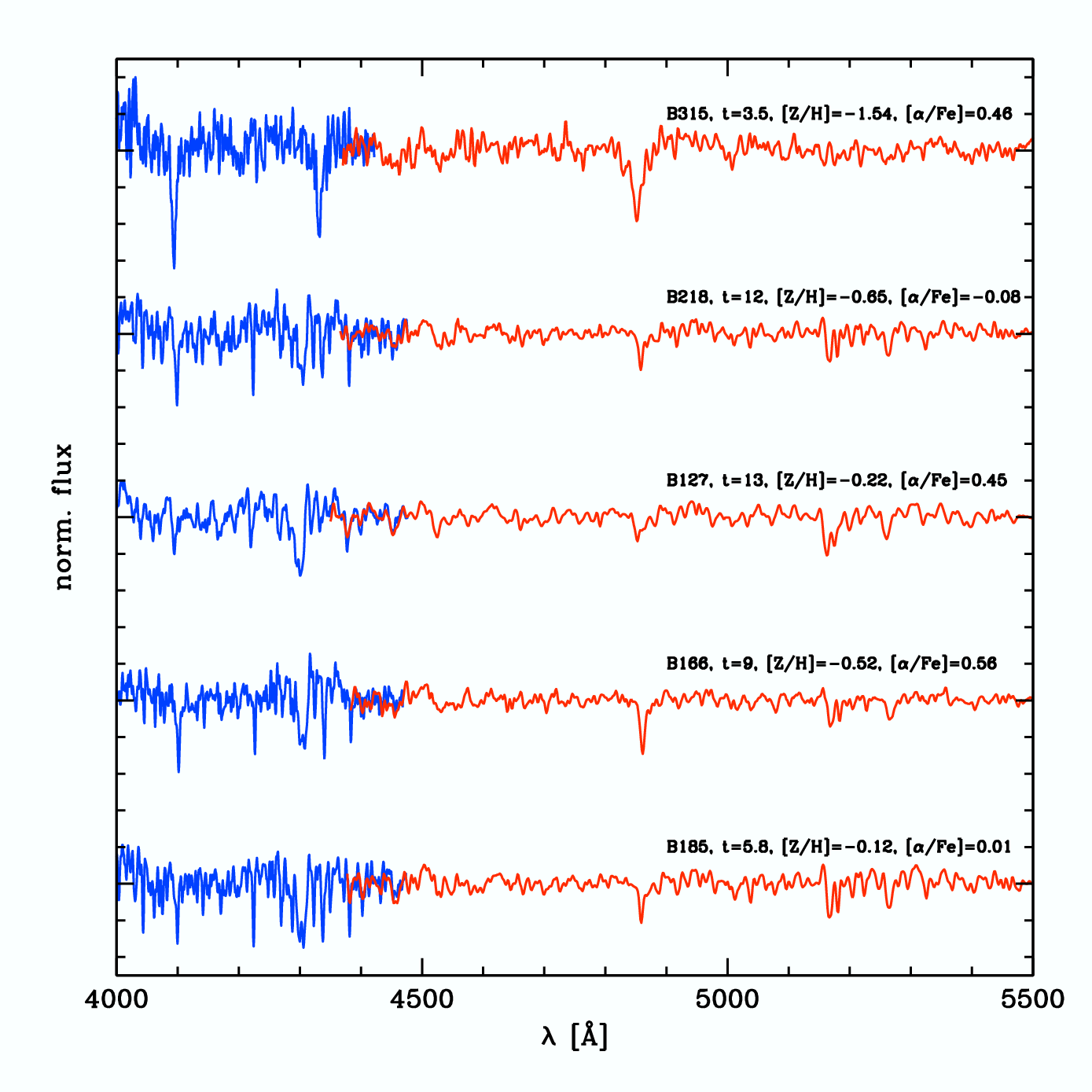}
\caption{Representative spectra from our M31 globular cluster sample 
for selected targets, sorted by metallicity (increasing from top to bottom). 
Blue and red spectra are plotted separately. Derived ages (in Gyr), metallicities, and
[$\alpha$/Fe] ratios from the subsequent analysis steps are indicated.}
\label{ps:spectra}
\end{center}
\end{figure*}

We find two matches between our globular cluster data and the Beasley et
al. dataset. The Lick index measurements for the two clusters, i.e. B225
and B243, are in very good agreement (within 1-$\sigma$), especially for
the widely-defined molecular-band indices (e.g. CN$_{2}$ and Mg$_{2}$, but
{\it not} for G4300\footnote{We find a $\sim1.5$ \AA\ systematic offset
towards higher values in the Beasley et al. sample, although one of our
measurements has a very large error of 1.2 \AA.}). Very good agreement is
also found for indices crucial to further analysis steps, such as Balmer,
iron, and magnesium indices:
H$\delta_{A}$, H$\gamma_{A}$, H$\beta$, Mg$_{2}$, Mg$b$, Fe5270, Fe5335,
etc. Because of the smaller statistical errors of the Beasley et al. data
we adopt the latter for the following analysis. Hence our final dataset
contains high-quality spectra (S/N~$\ga25$ \AA$^{-1}$) for 70 globular
clusters in M31. The full set of index measurements is presented in
Tables~\ref{tab:m31indices} and \ref{tab:m31indexerr} in the Appendix.
Note that object B166 is suspected to be a foreground star
\citep{galleti04}. However, since this is not certain, and because we do
not find any features in our spectrum that would be characteristic of a
foreground star, we mark this object as potentially misclassified but keep
it in our globular cluster sample. Note that based on their colors and
radial velocities, all other objects were classified as genuine globular
clusters by several independent studies \citep[see][]{galleti04}.
Figure~\ref{ps:spectra} shows representative spectra from our sample.

\begin{figure}[!t]
\begin{center}
\includegraphics[width=7.3cm, bb=40 130 600 700]{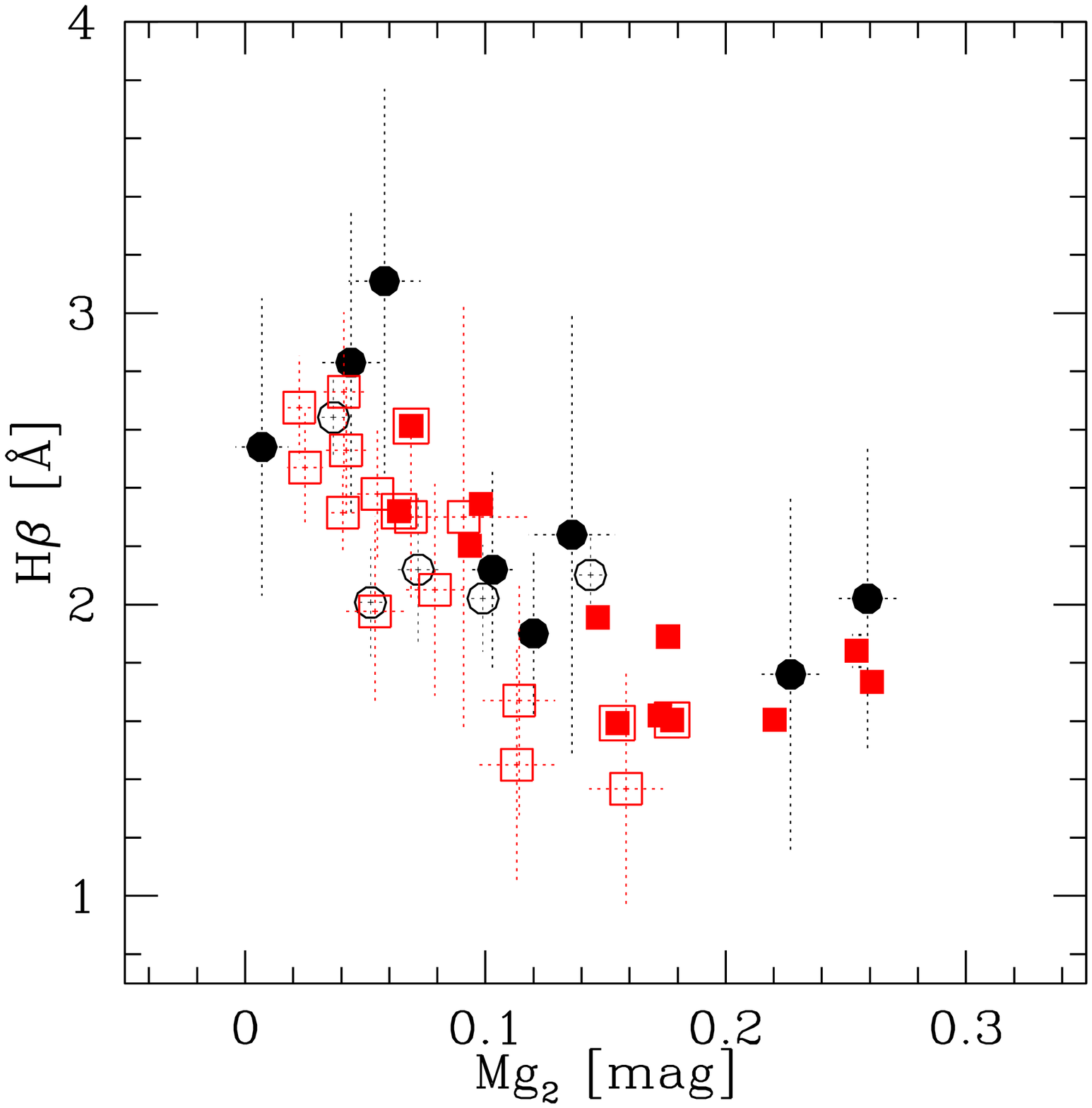}
\includegraphics[width=7.3cm, bb=40 130 600 700]{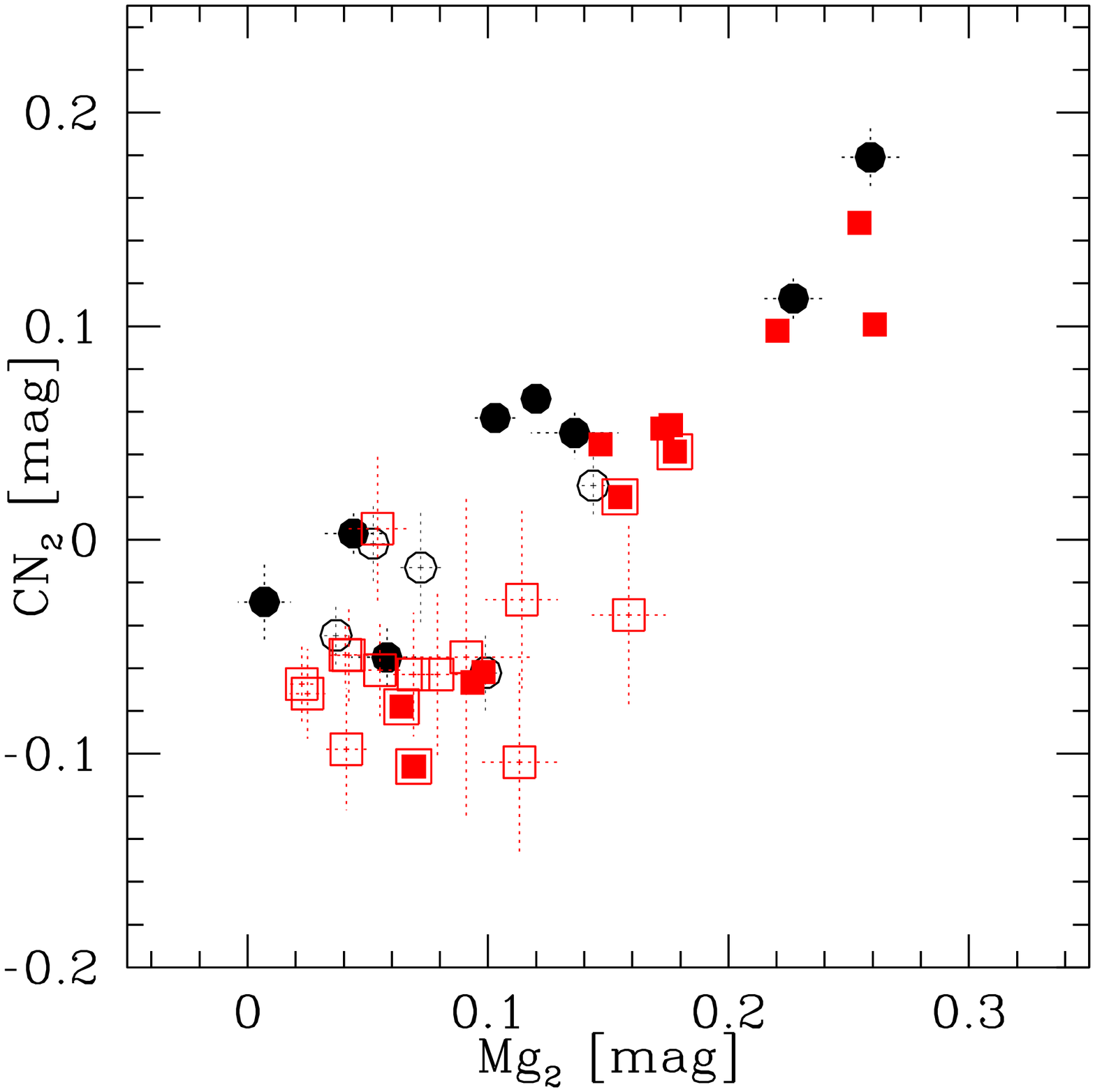}
\caption{H$\beta$ and CN$_{2}$ vs. Mg$_{2}$ plots for M31 and Galactic globular
clusters. Milky Way globular clusters are indicated as squares, while M31
globular clusters are marked by circles. Solid squares show data taken
from \cite{puzia02}. Open squares are Galactic globular clusters from
\cite{trager98}. Large open squares illustrate matches between the
\citeauthor{puzia02} and \citeauthor{trager98} dataset. Solid circles are
M31 globular clusters included in the \citeauthor{burstein84} dataset
that were re-measured in this study. Open circles add M31 data from
\cite{trager98}. See text for details.}
\label{ps:burstein}
\end{center}
\end{figure}

We do not include indices from \cite{burstein84} since they were measured
in the old Lick passband system, for which no model predictions with
variable abundance ratios are available. However, we find eight matched
M31 clusters in our and the \cite{trager98} data, who re-measured Lick
indices for M31 and Milky Way globular clusters from the
\citeauthor{burstein84} sample with slightly modified passband definitions
compared to the ones used in this work\footnote{The difference in passband
definitions between the ones adopted in this work \citep{worthey94,
worthey97} and the ones of \citep{trager98} is significant for some
indices. Linear transformations between the two systems were computed by
\cite{puzia04}. H$\beta$ has identical definitions in both systems. For
the CN$_{2}$ the offset is $\Delta$CN$_{2}({\rm Tr98-Wo94})0.014\pm0.020$
mag. Where applicable, these linear corrections are applied to the Trager
et al. data.}. We therefore briefly investigate the previously detected
H$\beta$ offset between Galactic and M31 globular clusters in the
\cite{burstein84} study. In Figure~\ref{ps:burstein} we reproduce the
H$\beta$ and CN$_{2}$ vs. Mg$_{2}$ plots from \cite{burstein84}. We
include only the eight M31 globular clusters that overlap with the
\citeauthor{trager98} data and were re-observed in this study (solid
circles) and augment the sample with clusters from the \cite{trager98}
study that meet our quality selection criteria (five open circles). Also
included are 12 Galactic globular clusters from \cite{puzia02} (solid
squares), and 13 clusters from \cite{trager98} (open squares).

We find four matches between the Galactic cluster samples of
\citeauthor{puzia02} and \citeauthor{trager98}, and plot for matched
objects the indices measured by \citeauthor{puzia02}, because of their
smaller errors. The systematic offsets for the four matches in the sense
\citeauthor{puzia02} -- \citeauthor{trager98} are
$\Delta$H$\beta=0.31\pm0.28$ \AA\ and $\Delta$CN$_{2}=-0.02\pm0.04$ mag.
The matched objects are marked with large open squares. From the eight
matched M31 globular clusters between our sample and that of
\citeauthor{trager98} we find in general very good agreement between the
two index measurements for all relevant indices, such as Balmer, Mg, and
Fe indices. In particular, the offsets, in the same sense as above, are
$\Delta$H$\beta=0.06\pm0.13$ \AA\ and $\Delta$CN$_{2}=0.026\pm0.023$ mag.
The systematic offset in H$\beta$ is clearly less significant than the one
found between the Galactic cluster samples.

Although the error bars of the M31 sample are relatively large, there is
no significant systematic offset between M31 and Milky Way globular
clusters in the newly acquired data (solid symbols). Fitting straight
lines to the new data in the H$\beta$ vs. Mg$_{2}$ plot yields the
following linear relations:
\[
\begin{array}{cc}
{\rm M31:\;\;}&{\rm H}\beta=(-4.05\pm1.03)\cdot{\rm Mg}_{2}+(2.58\pm0.18) \\
{\rm MW:\;\;}&{\rm H}\beta=(-3.95\pm1.50)\cdot{\rm Mg}_{2}+(2.78\pm0.22)
\end{array}
\]
The rms for both relations are 0.30 and 0.21 \AA\ for M31 and Galactic
globular clusters. Since the gradients are virtually identical, the
zero-point offset is a measure of the systematic offset. It is smaller
than the mean rms of both relations. If we split the datasets at
Mg$_{2}=0.2$ \AA, the systematic offsets for both high and low-Mg$_{2}$
subsets disappear, i.e. they are less significant than $\sim0.5\sigma$.
However, if the Galactic globular clusters from \citeauthor{trager98}
(small and large open squares) are considered, we confirm a systematic
offset $\Delta$H$\beta\approx0.5$ \AA\ ($\ga3\sigma$) at Mg$_{2}\ga0.1$
mag between M31 and Milky Way (see also Fig.~5k in \citealt{burstein84}).
In summary, a combination of sample selection, small number statistics,
and a potential calibration inconsistency might have introduced spurious
H$\beta$ offsets between Galactic and M31 globular clusters in the
\citeauthor{burstein84} work.

The new data do not appear to alter the conclusion of
\citeauthor{burstein84} on the CN offset. However, for metal-rich globular
clusters the offset in CN seems smaller than for metal-poor globular
clusters. In general, this exercise demonstrates that for drawing reliable
conclusions from a direct comparison of datasets, a homogeneous and
consistent sample selection of these datasets is crucial.

Later in this work we show that the speculation of \citeauthor{burstein84}
that the H$\beta$ offset was produced by a systematically younger mean age
of their observed M31 globular cluster sample is likely to be the result
of Milky Way globular cluster sample selection. From their sample of 19
M31 globular clusters, 16 had their indices on high-quality spectra
re-measured in subsequent studies. In fact, in this study we find that the
mean age of this M31 sub-sample is $\sim10$ Gyr with only four
globular clusters between 6 and 9 Gyr. However, their intuition that M31
might host young globular clusters turns out to be correct, if a larger
more representative sample of M31 globular clusters is considered.

\section{Analysis}
\label{ln:analysis}
\begin{figure*}[!th]
\begin{center}
\includegraphics[width=5.5cm, bb=40 130 600 700]{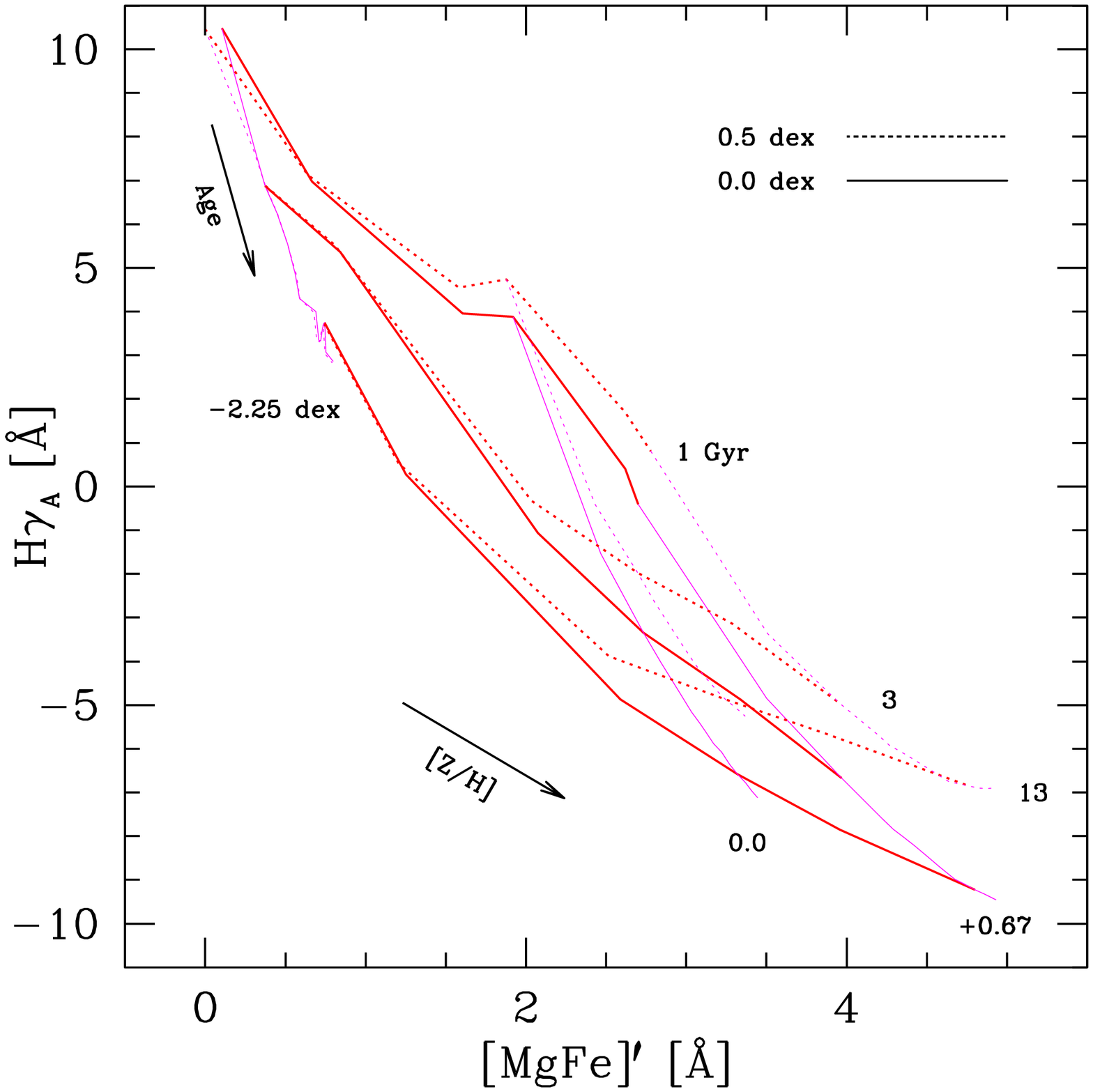}
\includegraphics[width=5.5cm, bb=40 130 600 700]{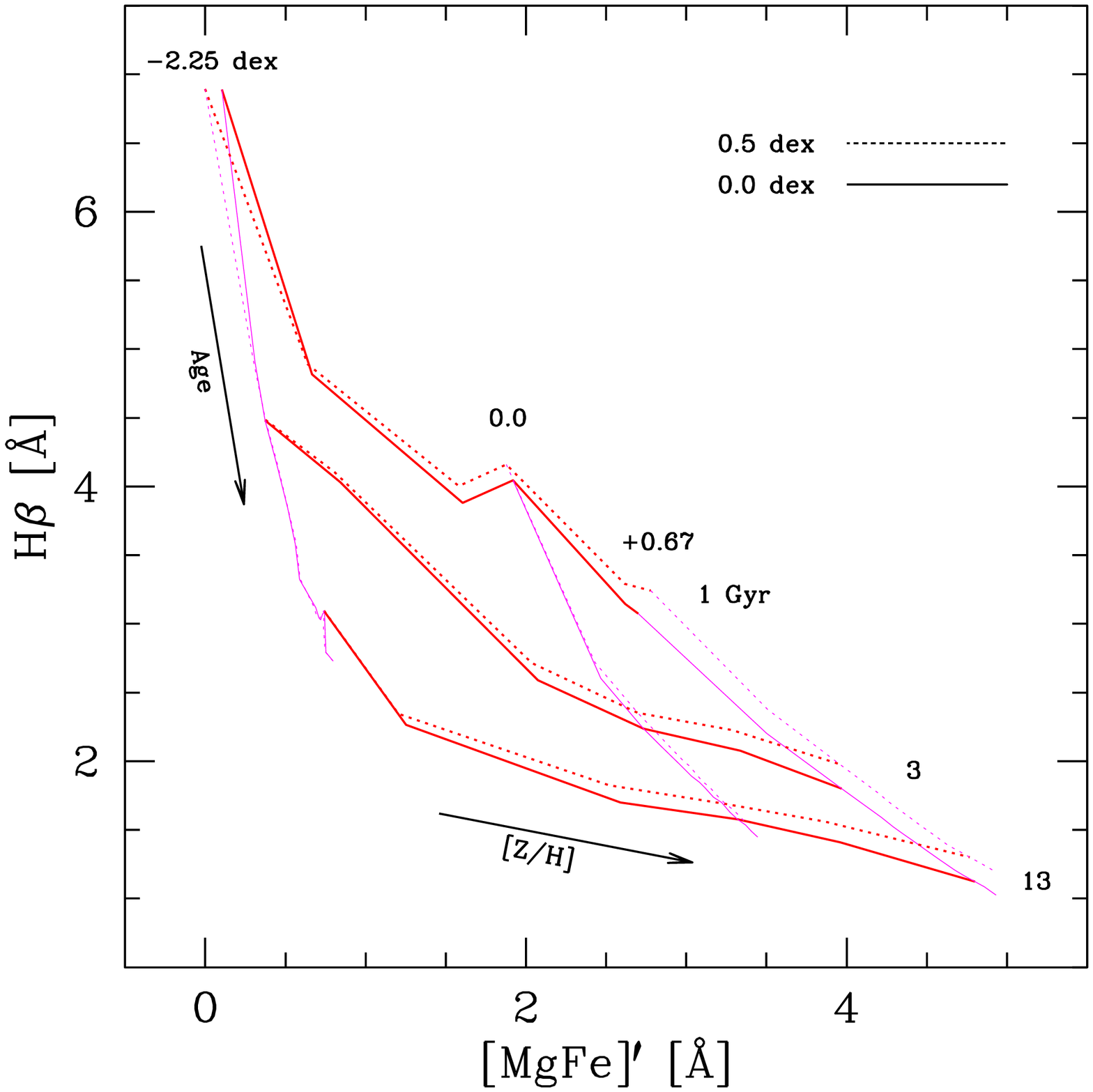}
\includegraphics[width=5.5cm, bb=40 130 600 700]{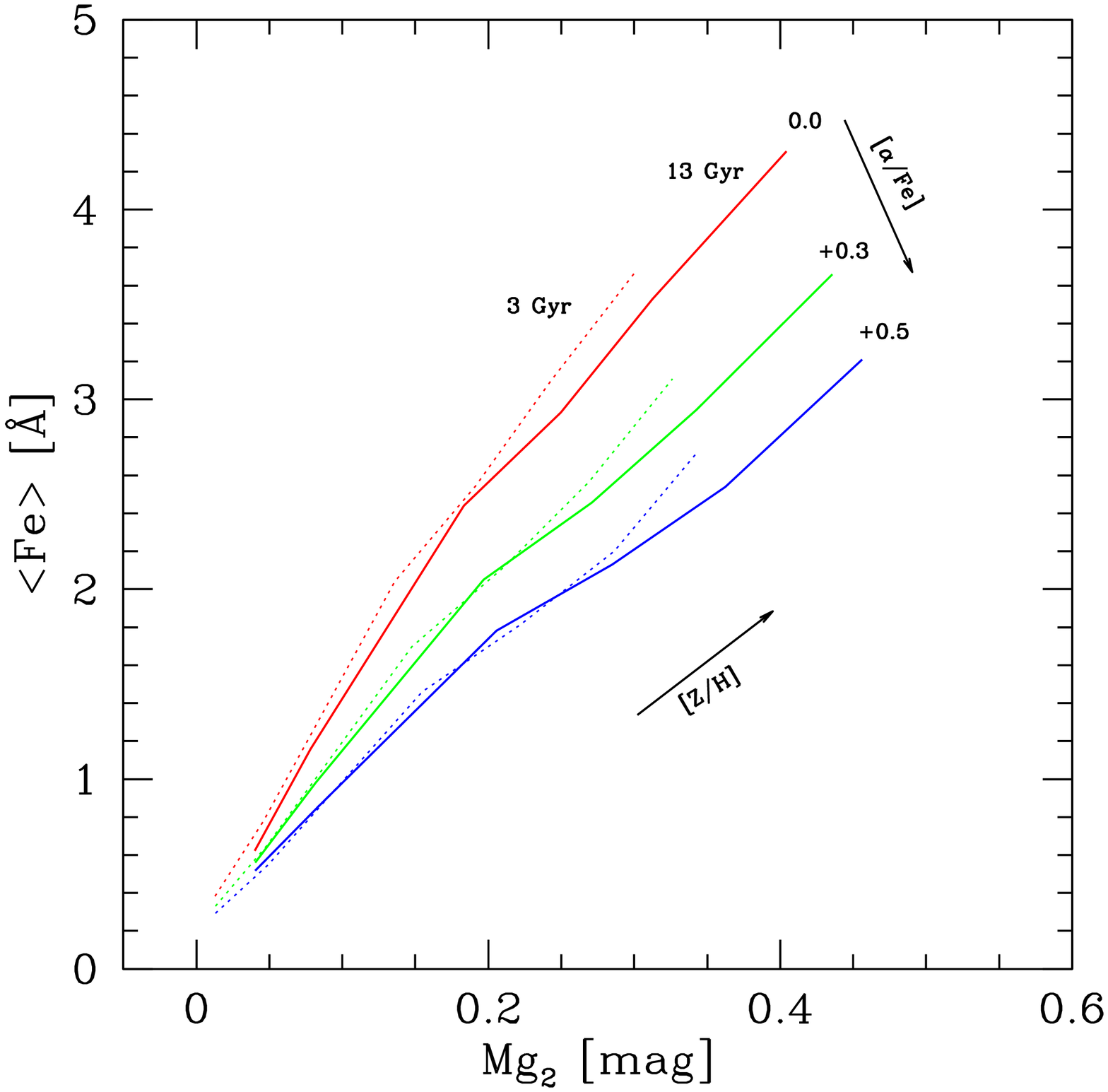}
\caption{Comparison of diagnostic grids for Lick index measurements, 
parameterized for different [$\alpha$/Fe] abundance ratios and ages. Model
predictions were taken from \cite{tmb03, tmk04}. The left and
middle panel show two grids constructed from [MgFe]\arcmin\ as a
metallicity indicator and two different Balmer line indices as age
indicators, i.e. H$\gamma_{\rm A}$ and H$\beta$, respectively. The plots
show models grids for two different values of [$\alpha$/Fe]~$=0.0$
(solid lines) and 0.5 dex (dashed lines). Note the
increasing differential offset between the grids towards higher
metallicities. The right panel shows the [$\alpha$/Fe] diagnostic diagram,
which features the Mg$_{2}$ index as a representative $\alpha$-element
indicator and $<$Fe$>$ as a iron-abundance proxy. The diagram shows the
dependence of iso-[$\alpha$/Fe] as a function of age. Dotted lines show
the model predictions for 3 Gyr old populations and solid lines for a 13
Gyr counterpart.}
\label{ps:comp}
\end{center}
\end{figure*}

\subsection{Comparison with SSP Models}
In the following Section we compare line index measurements for M31
globular clusters and background fields with predictions of SSP models,
which were calculated for stellar populations with well-defined
[$\alpha$/Fe] abundance ratios \citep{tmb03,tmk04}. These models are based
on \cite{worthey94} and \cite{worthey97} fitting functions, response
functions from \cite{tripicco95} and Korn et al. (2004, in preparation),
solar-scaled stellar evolutionary tracks, and employ the principles of the
fuel consumption theorem \citep{renzini86}.

Spectroscopic ages and metallicities of unresolved globular clusters are
traditionally derived from Balmer line indices and a combination of metal
indices. According to \cite{puzia03proc} and \cite{puzia04b} the best
combination of Lick indices to construct an age-metallicity diagnostic
diagram depends on internal uncertainties of the Lick system and the data
quality. In our case the most reliable combination is H$\gamma_{\rm A}$
and [MgFe]\arcmin \footnote{[MgFe]\arcmin~$=\sqrt{{\rm Mg}b\,
(0.72\cdot{\rm Fe5270}+0.28\cdot{\rm Fe5335})}$.}. The combination of
magnesium and iron-sensitive indices, in the form of the composite index
[MgFe]\arcmin, is an excellent metallicity tracer \citep{puzia02} and
shows the weakest sensitivity to [$\alpha$/Fe] abundance ratio variations
\citep{tmb03}. Among all Balmer Lick indices, H$\gamma_{\rm A}$, followed
by H$\beta$ and H$\delta_{\rm A}$, are least affected by internal
systematics, such as systematic uncertainties in the fitting functions.
However, we note that all Balmer indices are influenced by abundance ratio
variations (e.g. $\alpha$/Fe), in particular the higher-order Balmer
indices. Although response functions exist only for a selected set of
elements \citep{tripicco95}, with the new SSP model predictions these
variations are under control to first order.

Considering the signal-to-noise of a dataset, the accuracy of the calibration to
the Lick system, and internal systematic uncertainties of Balmer Lick
indices, \cite{puzia04} computes a reliability parameter ${\cal R}$, that
indicates the quality of a Balmer Lick index as an age indicator. Based on
this classification we have decided to derive ages and metallicities from
diagnostic diagrams constructed from the Balmer indices H$\gamma_{A}$,
H$\beta$, and H$\delta_{A}$, only. The narrowly-defined Balmer indices,
H$\gamma_{F}$ and H$\delta_{F}$, have low signal-to-noise and measure the
same absorption features as H$\gamma_{A}$ and H$\delta_{A}$ (hence, do not
provide additional information). They have the lowest reliability
parameter ${\cal R}$ among all Balmer indices \citep[see][]{puzia04}, so
that we refrain from including them in subsequent analyses.

\subsection{Fitting Routine}
To derive reliable ages and metallicities, we need to utilize a diagnostic
grid from which the [$\alpha$/Fe] ratio of the observed stellar population
can be derived. We tested several combinations of Lick indices and, based
on the internal systematics, the relative separation of iso-[$\alpha$/Fe]
tracks, and the mean measurement error, decided on an [$\alpha$/Fe]
diagnostic grid that is constructed from the indices Mg$_{2}$ and
$<$Fe$>$. The Mg$_{2}$ index measures the strength of the Mg$b$ and MgH
features at $\sim5100$ \AA. The mean iron index $<$Fe$>$ is a composite
index built from the indices Fe5270 and Fe5335, i.e. $<$Fe$>=\sqrt{{\rm
Fe5270}+{\rm Fe5335}}$, which trace two strong iron lines at similar
wavelengths ($\lambda\ 5270, 5335$ \AA) as Mg$_{2}$. However, since the
[$\alpha$/Fe] diagnostic grid itself is degenerate in age and metallicity,
the best possible determination of ages, metallicities, and [$\alpha$/Fe]
ratios requires an iterative fitting of the data, feeding information
acquired from the $\alpha$/Fe-grid into the age/metallicity-grid.

Extensive tests have shown that the values derived from both diagnostic
diagrams converge after a few iterations. Our routine starts with the
determination of [$\alpha$/Fe] ratios from the Mg$_{2}$ vs. $<$Fe$>$
diagram, assuming a 13 Gyr old population, and is based on an $\chi^{2}$
minimization technique. The age and metallicity of each globular cluster
is then determined with the same $\chi^{2}$ minimization approach on an
individually interpolated\footnote{We linearly interpolate the grid in age
and Z-space (not metallicity log-space).} H$\gamma_{\rm A}$, H$\beta$, and
H$\delta_{\rm A}$ vs. [MgFe]\arcmin\ grid for a given [$\alpha$/Fe]. For
globular clusters with index measurements that do not fall on the
diagnostic grid, we assign the most likely value, extrapolated along the
most likely error vector which is constructed from the square sum of
individual index uncertainties for each globular cluster in a given
diagnostic grid. The age and metallicity values are then fed back into the
[$\alpha$/Fe] diagram and the process is iterated. In most cases this
process requires only a few cycles for the values to converge. This method
is superior to any age-metallicity determination based on the
age-metallicity grid {\it only}. As shown in Figure \ref{ps:comp}, the
isochrones of the age-metallicity grid can change significantly up to a
few typical 1-$\sigma$ measurement errors as a function of [$\alpha$/Fe]
\citep[see also][]{tmb03, tmk04}. To control these effects and obtain
meaningful age/metallicity results it is imperative to include abundance
ratio variations in each age-metallicity determination based on Lick index
measurements.

Ages, metallicities, and [$\alpha$/Fe] ratios for all M31 sample
globular clusters are summarized in Table~\ref{tab:ama} in the Appendix.

\subsection{Consistency Check}
\begin{figure*}[!th]
\begin{center}
\includegraphics[width=5cm, bb=40 130 600 700]{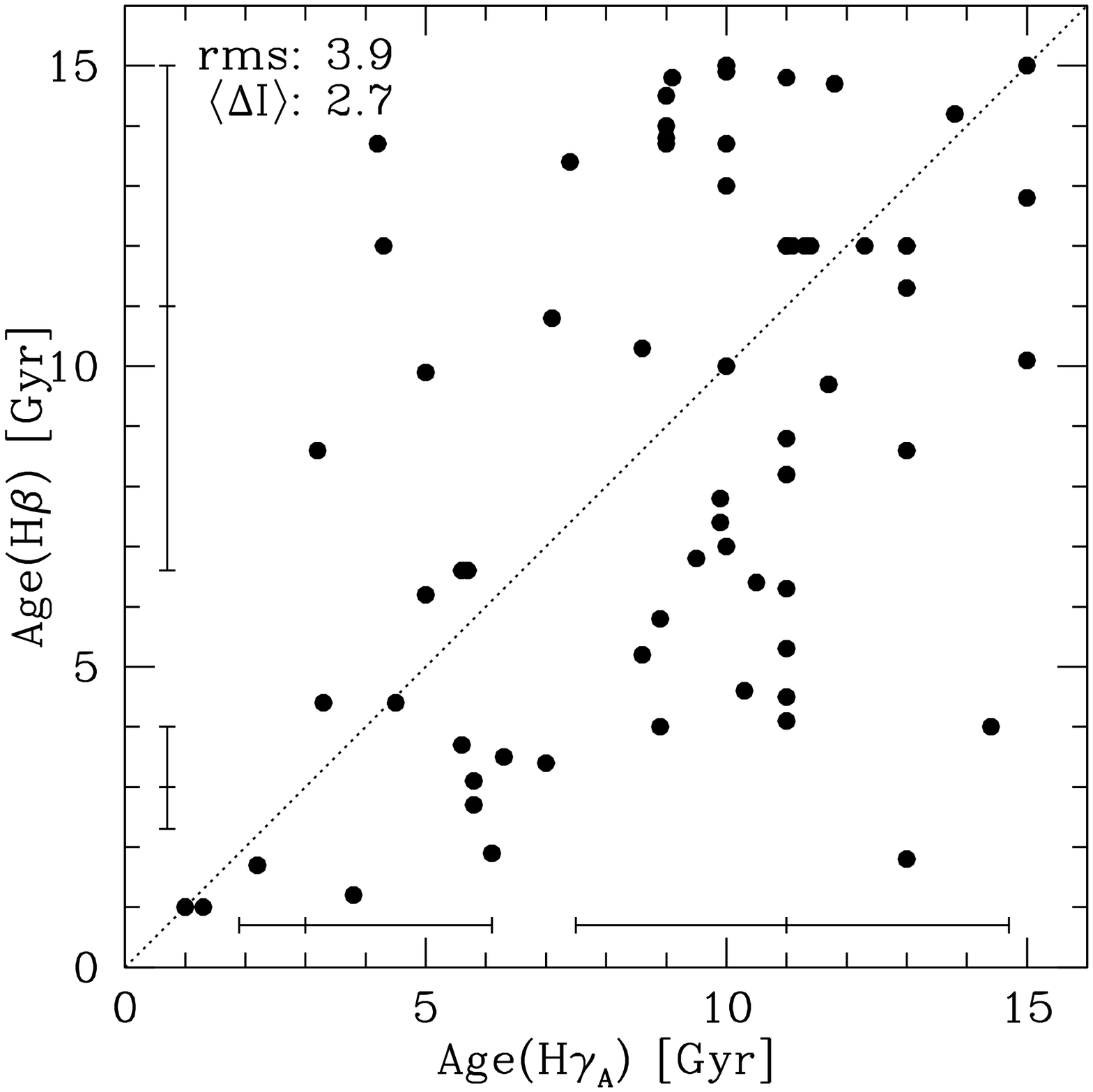}
\includegraphics[width=5cm, bb=40 130 600 700]{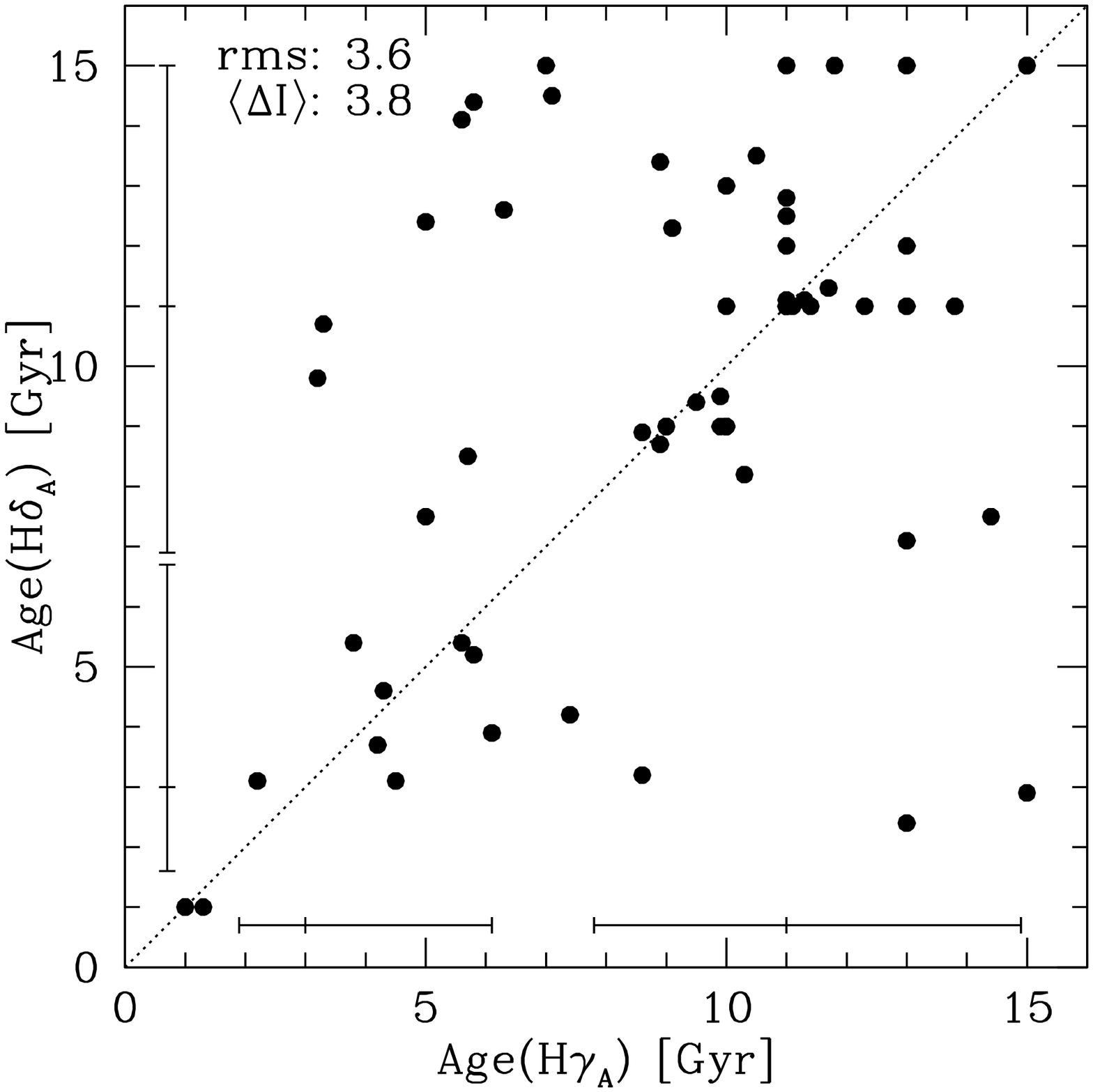}
\includegraphics[width=5cm, bb=40 130 600 700]{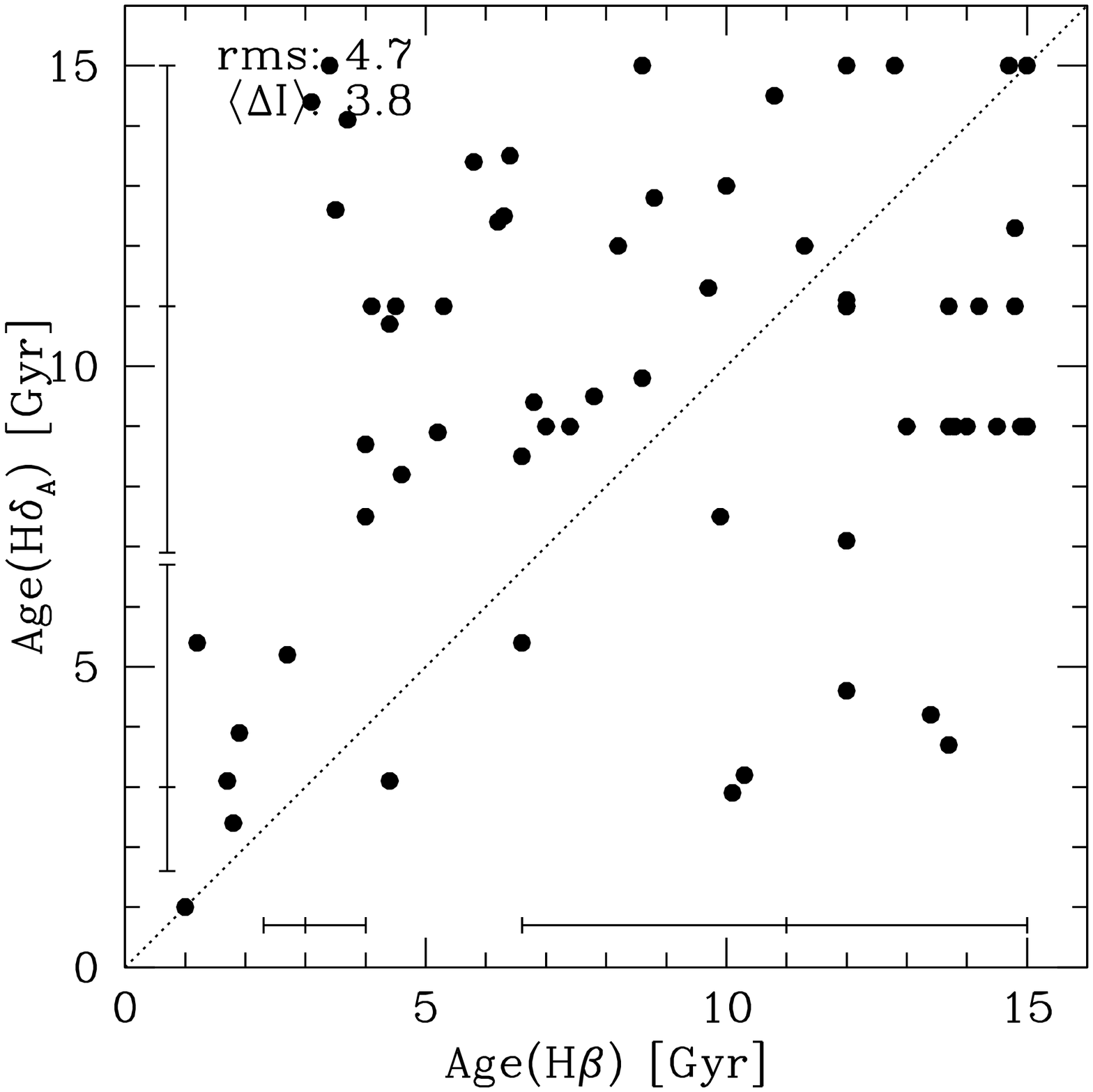}
\includegraphics[width=5cm, bb=40 130 600 700]{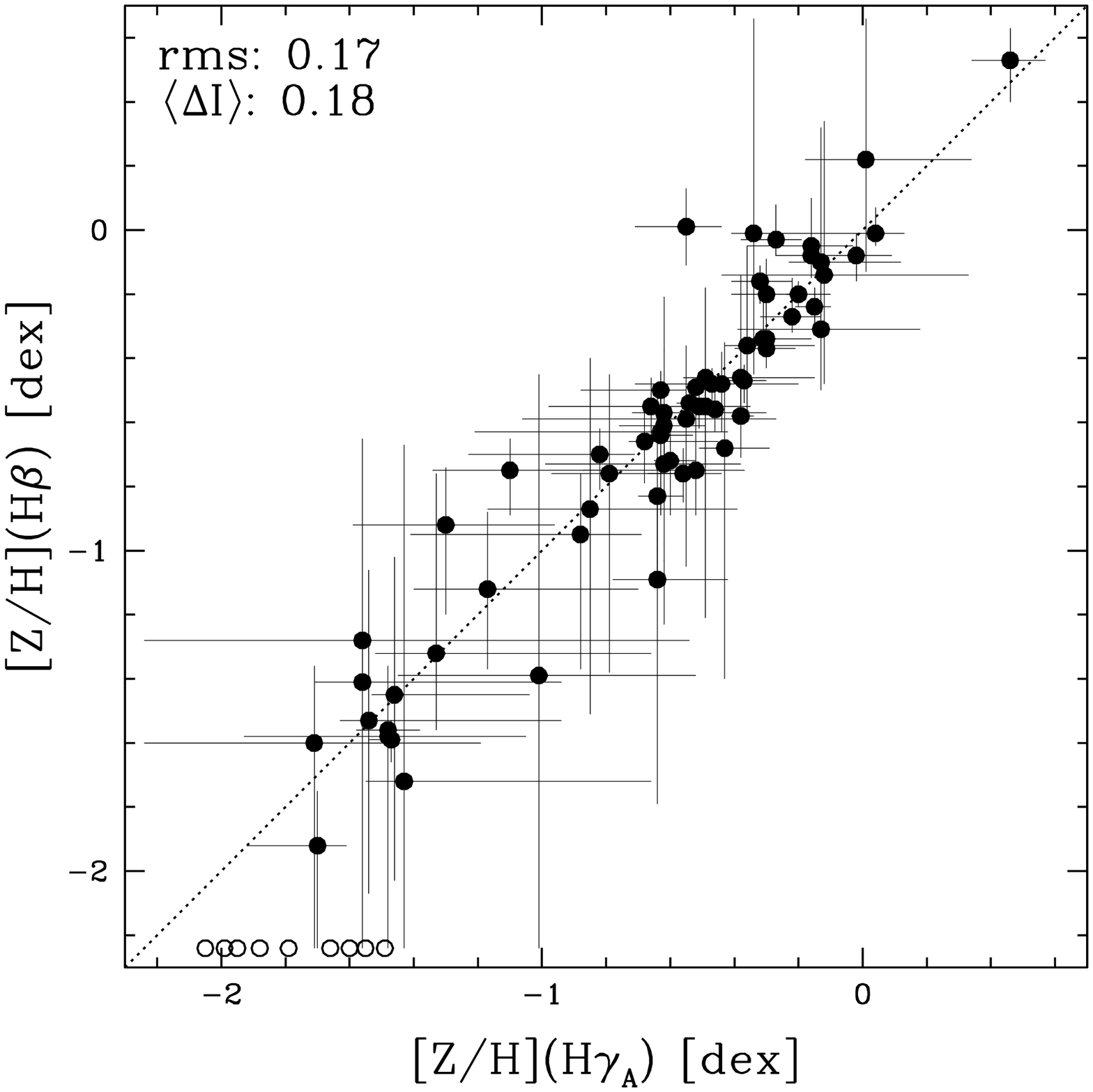}
\includegraphics[width=5cm, bb=40 130 600 700]{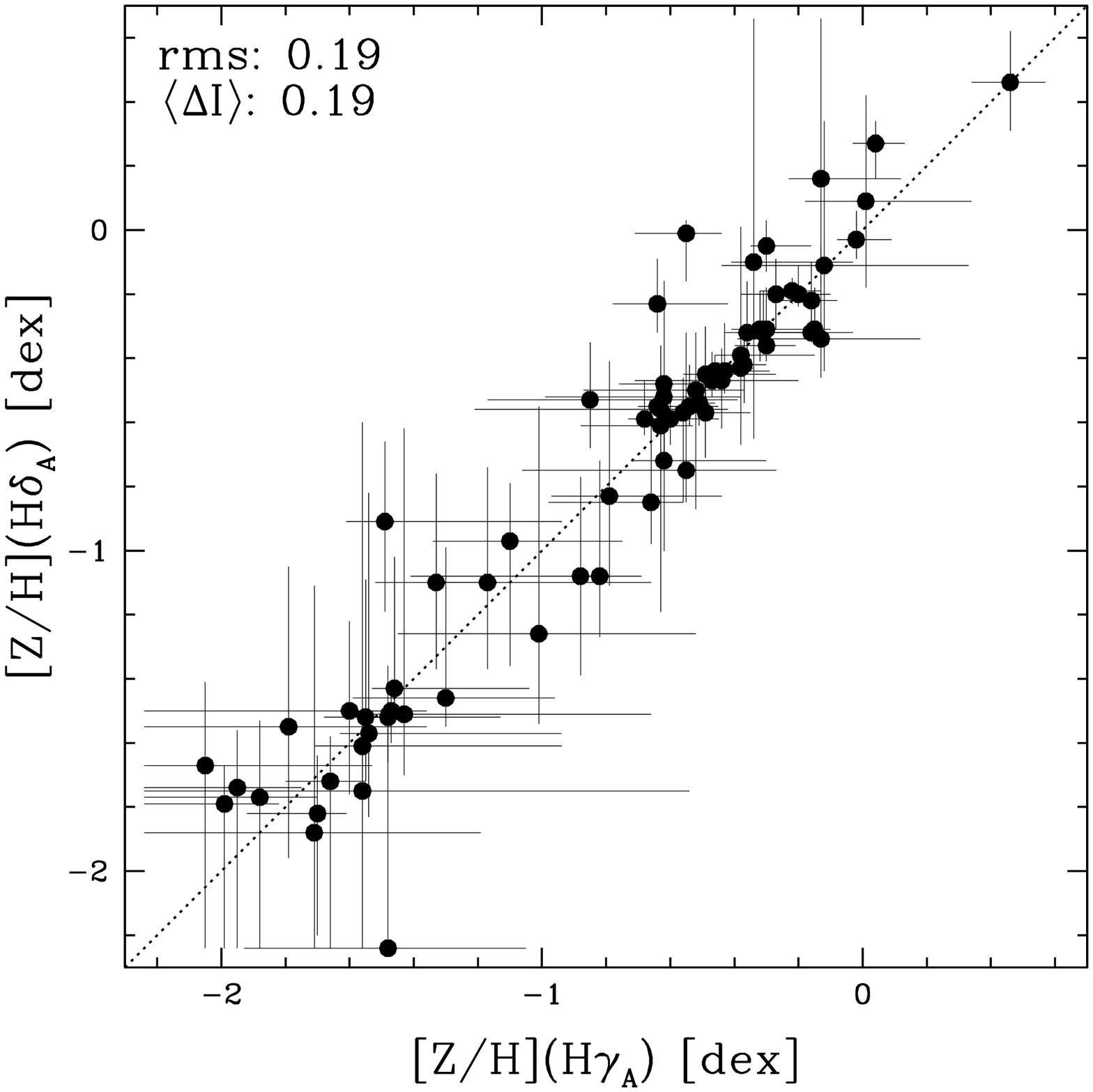}
\includegraphics[width=5cm, bb=40 130 600 700]{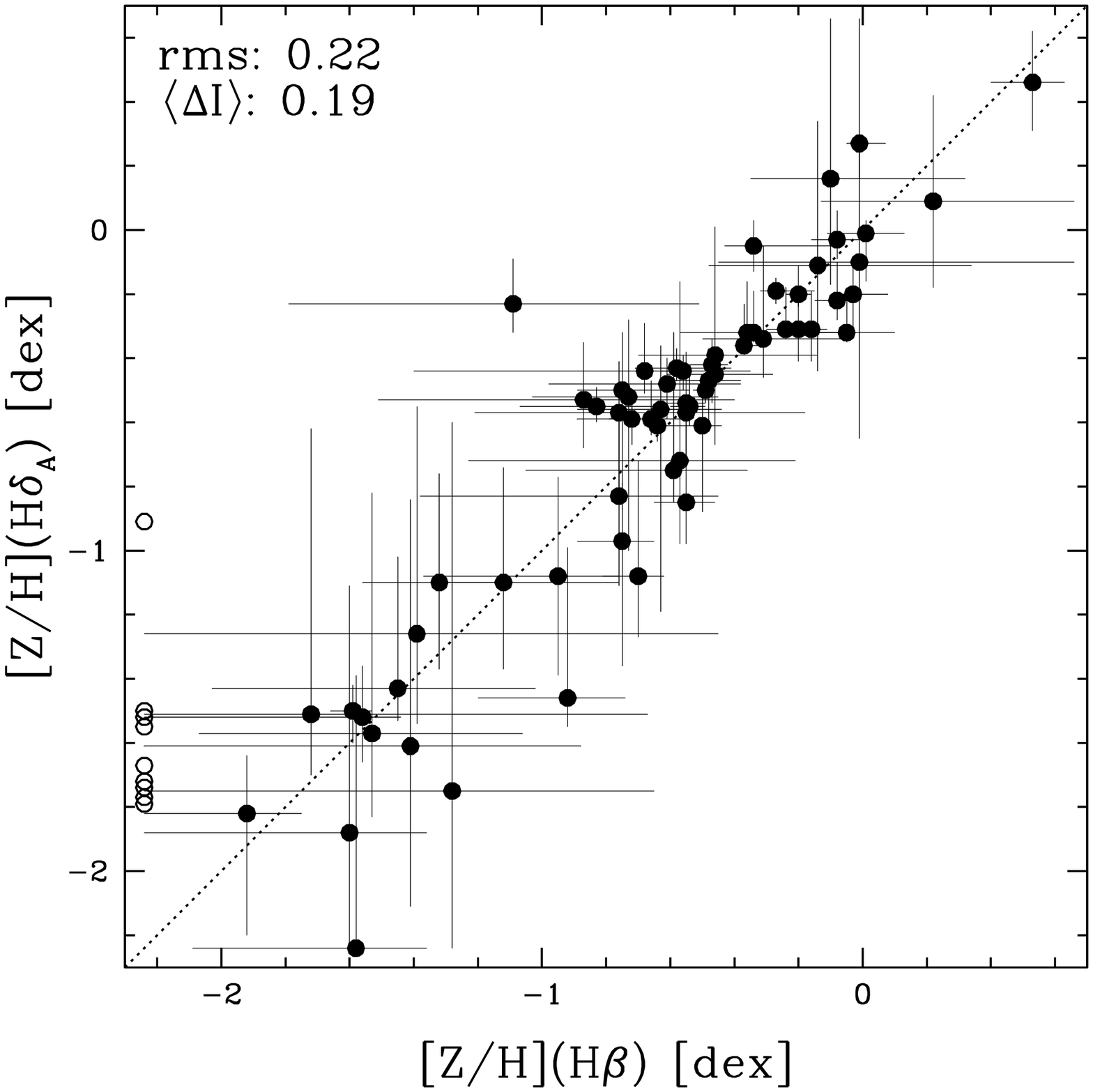}
\caption{Comparison of age and metallicity predictions from the three different
age-metallicity diagnostic grids shown in Figure~\ref{ps:nmgfehga}. Upper
panels show the comparison of ages; lower panels illustrate the comparison
of metallicity values. The rms scatter is indicated in the upper left
corner of each plot and is the scatter around the one-to-one relation.
$\langle\Delta I\rangle$ indicates the median error for this panel. In the
upper panels median error bars are indicated at the abscissa and ordinate,
calculated as the sample median of the averaged globular cluster
asymmetric error. Open circles mark outliers which were not included in
the rms determination.}
\label{ps:agemetcomp}
\end{center}
\end{figure*}

In Figure~\ref{ps:agemetcomp} we compare age and metallicity
determinations from the three diagnostic grids using different Balmer-line
indices, H$\gamma_{A}$, H$\beta$, and H$\delta_{A}$, as age indicators
(see Fig.~\ref{ps:nmgfehga}). The rms scatter in the comparison of ages
derived from two different grids varies between 3.6 and 4.7 Gyr. Note that
the age scatter increases from young to old ages, which is an effect of an
increasing age resolution with decreasing absolute age. Given the quality
of our dataset, the rule of thumb for the average age uncertainty is
$\Delta t/t\approx 1/3$. The consistency check shows that the rms-scatter
in all upper panels of Figure~\ref{ps:agemetcomp} is in most cases
comparable or larger than the mean measurement error. This is an
indication for the presence of systematic offsets in the absolute age
scale between two diagnostic grids. The rms-scatter in all metallicity
consistency plots agrees very well with the mean measurement error, and
shows the good absolute metallicity calibration of individual diagnostic
grids. It is not too surprising that the metallicity values derived from
different grids agree very well for individual globular clusters, since
all plots use [MgFe]\arcmin\ as metallicity proxy. A good estimate for the
mean metallicity uncertainty is 0.2 dex.

The consistency shown between the model predictions should be taken as a
demonstration of the quality of age/metallicity predictions that are
feasible with today's high-quality datasets and SSP models. Rather than
being an endorsement of the particular model used in this study, we point
out that SSP model predictions for stellar populations with well-defined
abundance ratios are essential to reach the same quality of
age/metallicity determinations as presented here. Other SSP models which
do not take the systematics of abundance-ratio variations into account may
lead to similarly good fits for a given diagnostic grid, but the results
will suffer from systematic uncertainties described earlier. However, even
for state-of-the-art SSP models the litany of systematic uncertainties is
long\footnote{Systematic uncertainties are still hard-wired into {\it
all} current SSP Lick-index predictions and might, if accounted for,
significantly improve future age, metallicity, and abundance ratio
estimates. To name a few ''to-dos'': control of systematic uncertainties
introduced by variations in horizontal branch morphology, self-consistent
use of enhanced stellar evolutionary tracks/isochrones in SSP models,
extension of response functions to more elements, self-consistent
calculation of internal energy production for various chemical
compositions, etc.}.

In general, we find good congruence between age and metallicity
predictions between diagnostic plots using the three indices
H$\delta_{A}$, H$\gamma_{A}$, and H$\beta$. The relative age accuracy
($\sim3-5$ Gyr) allows us to disentangle old from intermediate-age and
young globular cluster sub-populations within a wide range of metallicities.

\section{Results}
\label{ln:results}
\subsection{Ages and Metallicities}
\label{ln:am.res}

\begin{figure*}[!t]
\begin{center}
\includegraphics[width=7.8cm, bb=40 130 600 700]{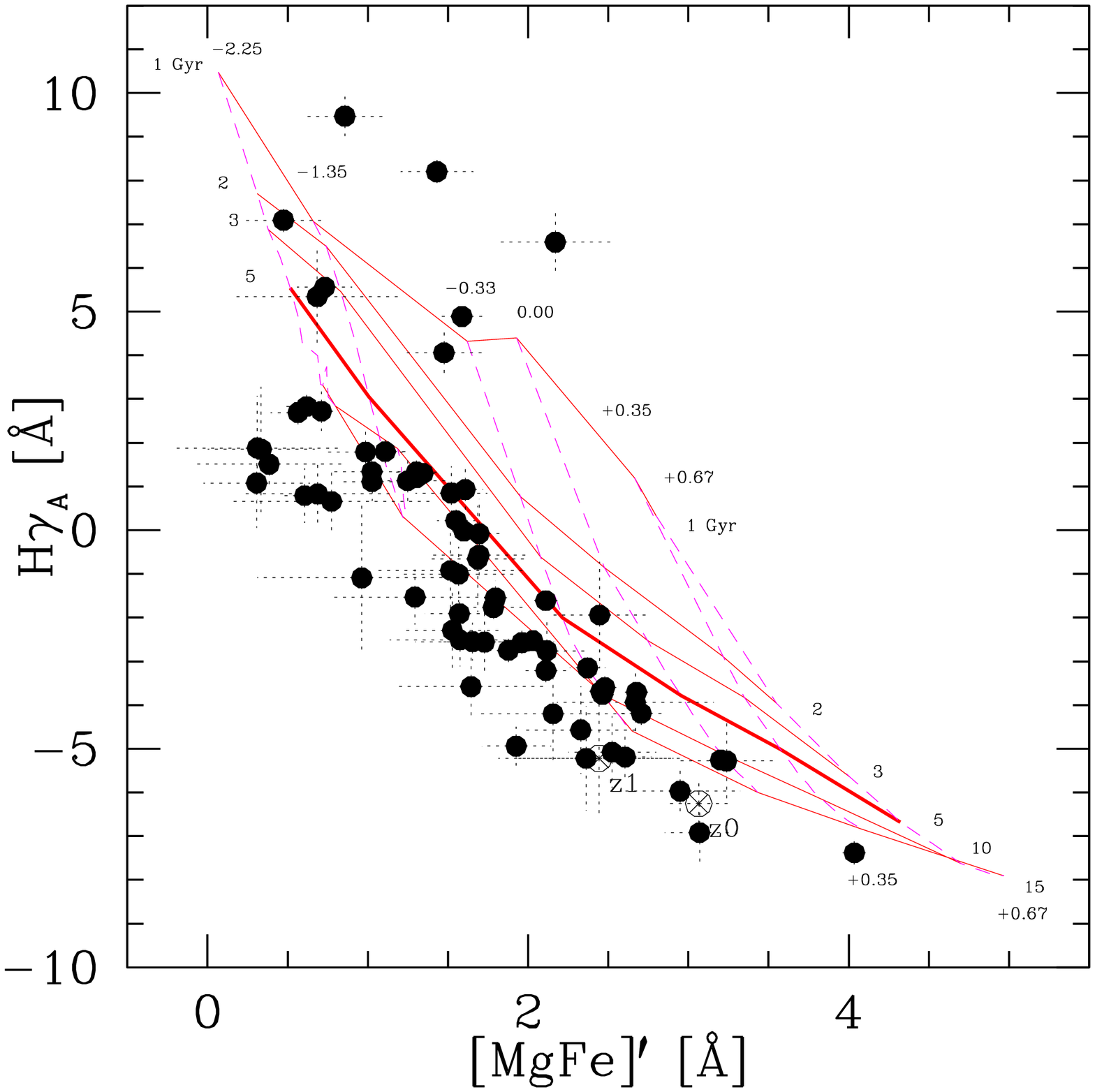}
\includegraphics[width=7.8cm, bb=40 130 600 700]{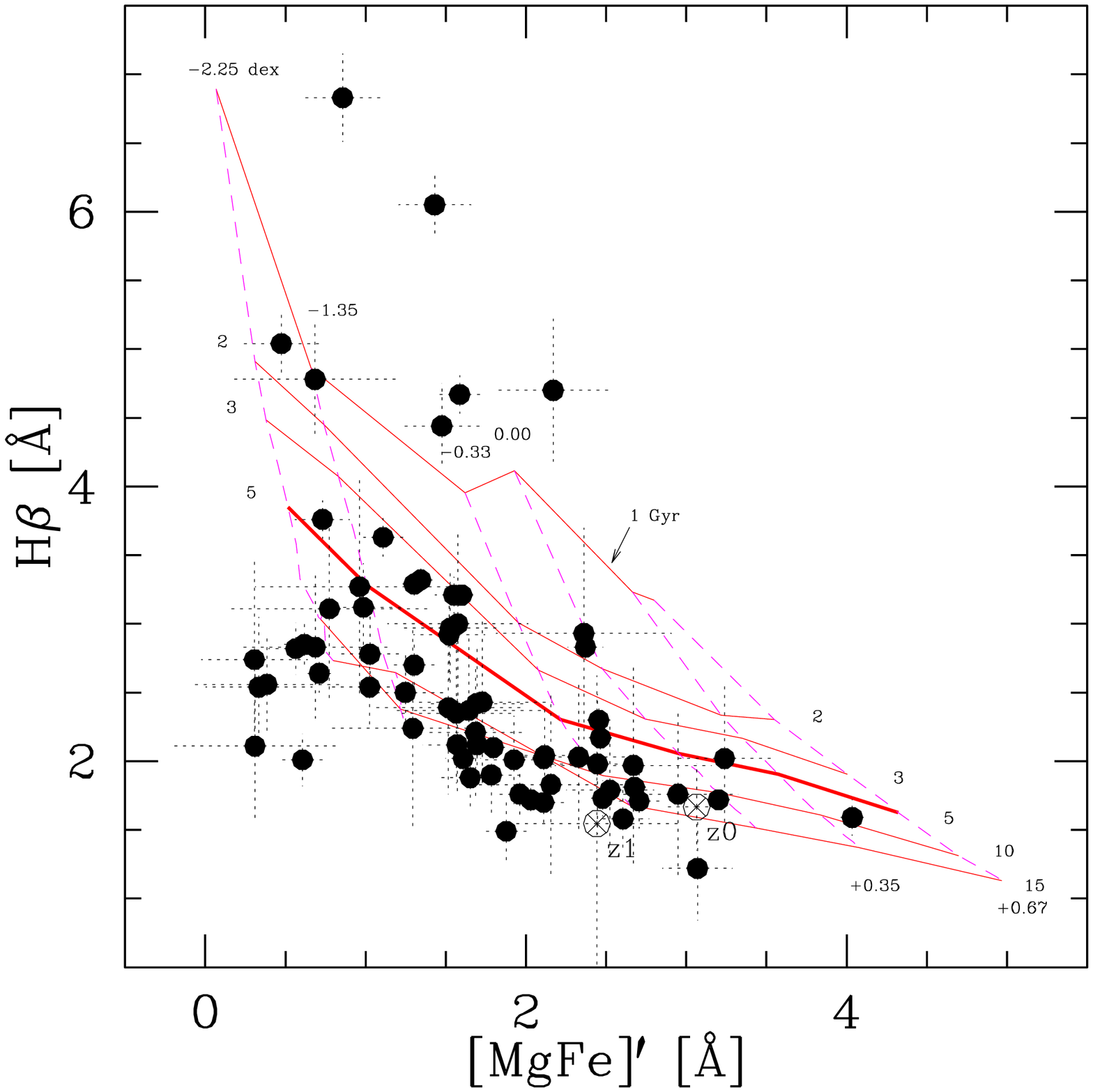}
\includegraphics[width=7.8cm, bb=40 130 600 700]{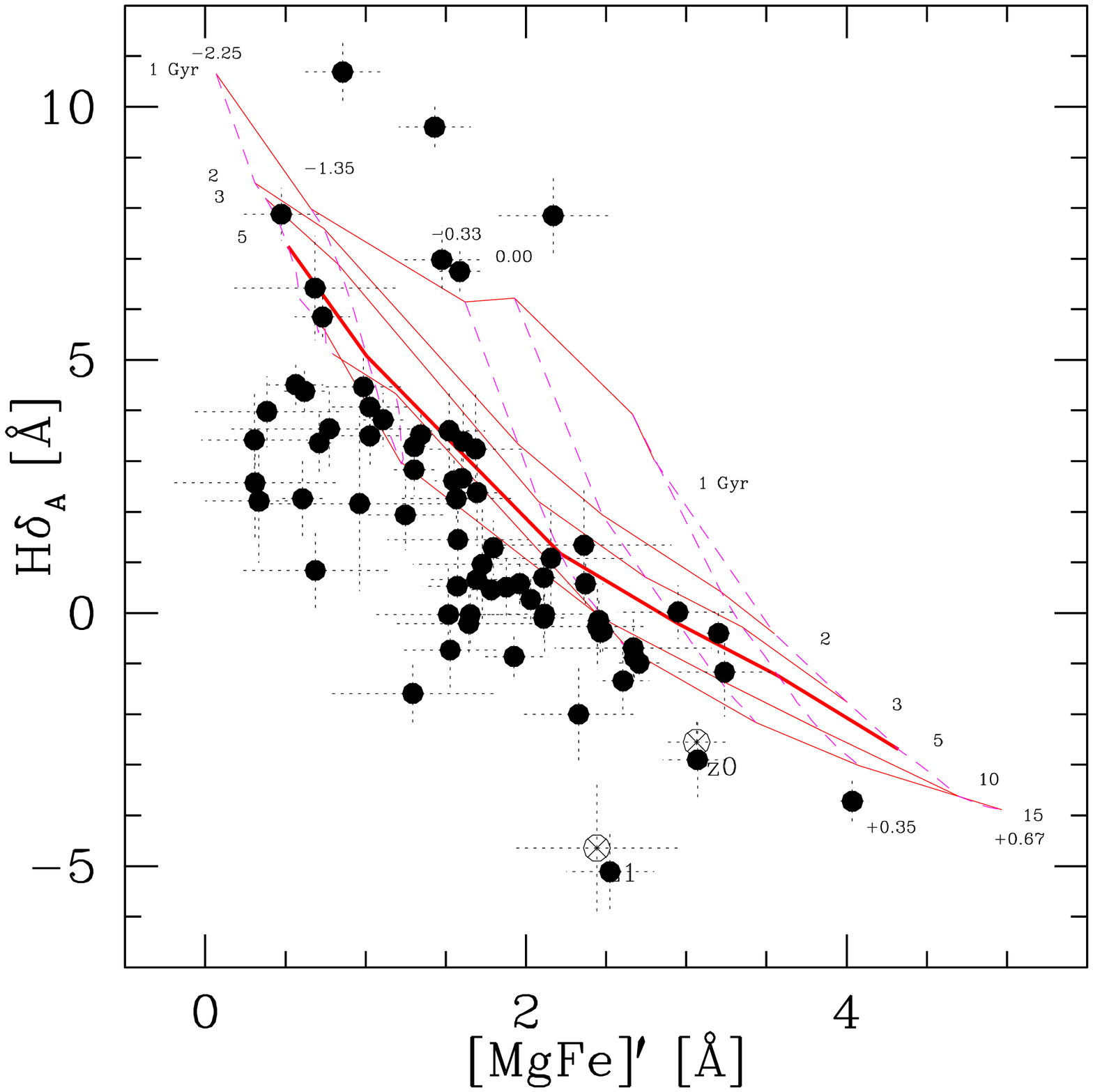}
\includegraphics[width=7.8cm, bb=40 130 600 700]{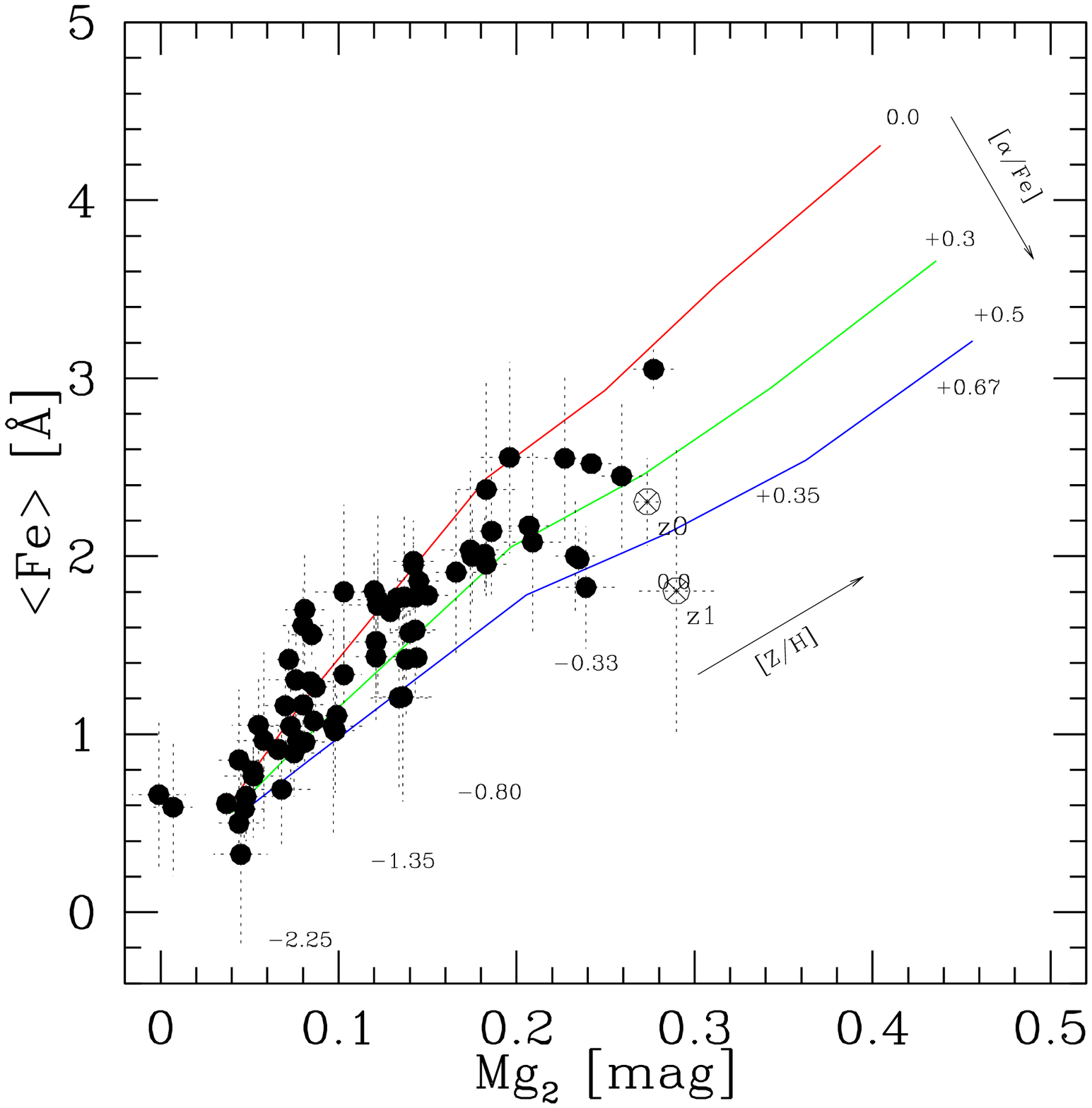}
\caption{Diagnostic Plots: H$\gamma_{A}$, H$\beta$, and H$\delta_{\rm A}$ vs.
[MgFe]\arcmin\ age-metallicity diagnostic plot for stellar populations
with [$\alpha$/Fe]~$=0.3$ dex. The lower right panel shows the
[$\alpha$/Fe] diagnostic plot with Mg$_{2}$ vs. $\langle$Fe$\rangle$.
Solid circles are M31 globular clusters. SSP model predictions were taken
from \cite{tmb03, tmk04} and are shown in the age-metallicity grids as solid
isochrones for stellar populations with ages 1, 2, 3, 5, 10, and 15 Gyr
and as dashed iso-metallicity tracks for metallicities [Z/H]~$=-2.25,
-1.35, -0.33, 0.0, +0.35$, and $+0.67$ dex, for a stellar population with
[$\alpha$/Fe]~$=0.3$ dex. The SSP model in the lower right panel shows
iso-[$\alpha$/Fe] tracks for metallicities [Z/H]~$=-2.25$ to $+0.67$ and an
age of 13 Gyr. Open circles mark M31 bulge fields at different
galactocentric radii, see Table~\ref{tab:bkgrad}. NB87, which has the
highest metallicity among our sample globular clusters, is the only object
with [MgFe]\arcmin\ $\approx 4$ \AA\ and $\langle$Fe$\rangle\approx 3$
\AA. It should be stressed again that the shown diagnostic grids are not
suited to compare individual globular clusters, because of the
$\alpha$/Fe-sensitivity of Balmer indices. All grids have
to be interpolated for each individual globular cluster according to its
[$\alpha$/Fe]. However, H$\beta$ shows the weakest response to
$\alpha$/Fe variations (see Fig.~\ref{ps:comp}).}
\label{ps:nmgfehga}
\end{center}
\end{figure*}

Figure \ref{ps:nmgfehga} shows the M31 globular cluster data, indicated by
solid circles, together with theoretical SSP model predictions taken from
\cite{tmb03, tmk04}. The model grid covers ages from 1 to 15 Gyr and
metallicities [Z/H] from $-2.25$ to $+0.67$ dex, indicated by solid and
dashed lines respectively. The 5 Gyr isochrone is indicated as a thicker
line to guide the eye when comparing diagnostic plots constructed with
different Balmer-line indices. We note that direct comparisons of data and
diagnostic grids are misleading when the [$\alpha$/Fe] abundance of each
individual globular cluster is not taken into account (see previous
Section). However, to allow an assessment of the {\it mean} age and
metallicity distribution of all globular clusters and to check for
consistency of these predictions from all three diagnostic diagrams, we
plot the data together with diagnostic grids for stellar populations with
[$\alpha$/Fe]~$=0.3$ dex.

The majority of our sample of M31 globular clusters are consistent with
old ages, However, the sample spans a wide range of ages, with $\sim10$\%
of the entire sample younger than 5 Gyr. Several globular clusters have
index values which are consistent with formal ages similar or less than
$\sim1$ Gyr at an average metallicity $\langle$[Z/H]$\rangle\approx -0.4$
dex. Unfortunately, the models do not cover ages below 1 Gyr and we do
{\it not} attempt to extrapolate the grid, being aware of the potential
uncertainties introduced by the AGB-phase transition which occurs at ages
at or below $\sim1$ Gyr. Depending on the total metallicity, the AGB can
contribute $\sim50$\% to the total bolometric light at these young ages
\citep{renzini86, ferraro95}. Theoretical models show that, although the
fractional contribution of the AGB to the total light is subject to only
small model-to-model fluctuations, the exact onset and duration of the
AGB-phase transition depends on the treatment of classical mixing vs.
overshooting \citep[e.g.][]{alongi93}.

Clusters with such young ages are consistent with the findings of
\cite{beasley04} who derive ages for their sub-sample of young globular
clusters from a comparison with empirical spectra of young LMC globular
clusters. This comparison is backed by their H$\beta$ measurements being
consistent with ages between 100 and 800 Myr within the latest \cite{bc03}
SSP models\footnote{Note that the \cite{bc03} models do {\it not} include
predictions for stellar populations with a given [$\alpha$/Fe] ratio,
contrary to the models of \cite{tmb03, tmk04}. However, given the young
ages, the sensitivity of the H$\beta$ index to [$\alpha$/Fe] variations is
relatively weak, assuming the changes in chemical composition are
correctly modeled. Hence, to first order the age-metallicity predictions
of \cite{beasley04} should be relatively robust. This is not the case for
higher-order Balmer Lick indices \citep{puzia04b}.}.

We find several M31 globular clusters with near-solar metallicities and
one object (NB87) reaching a formal [Z/H]~$\approx+0.5\pm0.1$ dex at a
formally derived age of $9\pm2$ Gyr, and a [$\alpha$/Fe]~$=0.09\pm0.05$.
It is the most metal-rich globular cluster in our sample. Because of its
relatively high metallicity, this object might host a stellar population
representative of those in massive early-type galaxies, although it is
slightly under-abundant in $\alpha$-elements.

\subsubsection{Age-Metallicity Distributions and Correlations}
In the following we compute mean ages and metallicities for our sample
globular clusters using the information acquired from all three diagnostic
grids. Based on the ${\cal R}$-parameter ranking and the sensitivity
of Balmer lines to abundance variations \citep[see][]{puzia04}, we assign
a weighting of 1.0 to values obtained from the H$\gamma_{A}$ grid and a
weighting factor of 0.5 to results obtained from the H$\beta$ and
H$\delta_{A}$ grids. The reason for this weighting pattern are internal
systematics in the age-metallicity diagnostic grids utilizing different
Balmer-line indices \citep[for details see][]{puzia04}. 

\begin{figure}[!ht]
\begin{center}
\includegraphics[width=8.2cm, bb=40 150 560 700]{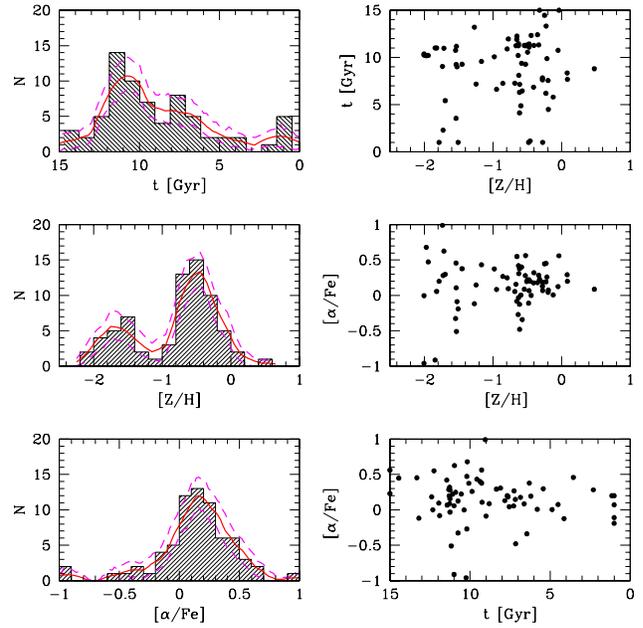}
\caption{Ages, metallicities, and [$\alpha$/Fe] ratios for M31 globular clusters.
Left panels show histograms for age, metallicity, and [$\alpha$/Fe]
ratios for our sample globular clusters. Solid lines represent non-parametric
probability density estimates using a variable-width Epanechnikov kernel
\citep{silverman86} with their 90\% confidence limits (dashed lines). Right 
panels show correlations between the three parameters. Error bars are
not shown for clarity; the mean error in age is $\Delta t/t\approx 1/3$, the
mean uncertainty for metallicity and [$\alpha$/Fe] is $\sim0.2$ dex.}
\label{ps:amahistos}
\end{center}
\end{figure}

Using this averaging prescription, we construct age and metallicity
histograms for the M31 globular cluster sample, which are shown in
Figure~\ref{ps:amahistos}. The upper left panel shows the age distribution
of observed M31 globular clusters. A clear peak around $10-12$ Gyr
dominates the age structure. The majority of the globular cluster sample
($\sim60$\%) has formal ages older than 9 Gyr. A surprising feature of the
age structure in the observed M31 globular cluster system is a population
of intermediate-age objects with formal ages between 6 and 8 Gyr. The mean
uncertainty at those ages varies between 2 and 3.5 Gyr, so that it is
unlikely that these objects have much older or younger ages. In
particular, this feature is not reproduced by the overlap of old iso-age
tracks, which is the result of increasing flux from hot horizontal-branch
stars with increasing age, at metallicities below $\sim-0.6$ dex. Only six
globular clusters are affected by the region where the age determination
is ambiguous. For these clusters the code randomly assigns either old or
young age values.

After clipping globular clusters with ages $\leq1$ Gyr, a heteroscadastic
KMM test\footnote{Two Gaussian modes with independent mean values and
dispersion are fit to the data, as opposed to a homoscedastic fit where
two modes with independent mean values but same dispersions are adopted
during the fit.} \citep{ashman94} shows that a bimodal age distribution is
more likely than a single-peak Gaussian distribution with a mean at
$\sim11$ Gyr. We find a 0.1\% likelihood for the single-peak distribution
and point out that a bimodal distribution is also more likely than a
tri-modal fit. For the bimodal distribution the KMM test gives peaks at
$t_{1}=10.9\pm0.3$ and $t_{2}=6.5\pm0.4$ Gyr\footnote{The given errors
are errors of the mean. Systematic errors are larger and of the order of
2--3 Gyr. For a homoscedastic KMM test the peak values change only
marginally within the uncertainties. The mean dispersion of both peaks is
$2.4$ Gyr.} with dispersions 1.7 and 3.5 Gyr, respectively. The integrity
of the intermediate-age peak persists when we overplot a non-parametric
probability density estimate with its 90\% confidence limits. We use a
non-parametric variable-width Epanechnikov kernel \citep{silverman86} that
is consistent with the mean age uncertainty of each individual globular
cluster.

In the left middle-row panel of Figure~\ref{ps:amahistos} we present the
metallicity distribution of studied M31 globular clusters. The
distribution is clearly bimodal stretching from around $-2.2$ dex to
values slightly above solar. A heteroscedastic KMM-test gives peaks at
$-1.66\pm0.05$ and $-0.45\pm0.04$ dex with dispersions 0.23 and 0.29 dex,
respectively. From an empirical calibration of optical/near-IR colors
with Milky Way globular clusters, \cite{barmby00} find peaks around
$-1.4\pm0.05$ and $-0.6\pm0.05$ dex\footnote{These metallicity values are
on the Zinn-West scale, which is generally adopted throughout this work.}.
We note that our sample is biased towards metal-rich globular clusters
which is a consequence of the sample selection. The latter was driven by
the observational setup and focused mainly on M31's disk and bulge
\citep[see][]{perrett02, beasley04}. Hence, the mean of each peak and its
dispersion are subject to change for larger samples. In particular, the
ratio between metal-poor and metal-rich globular clusters is not expected
to be representative for the entire globular cluster system. This is at
least partly the reason for the offset between our spectroscopic and
\citeauthor{barmby00}'s photometric mean metallicity in the metal-poor
peak. However, it is less likely to explain the $\sim3\sigma$ offset for
the metal-rich peak entirely, which might be a consequence of different
absolute metallicity scales of the used SSP models.

In the upper right panel of Figure~\ref{ps:amahistos} we present the
age-metallicity correlation plot. We find old globular clusters at all
metallicities. For globular clusters with formal ages below $\sim8$ Gyr,
the evidence for metallicity bimodality remains. A group of five clusters
(B315-038, B321-046, B322-049, B327-053, and B380-313) with ages below
$\sim5$ Gyr and metallicities around $-1.8$ dex appears as a very distinct
feature in the plot. The other metal-rich young objects are clustered
around [Z/H]~$\approx -0.6$ to $-0.2$ dex. If the group of five young
metal-poor globular clusters is excluded, the overall age dispersion
increases from $\sigma_{\rm mp}\approx 1.3$ Gyr for objects with
metallicities below $-1.0$ dex to $\sigma_{\rm mr}\approx 3.4$
Gyr for metal-rich counterparts. The mean
metallicity of intermediate-age globular clusters with ages between 3 and
8 Gyr is $-0.61\pm0.11$ dex with a dispersion of $0.48$ dex. Without two
clusters from the young metal-poor population (B321-046 and B380-313), the
mean changes to $-0.49\pm0.08$ dex and a dispersion $0.34$ dex. There is
tentative evidence that this relatively metal-rich intermediate-age
globular cluster sub-population might have a bimodal metallicity
distribution in itself with peaks around $-0.6$ and $-0.2$ dex. Such a
multi-modality in the metallicity distribution of M31 globular clusters
was already discovered by \cite{barmby00} in their photometric study (see
their Fig.~19). It is also worthy of notice that the metal-poor peak of
this intermediate-age sub-sample coincides with the metal-rich peak of old
globular clusters, while the metal-rich intermediate-age clusters appear
to have $\sim0.4$ dex higher mean metallicities than the average
metal-rich old globular cluster. However, these results need follow-up
analyses with larger samples.

Irrespective of the completeness and metallicity coverage of our sample,
we do find subpopulations of old ($\sim10-12$ Gyr), intermediate-age
($\sim6-8$ Gyr), and few young ($\la2$ Gyr) globular clusters in M31. Our
results are fully consistent with the recent spectro-photometric study of
172 globular clusters by \cite{jiang03}, who find a significant population
of intermediate-age and young globular clusters (see their Fig.~6). Their
results also indicate that metal-rich globular clusters are on average
younger than metal-poor globular clusters. It would be presumptious to
speculate on the significance of this result {\it ex parte}, given our
limited sample size (70 out of $460\pm70$ globular clusters). With the
current dataset we cannot conclusively estimate the fractions of these new
types of intermediate-age and young globular clusters and assess whether
these sub-populations are small components or fundamental ingredients
of the M31 globular cluster system. In any case, the distribution of
globular cluster ages suggests that M31 managed to form star clusters
that survived as globular clusters, until a few Gyr ago.

\subsubsection{[$\alpha$/Fe] Ratios}
\label{ln:afe}
The determination of [$\alpha$/Fe] ratios yields very stable values within
a few iterations of our grid-interpolation routine, which is the result of
the small dependence of the [$\alpha$/Fe] diagnostic grid on age and
metallicity. The left bottom-row panel of Figure~\ref{ps:amahistos} shows
the distribution of [$\alpha$/Fe] ratios, for which a single-peak
distribution appears to be a good approximation. The mean of the
single-peak distribution is $0.14\pm0.04$ dex\footnote{Median: $0.18$ dex
with $0.15$ dex semi-interquartile range.} with a dispersion $\sigma=0.32$
dex. However, using the KMM test we find that a two-Gaussian distribution
is more likely (99.4\%) than the single-peak distribution. The test yields
a relatively narrow peak at [$\alpha$/Fe]~$=0.20\pm0.02$ dex with a
dispersion $\sigma=0.19$ dex, and a relatively broad peak at
$-0.03\pm0.16$ with a factor $\sim2.5$ larger dispersion of $\sigma=0.50$
dex. The mixture of these modes is six to one, where the super-solar
[$\alpha$/Fe] peak is more populated.

We can now correlate [$\alpha$/Fe] ratios of our sample globular clusters
with their ages and metallicities. The right middle-row and bottom-row
panels of Figure~\ref{ps:amahistos} illustrate the metallicity vs.
[$\alpha$/Fe] and age vs. [$\alpha$/Fe] plots. We find no correlation in
the metallicity vs. [$\alpha$/Fe] plot. Metal-poor and metal-rich clusters
have similar super-solar mean [$\alpha$/Fe] ratios within the uncertainty
of $0.1-0.2$ dex. We find an increasing dispersion towards lower
metallicities, which is mainly due to a decreasing [$\alpha$/Fe]
resolution at lower [Z/H]. The sample of the five clusters with ages
$\sim1$ Gyr and metallicities around $-1.6$ dex shows super-solar
[$\alpha$/Fe] ratio $+0.21\pm0.11$. Metal-rich ([Z/H]~$\geq-1.0$ dex)
globular clusters with ages below 8 Gyr show solar-type $\alpha$/Fe
ratios, i.e. [$\alpha$/Fe]~$=0.05\pm0.05$ dex. Their old counterparts, on
the other hand, have a higher mean [$\alpha$/Fe]~$=0.23\pm0.03$. Hence, we
find evidence for an age-[$\alpha$/Fe] relation in the metal-rich globular
cluster system.

An eye-catching feature of the age-[$\alpha$/Fe] correlation plot (right
bottom-row panel of Figure~\ref{ps:amahistos}) is the decreasing scatter
in [$\alpha$/Fe] with decreasing age. Globular clusters with ages above 8
Gyr have a mean [$\alpha$/Fe]~$=0.18\pm0.05$ and a dispersion of 0.37
dex\footnote{These values exclude two outliers at [$\alpha$/Fe]~$-1.0$ and
1.0 dex.}. Younger objects have a smaller [$\alpha$/Fe] ratio of
$0.09\pm0.09$ and a dispersion of 0.21 dex. The sub-sample of globular
clusters with ages $\la2$ Gyr is consistent with solar abundance ratios
[$\alpha$/Fe]~$=0.03\pm0.08$ dex and a $\sigma=0.18$ dex.

\subsection{CN Enhancement}
\label{ln:cn}
In the following section, we investigate correlations between ages,
metallicities, and [$\alpha$/Fe] ratios and other Lick indices for our
sample of M31 globular clusters. Owing to previously observed CN anomalies
of M31 globular clusters \citep{vdB69, burstein84, tripicco89, davidge90,
brodie91, beasley04}, we focus in the following analysis on correlations
of the CN index with globular cluster age, metallicity, and [$\alpha$/Fe]
ratio.

\begin{figure*}[!t]
\begin{center}
\includegraphics[width=7.8cm, bb=40 130 600 700]{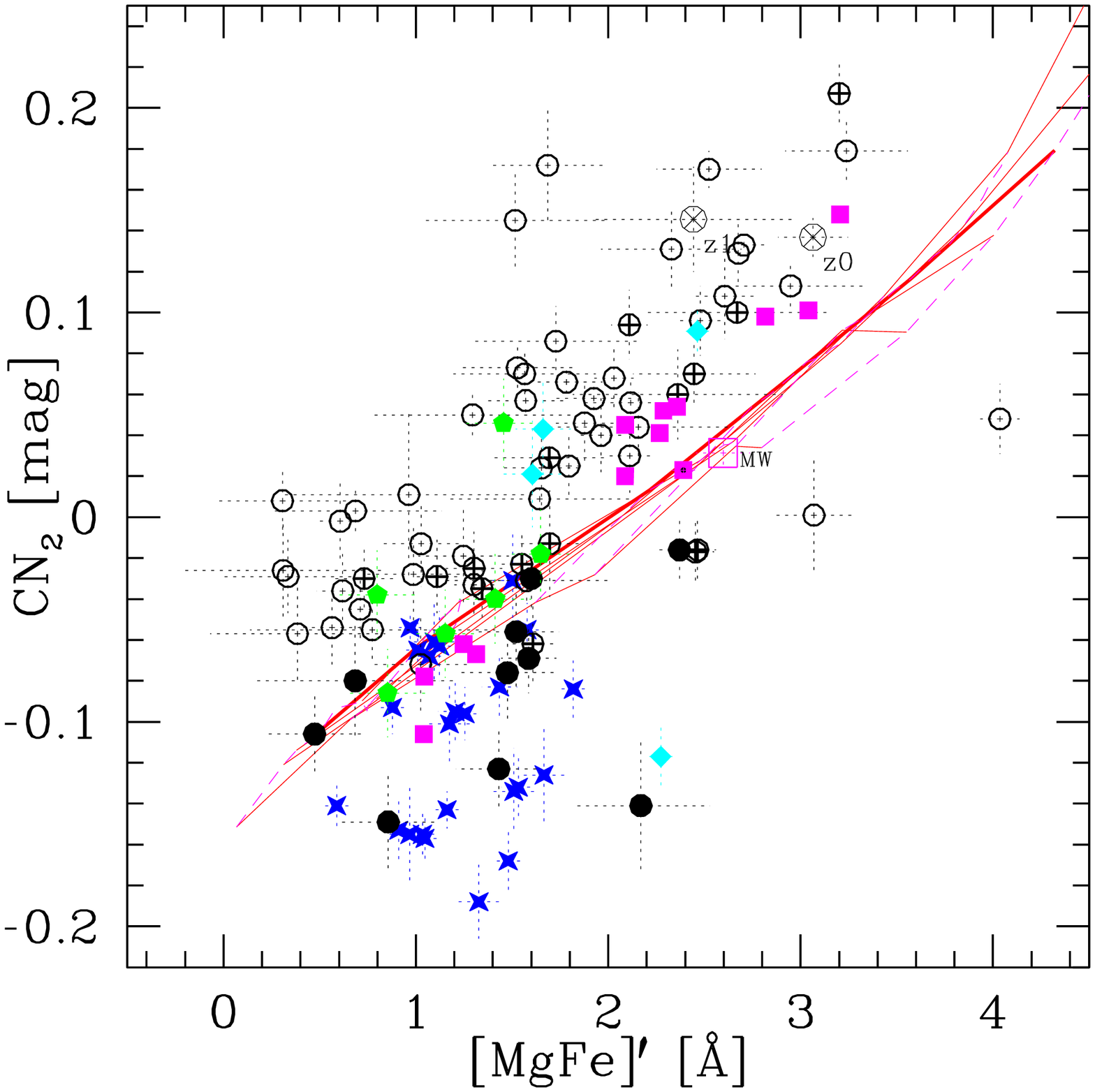}
\includegraphics[width=7.8cm, bb=40 130 600 700]{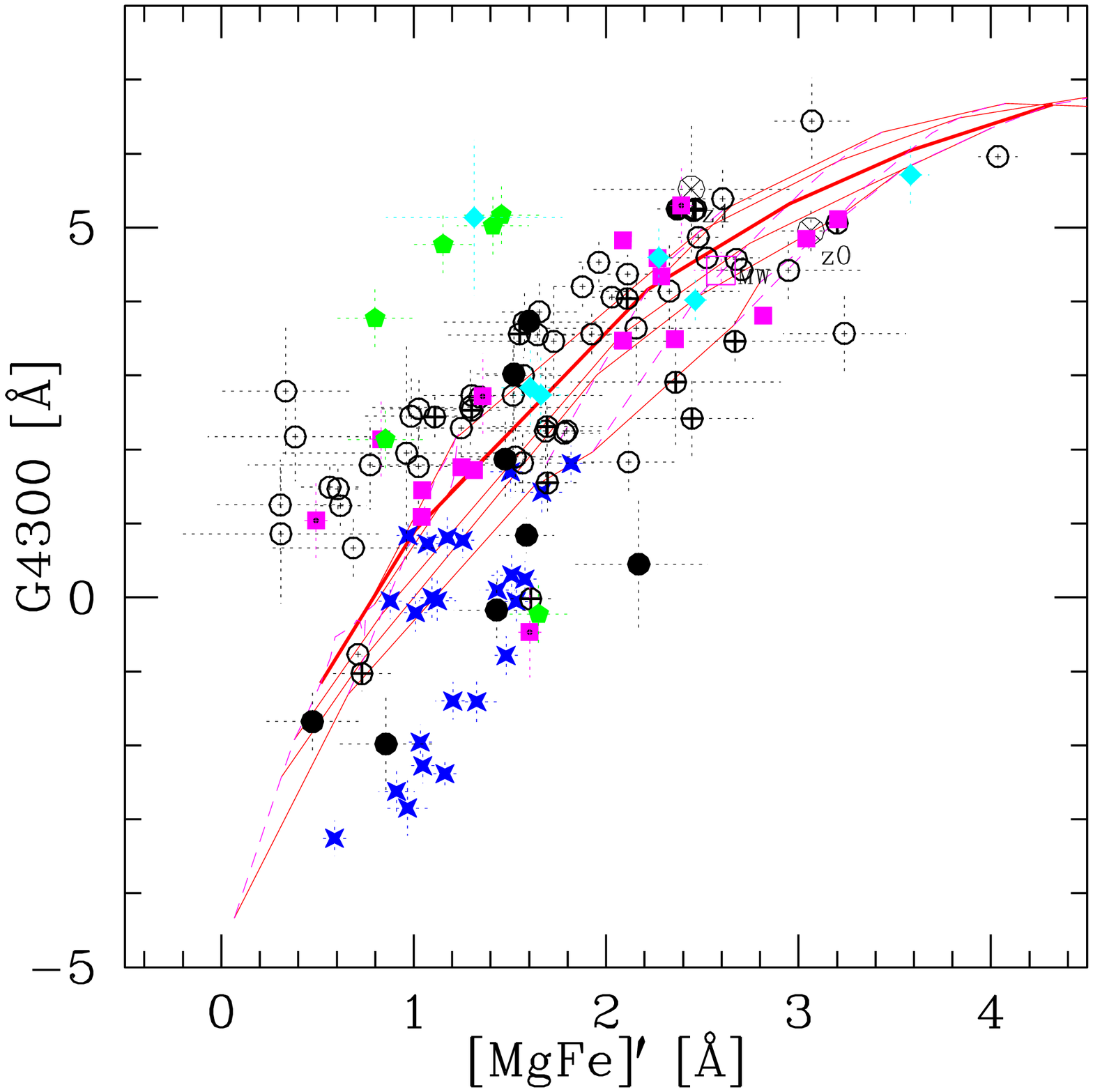}
\includegraphics[width=7.8cm, bb=40 130 600 700]{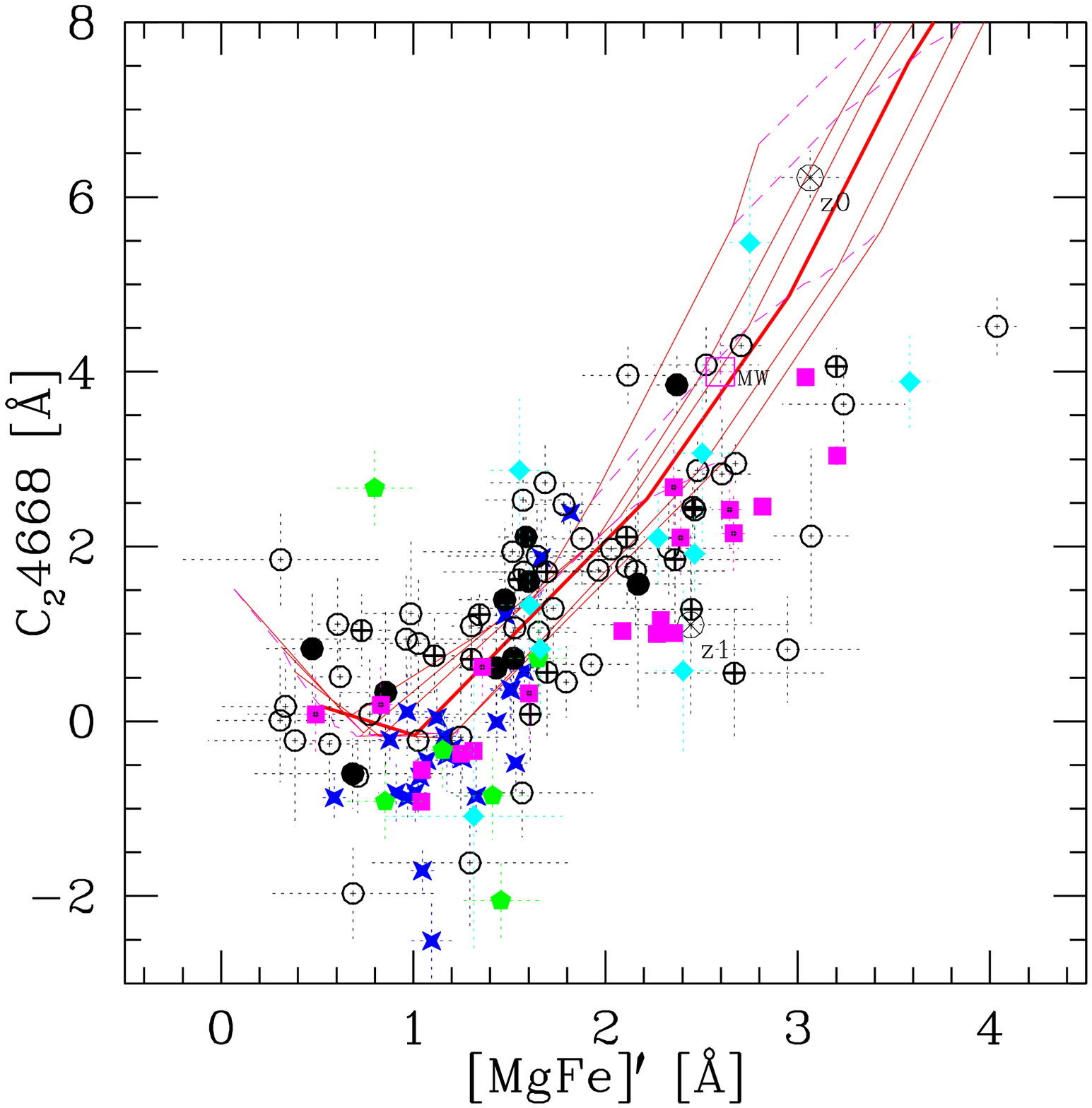}
\includegraphics[width=7.8cm, bb=40 130 600 700]{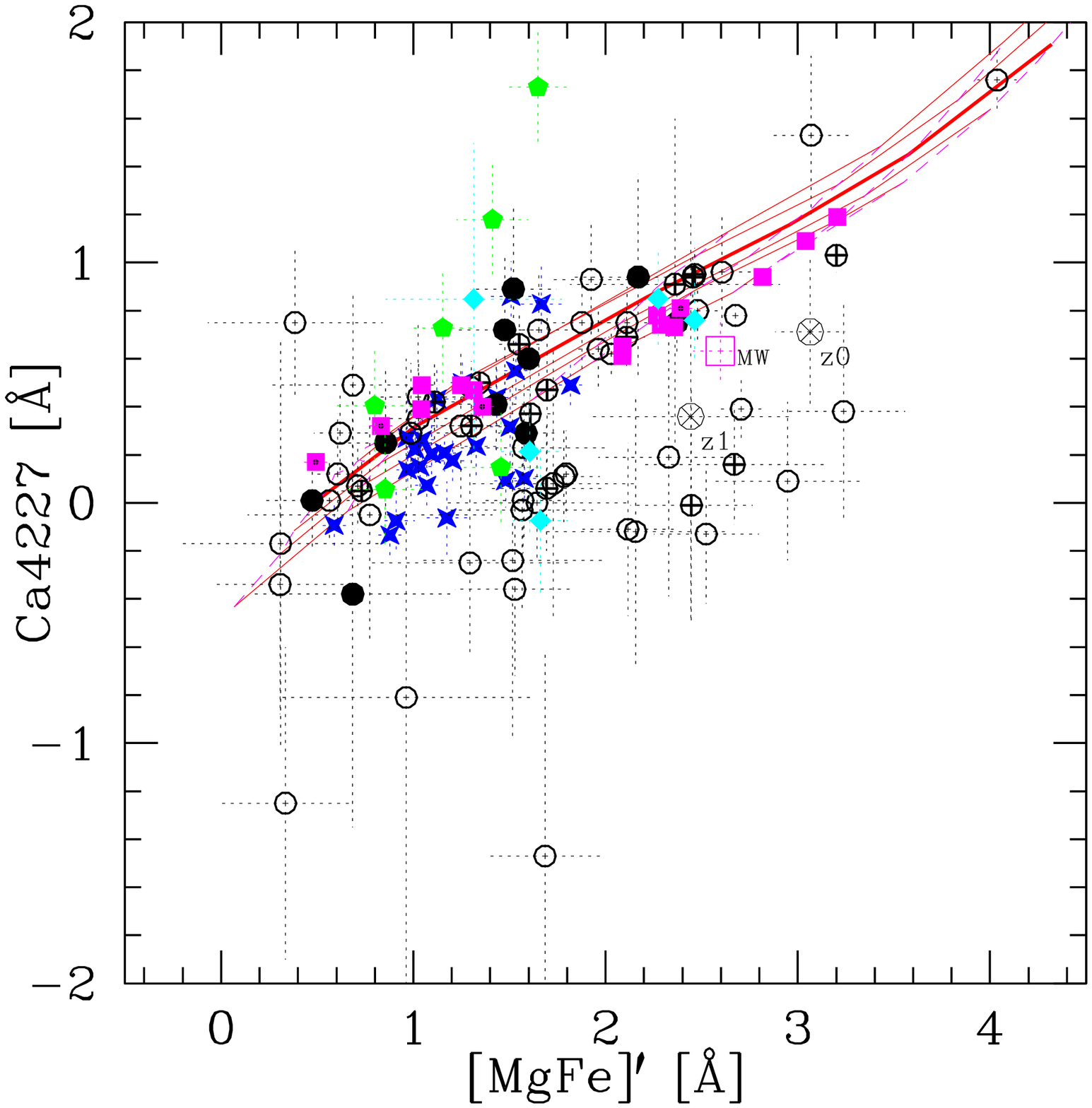}
\caption{CN$_{2}$, G4300, C$_{2}$4668, and Ca4227 vs. [MgFe]\arcmin\ 
for M31 ({\it circles}), Milky Way ({\it squares}), M81 ({\it diamonds}),
LMC ({\it stars}), and Sculptor-group spiral galaxy globular clusters
({\it pentagons}). The individual datasets are described in 
Section~\ref{ln:compm31mwlmc}. The M31 symbols are parameterized by the globular
cluster age: filled circles show globular clusters younger than 5 Gyr,
circles with a central plus sign are intermediate-age clusters with ages
between 5 and 8 Gyr, while open circles show globular clusters older than
8 Gyr. Note the offsets in CN and Ca4227 indices between old ({\it open
circles}) and young ({\it filled circles}) M31 globular clusters which is most
likely a result of a factor three relative nitrogen enhancement of the old
cluster system.}
\label{ps:nmgfecn2}
\end{center}
\end{figure*}

Since CN$_{2}$ has its blue background passband designed to avoid the
H$\delta$ \citep[see Fig.~3 in][]{puzia04b} feature and has a calibration
uncertainty of the transformation to the Lick system which is smaller than
for CN$_{1}$, we favor CN$_{2}$ throughout the subsequent analysis and
refer to it as the CN index. The CN index is a measure of the strength of
the molecular CN band at 4150 \AA.

In Figure~\ref{ps:nmgfecn2} we plot CN vs. [MgFe]\arcmin\ along with SSP
models for ages from 1 to 15 Gyr (isochrones are solid lines),
metallicities $-2.25$ to $+0.67$ dex (iso-metallicity tracks are dashed
lines), and a [$\alpha$/Fe] ratio of $0.3$ dex. The model grid is
completely degenerate in age, metallicity, and [$\alpha$/Fe] and serves
only as a reference line. In the plot we parameterize M31 globular
clusters by their age: open circles mark old clusters with formal ages
$>8$ Gyr, open circles with a central plus sign are intermediate-age globular
clusters with ages between 5 and 8 Gyr, and solid circles show clusters
formally younger than 5 Gyr. We find a clear offset in the
CN-[MgFe]\arcmin\ sequence between old and young M31 globular clusters.
Young objects tend to have weak CN indices and fall below the model
tracks, while old globular clusters show strong CN indices and populate
the region above the model tracks. The CN index offset in the range
[MgFe]\arcmin$=0-1.8$ \AA\ ([Z/H] between $-2.3$ and $-0.6$ dex) between
these two groups is $0.13\pm0.02$ \AA. The intermediate-age group falls in
between the young and old sub-population. However, a $\Delta{\rm
CN}=0.11\pm0.03$ with respect to young globular clusters makes it rather
consistent with CN indices of old globular clusters. An interesting
observation is the scarcity of globular clusters with small CN indices at
[MgFe]~$>2$ \AA. The mechanism which drives the CN difference between
young and old globular clusters may be a function of metallicity.
Unfortunately, our sample is still too small to explore any metallicity
dependence.

\subsubsection{Is the CN-offset a Consequence of Age, Metallicity, 
and/or [$\alpha$/Fe] Differences?} 
The CN index is mainly sensitive to the carbon and nitrogen abundance. In
relatively cool giants that dominate the integrated light of old stellar
populations, most of the carbon is locked in CO, so that the nitrogen
abundance affects the CN index more directly. Main-sequence dwarfs do not
contribute to the CN absorption \citep{burstein84}, since molecules
dissociate at higher $\log g$. Smaller CN absorption is also consistent
with higher effective temperatures, which are expected in young and/or
metal-poor stellar populations. The SSP models of \cite{tmb03, tmk04}
predict an offset of $\Delta{\rm CN}\approx0.02-0.04$ mag between 13 Gyr
and 3 Gyr\footnote{The mean age of the sub-population of globular clusters
with ages below 5 Gyr is $2.5\pm1$ Gyr.} for stellar populations with
metallicities between [Z/H]~$\approx-1.35$ and 0.0, and solar-type
abundance ratios. Since the models were calculated for defined
[$\alpha$/Fe] ratios, we can test the influence of $\alpha$-element
enhancement on the CN index. The maximum CN index offset between stellar
populations with [$\alpha$/Fe] 0.5 and 0.0 dex, for ages up to 13 Gyr and
metallicities up to solar, is $\sim0.02$ mag. For super-solar
metallicities the difference does not exceed 0.025 mag. Age and
metallicity are therefore excluded as primary parameters that drive the CN
offset.

Since we find evidence for an age-[$\alpha$/Fe] relation, we have reason
to assume that the CN offset is produced by varying [$\alpha$/Fe] ratios.
For an [$\alpha$/Fe] ratio of 0.2 dex the models predict very similar
maximum CN offsets of $\sim0.045$ mag between 3 and 13 Gyr for solar and
super-solar metallicites, still a factor 3 too small for the measured
offset. The models predict smaller offsets for smaller metallicities.
Going to highly super-solar [$\alpha$/Fe] ratios $\sim0.5$ dex, we find CN
offsets of the order of 0.1 mag. However, such high $\alpha$-element
enhancements are inconsistent with the measured [$\alpha$/Fe] ratios of
the young sub-sample. In summary, any combination of differences in age,
metallicity, and [$\alpha$/Fe] ratio between young and old M31 globular
clusters cannot explain the offset in their mean CN index.

\subsubsection{Is the CN-offset a Consequence of Carbon and/or 
Nitrogen Abundance Differences?}
Because \cite{tmb03} provide index predictions for stellar populations with
an enhanced  nitrogen abundance, we can compute a zeroth-order estimate of
the increase in nitrogen abundance that is consistent with the observed CN
index offset. Table~\ref{tab:cnoffset} summarizes offsets in the CN index
as a response to a factor three increased nitrogen abundance for different
ages, metallicities, and [$\alpha$/Fe] ratios. Note that the abundances of
other species remain unchanged. While other indices do not respond
significantly, the increase in nitrogen abundance is expected to decrease
the Ca4227 index (see \citealt{tripicco95}). Indeed, we observe such a
decrease which is illustrated in Figure~\ref{ps:nmgfecn2}. However, we
refrain from a more quantitative analysis due to the poor internal
calibration quality of this index \citep[see][]{puzia02}. For the mean age
$\sim 13$ Gyr, a metallicity $\sim-0.4$ dex, and [$\alpha$/Fe] ratio
$\sim0.2$ dex, we find a predicted CN offset of the order $\sim0.07$ mag,
consistent with a factor three increase in nitrogen abundance. Together
with the CN offset resulting from the age difference and under the
assumption that the young globular cluster sub-population is not enhanced
in nitrogen, the predicted CN offset amounts to $\sim0.09-0.11$ mag, which
is in the right ballpark as the measured value. In fact, the nitrogen
enhancement appears to be even larger than a factor of three.

Since the Lick system provides indices which are primarily sensitive to
carbon abundance, we attempt to address the question of whether a variation
in carbon abundance can explain the CN-index offset between young and old
M31 globular clusters. For this purpose we plot G4300 and C$_{2}$4668 vs.
[MgFe]\arcmin\ in the middle and right panel of Figure~\ref{ps:nmgfecn2}.
G4300 measures the strength of the G-band which is dominated by CH-band
absorption and a few Fe and Cr features; C$_{2}$4668, on the other hand,
measures the C$_{2}$ Swan band at 4700 \AA, but its index passbands also
include many weak Fe, Ti, Cr, and Sc lines. Both indices are primarily
sensitive to carbon abundance \citep{tripicco95}. To first order we do not
see any significant offset between young and old globular clusters in the
CN and C$_{2}$4668 index plot, although there are indications for
differences in G-band strength at very low [MgFe]\arcmin\ (i.e.
metallicities). We therefore conclude that a difference in carbon
abundance is not the driving factor for the observed CN offset.

\begin{table}[!t]
\centering
\caption{Index response of the CN index to a factor three increase in nitrogen
abundance. The values were calculated with model predictions taken from
\cite{tmb03}.}
\label{tab:cnoffset}
\begin{tabular}{lrrrrrr}
\hline\hline
\noalign{\smallskip}
&\multicolumn{6}{c}{$\Delta$CN (3$\times$N -- solar)} \\
{\rm [$\alpha$/Fe]} & 0.0 & 0.0 & 0.0 & $+0.5$ & $+0.5$ & $+0.5$ \\
{\rm [Z/H]} & $-1.35$ & $-0.33$ & 0.00 & $-1.35$ & $-0.33$ & 0.00 \\
\noalign{\smallskip}
\hline
\noalign{\smallskip}
3 Gyr   & 0.023 & 0.050 & 0.068 & 0.023 & 0.051 & 0.069 \\
13 Gyr & 0.029 & 0.059 & 0.082 & 0.030 & 0.060 & 0.083 \\
\noalign{\smallskip}
\hline
\end{tabular}
\end{table}

A more detailed look at the red background passband of the CN index
reveals that the index is contaminated by two iron features at
$\lambda\lambda 4260$ and 4272 \AA\ while the blue background passband has
a weak Mn\,{\sc i} feature located at its blue edge at $\lambda4084$ \AA.
Although this Mn feature is fairly weak, changes in Mn abundance might
have a significant influence on the CN index. The blue background passband
is relatively narrow ($12.5$ \AA) and located $\sim50$ \AA\ away from the
main CN feature, and small changes in Mn abundance can therefore translate
in a large leverage of the pseudo-continuum flux inside the feature
passband. Hence, complex, non-solar abundance ratio variations might
influence the CN index strength, as well. However, a quantitative analysis
of these effects is beyond the scope of this work as detailed modeling is
not available at the time.

We conclude that the offset in CN index between young and old M31 globular
clusters is likely due to a nitrogen enhancement of old globular clusters
by a factor three or more, compared to the young globular cluster system.
Our results are consistent with the studies of \cite{li03} and
\cite{burstein04} who find a systematically higher nitrogen abundance in
M31 globular clusters as compared to Galactic counterparts\footnote{We
note that the metallicity coverage of their globular cluster sample is not
consistent between M31 and Galactic globular clusters. This point is
discussed further below in the Discussion section.}.

\begin{figure*}[!th]
\begin{center}
\includegraphics[width=7.7cm, bb=40 130 600 700]{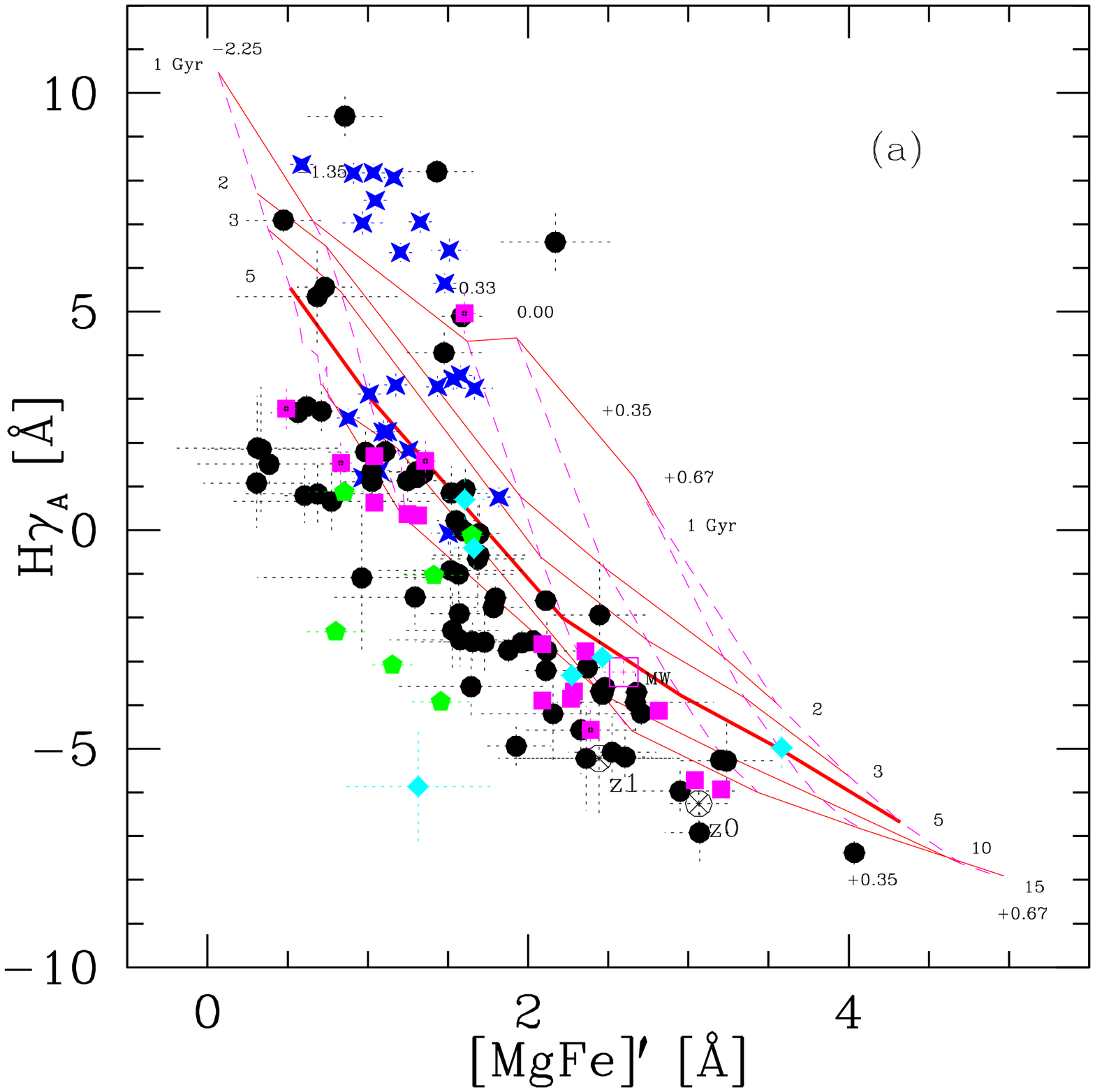}
\includegraphics[width=7.7cm, bb=40 130 600 700]{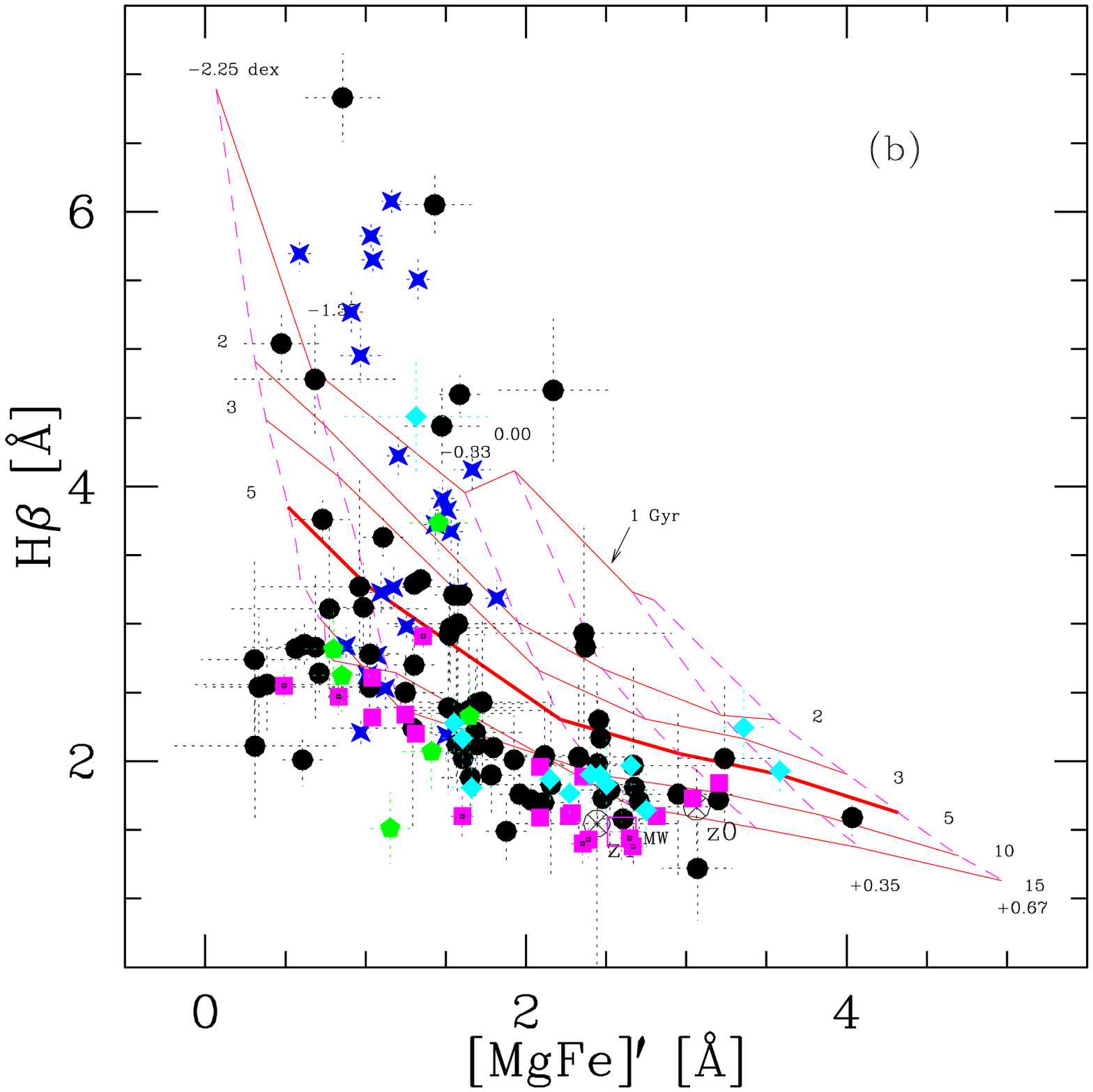}
\includegraphics[width=7.7cm, bb=40 130 600 700]{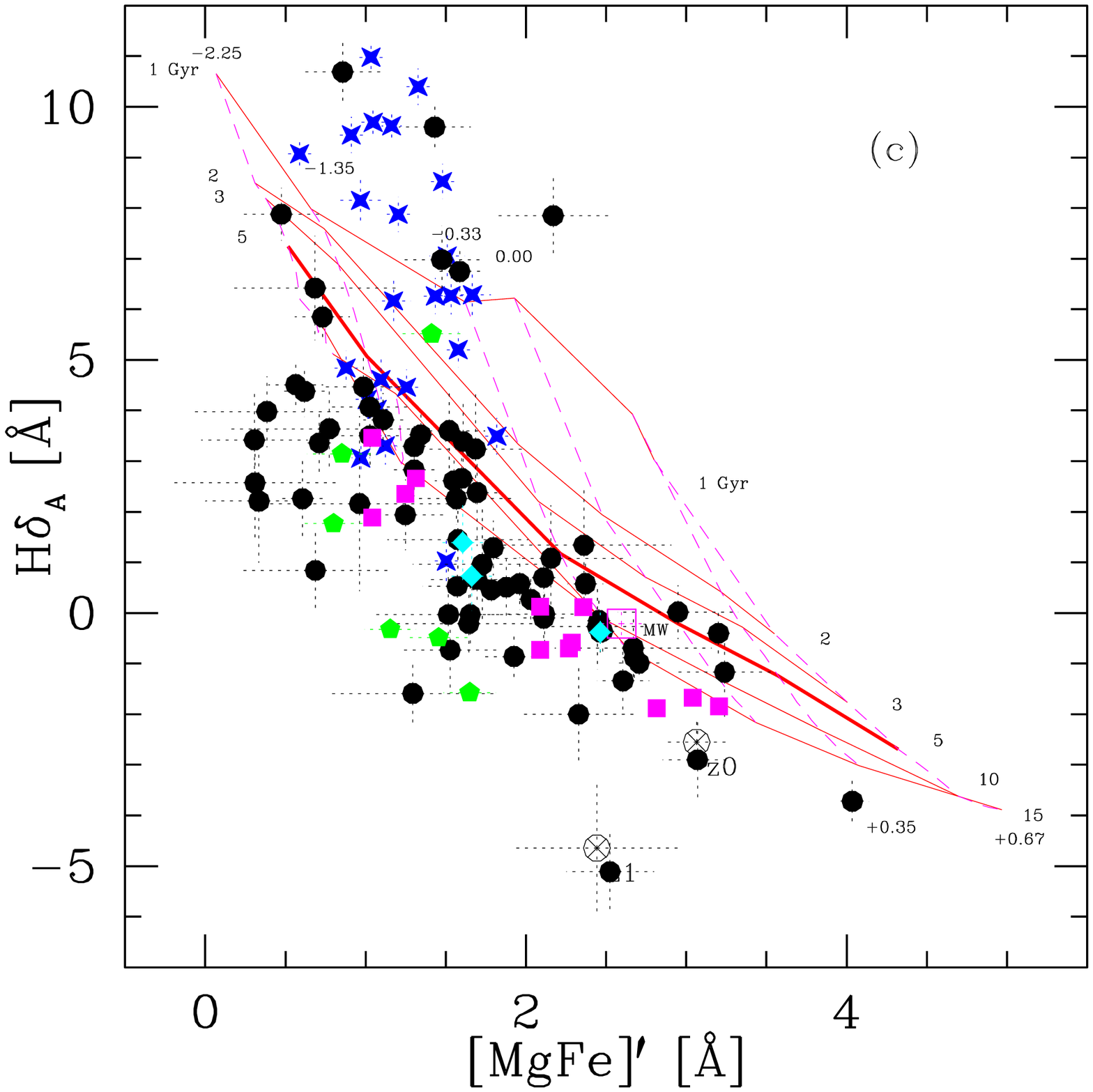}
\includegraphics[width=7.7cm, bb=40 130 600 700]{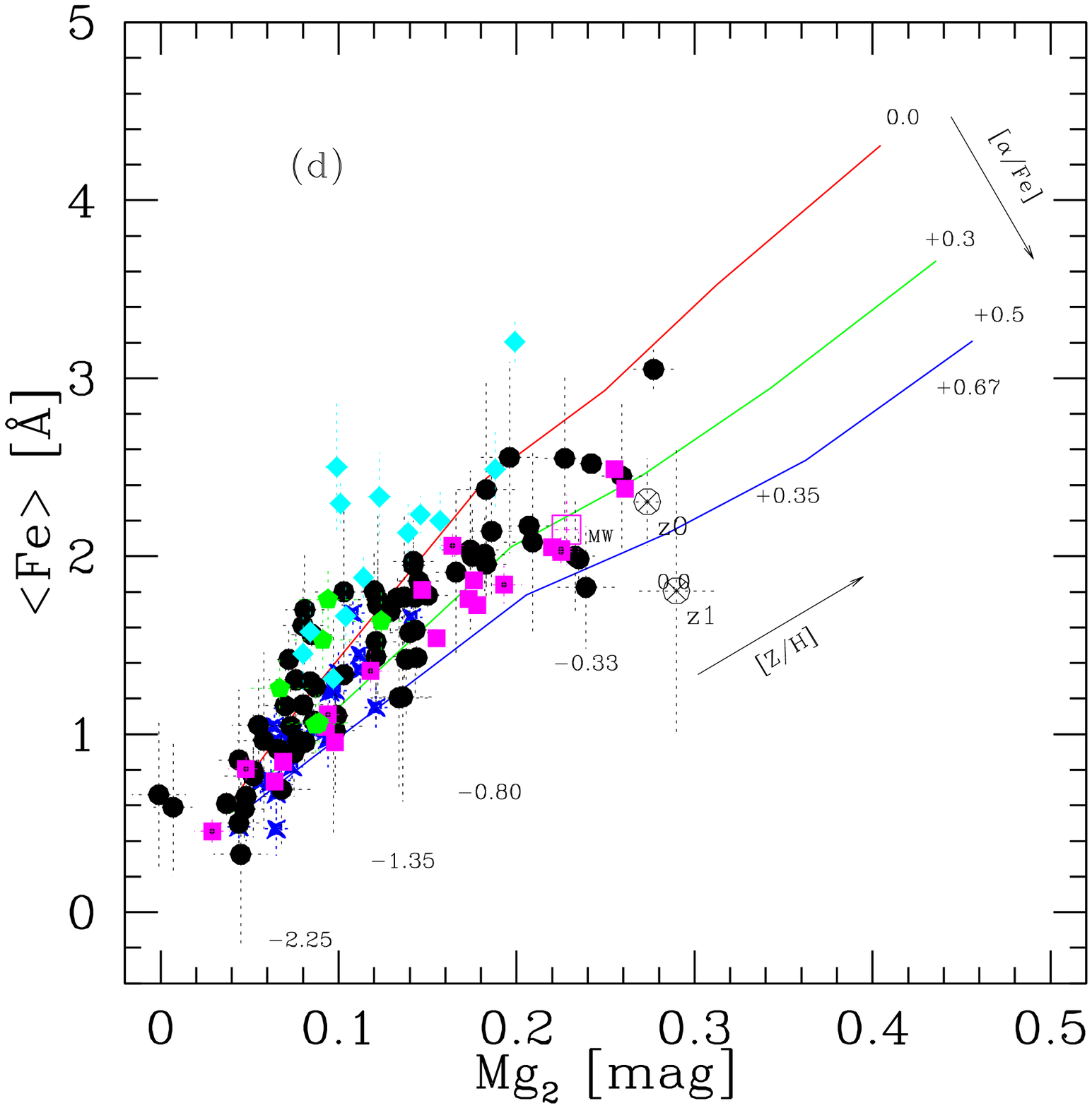}
\caption{Comparison of age-metallicity diagnostic diagrams 
{\it panels a--c}) for M31 ({\it solid circles}), M81 ({\it diamonds}),
Milky Way ({\it solid squares}), and LMC globular clusters ({\it stars}),
as well as globular clusters in Sculptor-group spiral galaxies ({\it
pentagons}). The panels show diagnostic plots using different Balmer
indices, H$\gamma_{A}$, H$\beta$, and H$\delta_{A}$. Also shown is the
$\alpha$/Fe diagnostic grid {\it panel d}). Milky Way globular clusters
are from \cite{puzia02}, while LMC globular clusters were taken from
\cite{beasley02}. Squares with a central dot indicate measurements of
Galactic globular clusters performed by \cite{cohen98} and calibrated by
\cite{beasley04}. Lick indices for M81 globular clusters were performed by
\cite{schroder02}, and \cite{olsen04} measured indices for globular
clusters in Sculptor-group spiral galaxies. Note that H$\delta_{\rm A}$
measurements are not available in the Cohen et al. dataset. Most of the
M81 data have no higher-order Balmer index measurements. Also plotted are
index measurements for the nuclear diffuse light of the Milky Way and M31.
A large open square marks the index measurements for the Galactic Bulge
light in Baade's Window, taken from \cite{puzia02}, and two crossed open
circles indicate measurements for the diffuse light of the M31 bulge at
two different galactocentric radii (labeled as z0 and z1, see
Table~\ref{tab:bkgrad}). Note, that all data and models use the same index
passband system. SSP model predictions are as in Figure~\ref{ps:nmgfehga}.} 
\label{ps:cnmgfehga}
\end{center}
\end{figure*}

\subsection{Comparison with Globular Clusters in other Spiral and Dwarf Galaxies}
\label{ln:compm31mwlmc}
In this Section we present a comparison of globular clusters in M31, Milky
Way, M81 (Sa), LMC, and Sculptor-group spiral (Sa-b) galaxies. To avoid
uncontrolled biases due to varying instrumental setups and spectral
characteristics (i.e. resolution, sky-subtraction, flux-calibration,
etc.), we strictly compare parameters derived from index measurements that
were obtained in the same index system. For conformity with SSP model
predictions we choose index passbands defined by \cite{worthey94} and for
higher-order Balmer lines the passbands of \cite{worthey97}. The size of
systematic uncertainties between different datasets is set by the
calibration quality of the transformation to the Lick system. For this
reason we collect only high-quality data from the literature with very
good index calibrations to keep the systematics small. For the Milky Way
sample we use data from \cite{puzia02}\footnote{Although we are aware of
the \cite{cohen98} data, we do not use these in the following analysis
because of the lack of H$\delta_{\rm A}$ index measurements, which might
introduce systematics in our analysis. However, the available index
measurements are included in following diagnostic plots and marked as
dotted squares. In general, they agree very well with the mean locus of
the other globular cluster data taken from \cite{puzia02}.}, and collect
index measurements for LMC globular clusters from \cite{beasley02}. M81
globular cluster Lick index measurements were taken from
\cite{schroder02}. For a handful of globular clusters in the
Sculptor-group spiral galaxies NGC~55, 247, 253, and 300, \cite{olsen04}
measured Lick indices. A list of index measurements was kindly provided to
us by Knut Olsen. Due to small sample sizes we merge the M81 and
Sculptor-group globular clusters into one sample with ten globular
clusters in total. We warn however that the individual globular cluster
samples can have very different age, metallicity, and abundance ratio
distributions. Diagnostic diagrams for all globular clusters are shown in
Figures~\ref{ps:nmgfecn2} and \ref{ps:cnmgfehga}.

\subsubsection{Ages and Metallicities}
\begin{figure}[!t]
\begin{center}
\includegraphics[width=9cm, bb=30 160 470 695]{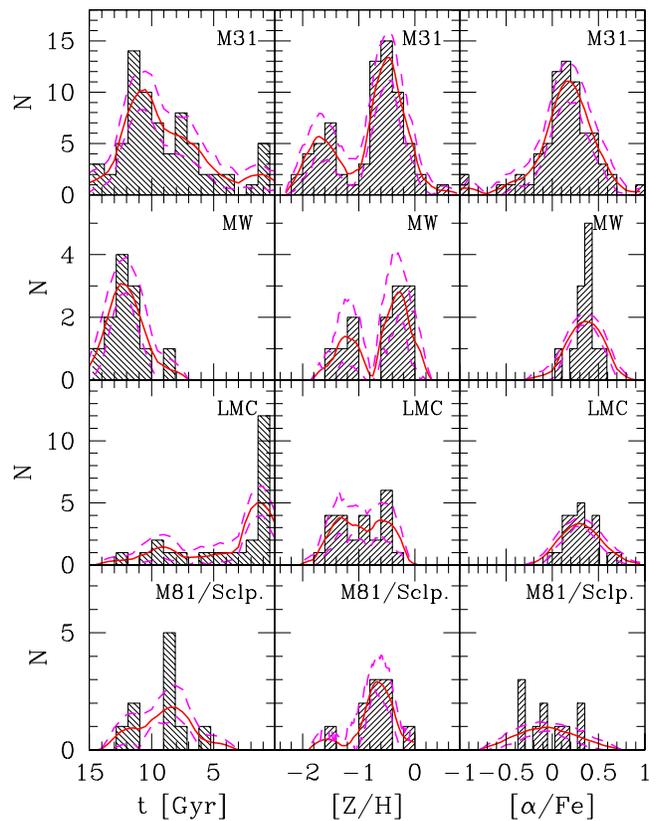}
\caption{Age, metallicity, and [$\alpha$/Fe] distributions for globular clusters in
M31, Milky Way, LMC, M81, and Sculptor-group spiral galaxies. Due to small
sample sizes, the M81 and Sculptor-group samples were merged. Solid lines
are non-parametric probability density estimates with their 90\%
bootstraped confidence limits shown as dashed lines \citep[see][ for details]{silverman86}.}
\label{ps:amaall}
\end{center}
\end{figure}

Figure \ref{ps:cnmgfehga} shows M31, Milky Way, M81, Sculptor-group spiral
galaxy, and LMC globular clusters in three different age-metallicity
diagnostic plots, as circles, squares, diamonds, pentagons, and stars,
respectively. After extracting ages, metallicities, and [$\alpha$/Fe]
ratios for globular clusters in other galaxies in exactly the same way as
described earlier for M31, we find several interesting systematic
differences between the observed globular cluster systems. Age,
metallicity, and [$\alpha$/Fe] distributions are shown in
Figure~\ref{ps:amaall}. Both the M31 and Milky Way globular cluster
systems are dominated by objects reaching oldest ages at $\sim11-13$ Gyr.
In addition to old clusters, the M31 system hosts populations of
intermediate-age (6--8 Gyr) and young ($\la2$ Gyr) globular clusters,
which are not observed in the Milky Way sample.

The mean age of the Milky Way sample is $12.0\pm1.3$ Gyr\footnote{This
uncertainty is the formal statistical uncertainty derived from our
approximation routine, while the systematic uncertainty derived from the
comparison of results from different diagnostic plots is $\sim3.5$ Gyr.
The median value is 12.1 Gyr.} which is in good agreement with previous
CMD results \citep[$11.5\pm1.3$ Gyr,][]{chaboyer98}. The metallicities
covered by M31 and Milky Way globular clusters are similar, which again
underlines the importance of careful sample selection when different
globular cluster systems are compared (see Sect.~\ref{ln:inputdata}). The
metallicity distribution of the Milky Way sample is bimodal with peaks
around $-1.2$ and $-0.3$ dex, although these estimates are uncertain due
to the small sample size of each peak.

LMC globular clusters span a wide range in age from objects as young as
$\sim1$ Gyr to old clusters resembling ages of old Galactic counterparts.
This is consistent with previous findings \citep[see also][ and references
therein]{beasley02}. The metallicity distribution of the LMC sample is
consistent with a broad single peak with a mean [Z/H]~$=-0.95\pm0.09$
dex\footnote{The median of the distribution is $-0.89$ dex.} and a
dispersion $\sigma=0.42$ dex.

We combine the samples of M81 and Sculptor-group globular clusters because
of the small number of studied objects in each galaxy. The combined age
distribution shows peaks around 12 and 8 Gyr with a clear indication for
intermediate-age globular clusters, which is similar to the M31 rather
than the Milky Way globular cluster age distribution. However, very young
objects are missing in this sample. The metallicity distribution is peaked
around $-0.6$ dex with only one cluster at roughly solar metallicity and
one metal-poor candidate at $-1.5$ dex. We warn that the distributions are
subject to change once larger and more representative samples are
available.

Although the samples of observed globular clusters in all galaxies are
still small, all cover the brightest fraction of each globular cluster
system. Keeping this in mind, we can draw the following conclusions. The
age structure of each globular cluster system suggests that luminous
globular clusters experienced significantly different formation histories
in all three galaxies. While the question about the presence of
intermediate-age globular clusters in the Milky Way will certainly be
answered by our ongoing spectroscopic survey of Local Group globular
clusters, photometric CMD studies suggest that a small fraction of
metal-rich Galactic globular clusters might be younger than $\sim9$ Gyr
\citep{salaris02}. However, among the studied bright clusters we find no
Galactic counterparts for the intermediate-age and/or young globular
clusters found in M31, M81, LMC, and Sculptor-group spiral galaxies. From
Figure~\ref{ps:amaall} it seems that the age structure of the M31 cluster
system is a composite of globular clusters in all other studied galaxies.
This aspect might suggest that M31 globular clusters with ages $\la9$
Gyr were accreted from satellite galaxies.

\subsubsection{[$\alpha$/Fe] Ratios}
\label{ln:afecomp}

The average [$\alpha$/Fe] ratios of LMC and Milky Way globular clusters
are $\sim0.1-0.2$ dex higher than the mean $\alpha$-element enhancement of
M31 globular clusters (see also Fig.~\ref{ps:cnmgfehga}d). The mean
$\alpha$-element enhancement of the Galactic sample is clearly super-solar
with a mean $0.35\pm0.04$ dex\footnote{The median value is $0.37$ dex.}
and a dispersion $\sigma=0.13$ dex. The mean [$\alpha$/Fe] ratio of the
LMC cluster sample is $0.32\pm0.04$ dex and has a dispersion of $0.17$
dex. The combined M81/Sculptor sample has a mean
[$\alpha$/Fe]~$=-0.05\pm0.08$ and a fairly broad distribution with a
dispersion of $\sigma=0.25$ dex.

Yet, all three globular cluster systems show super-solar [$\alpha$/Fe]
ratios, which implies enrichment dominated by type-II supernovae
\citep[e.g.][]{matteucci94}. Super-solar mean $\alpha$-element
enhancements around $\sim0.2-0.35$ dex are observed in high-resolution
spectra of single stars in Galactic halo and bulge globular clusters
\citep{barbuy99, cohen99, carretta01, carretta04, origlia02, origlia04}.
This is in good agreement with the values derived here and can be
considered as a zero-point calibration of our [$\alpha$/Fe] scale
\citep[see also][]{maraston03, tmb03, tmk04}. However, there are Galactic
globular clusters, such as Pal 12, which show peculiar abundance
signatures with roughly solar [$\alpha$/Fe] ratios \citep{cohen04}.
Judging from their kinematics, these objects appear to be stripped from
nearby satellite galaxies.

Compared to the mean [$\alpha$/Fe] of LMC clusters, the mean [$\alpha$/Fe]
ratio of intermediate-age M31 globular clusters is $\sim0.1-0.2$ dex
lower, and suggests that accretion of LMC-type globular cluster systems to
build up the intermediate-age globular cluster population in M31 is
unlikely. This low [$\alpha$/Fe] is rather consistent with the mean value
found for the M81/Sculptor cluster sample. It is also consistent with the
abundance pattern in thin/thick disk stars in the Milky Way
\citep[e.g.][]{edvardsson93, furhmann98}.

\subsubsection{Carbon and Nitrogen Abundance}
Beginning with the study of \cite{burstein84}, globular clusters in M31
were found to be unique in individual line indices compared to their
Galactic counterparts. Enhanced CN-band absorption in M31 clusters was
later found to be due to nitrogen enhancement \citep{ponder98, li03,
burstein04}. How do the CN indices of M31 globular clusters compare to
other globular cluster systems? In Figure~\ref{ps:nmgfecn2} we plot all
other globular clusters together with M31 globular clusters, parameterized
by their age.

Old M31 globular clusters appear on average CN-enhanced compared to the
other globular cluster systems. Intermediate-age clusters, on the other
hand, compare well in their mean CN index with the other cluster systems.
We find a striking coincidence in CN strength between LMC and young M31
globular clusters. The youngest M31 clusters populate the same locus in
the CN vs. [MgFe]\arcmin\ plot as most LMC globular clusters. Compared to
SSP models, which predict CN strength for solar-type abundance pattern,
this suggests that both LMC and young M31 globular clusters are
underabundant in carbon and/or nitrogen. Since no systematic difference in
the C$_{2}$4668 and G4300 indices, which are most sensitive to carbon
abundance, is found between LMC and young M31 globular clusters, the CN
offset suggest a difference in nitrogen abundance. Consistent with this
picture is the systematic Ca4227 offset between old M31 globular clusters,
on the one hand, and LMC and young M31 globular clusters, on the other
hand (see Sect.~\ref{ln:cn}).

\subsection{Kinematics}
\label{ln:kin}
\begin{figure}[!t]
\begin{center}
\includegraphics[width=8cm, bb=40 150 560 700]{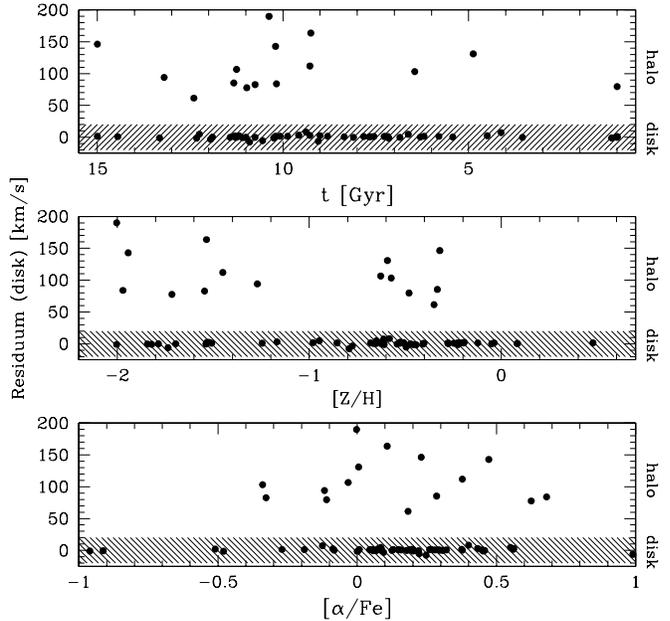}
\caption{Residual radial velocities relative to a disk model of
\cite{morrison04} as a function of age, metallicity, and [$\alpha$/Fe]
ratio for M31 globular clusters. Error bars are not shown for clarity; the
mean error in age is $\Delta t/t\approx 1/3$, the mean uncertainty in
metallicity and [$\alpha$/Fe] is $\sim0.2$ dex. The typical error in
radial velocity is $\leq 12$ km s$^{-1}$.}
\label{ps:amakin}
\end{center}
\end{figure}

In a recent radial-velocity study of the M31 globular cluster system,
\cite{morrison04} found a significant sub-population of globular clusters
showing thin-disk kinematics with metallicities down to $-2.0$ dex. The
presence of such a distinct disk globular cluster sub-system with
low-metallicity members implies  an early formation epoch of M31's thin
disk. We correlate the kinematics of \citeauthor{morrison04} with our
ages, metallicities, and [$\alpha$/Fe] ratios. A cut in normalized
residual radial velocity at $|0.75|$ km s$^{-1}$ relative to the thin-disk
model of \cite{morrison04} selects globular clusters which are likely
members of the thin-disk population, as suggested by the authors. This
selection leaves 18 objects with thin-disk kinematics in the sample.

This sub-sample covers ages between one and 11.9 Gyr. It has a mean age
$8.1\pm 0.9$ Gyr and a dispersion $\sigma=3.6$ Gyr. The average
metallicity is [Z/H]~$=-0.78\pm0.16$ dex with $\sigma=0.68$ dex; its mean
[$\alpha$/Fe] is slightly super-solar at $+0.12\pm0.07$ dex with a sample
dispersion of $0.30$ dex.  We confirm that thin-disk globular clusters
span a wide range in metallicity from about $-1.8$ dex up to solar values,
as reported by \cite{morrison04}.

If globular clusters with residuals smaller than $|0.25|$ km s$^{-1}$ are
selected, six objects remain in the sample and all but one\footnote{This
cluster (B171) has a formal age of 10.8 Gyr and might be a halo
interloper.} have intermediate to young ages $\la8$ Gyr. \cite{beasley04}
found also three young globular clusters with clear thin-disk kinematics.
Their median metallicity decreases to $-1.02\pm0.31$ dex with a higher
metallicity dispersion of 0.75 dex and their median [$\alpha$/Fe] ratio
increases to $0.26\pm0.05$ dex ($\sigma=0.12$ dex). This suggests that M31
globular clusters with clear thin-disk kinematics are members of the
intermediate-age cluster sub-population. Although their metallicities are
inconsistent with typical abundances found in the Milky Way's thin-disk
stellar population, their [$\alpha$/Fe] ratios are in accord with those of
Galactic bulge globular clusters (see Sect.~\ref{ln:afecomp}). It would be
interesting to obtain a larger sample of thin-disk globular clusters in
M31 (1) to test the maximum age of this globular cluster sub-population,
and (2) to check whether or not their chemical abundances are in contrast
with disk/bulge globular clusters in the Milky Way.

In Figure~\ref{ps:amakin} we plot the residual radial velocity, which was
calculated relative to the thin disk model of \citeauthor{morrison04} as a
function of globular cluster age, metallicity, and [$\alpha$/Fe] ratio.
There is a clear bimodality in the residuum distribution and we split the
sample into clusters that fall in the hatched region which indicates
globular clusters with normalized disk-model residuals $\leq|20|$ km
s$^{-1}$, and globular clusters with larger residuals indicating halo
membership. We consider the sub-population with small residuals as a
thin/thick disk component, being aware of potential contamination by halo
objects. About 70\% of our globular cluster sample falls into the
thin/thick disk category, excluding only halo members. Based on the number
counts of the halo and thin/thick-disk sample and assuming a Gaussian
distribution of velocities around the mean of the thin-disk model with
dispersions 30 and 150 km s$s^{-1}$, for the disk and halo population,
respectively, we expect $2\pm2$ halo interlopers in our thin/thick-disk
sample and less than one genuine disk globular cluster in the halo sample.

The most striking feature in the upper panel of Figure~\ref{ps:amakin} is
that most globular clusters of the halo population have ages older than 9
Gyr, with a median age of 10.4 Gyr. Only two halo globular clusters have
intermediate ages and we find one extremely young cluster (B324-S51) that
is inconsistent with disk kinematics. The thin/thick-disk population, on
the other hand, has a median age of 9.4 Gyr. The halo sub-sample shows a
clearly bimodal distribution of metallicities with peaks around $-1.6$ and
$-0.5$ dex.

\begin{figure*}[!t]
\begin{center}
\includegraphics[width=15.cm, bb=10 150 630 700]{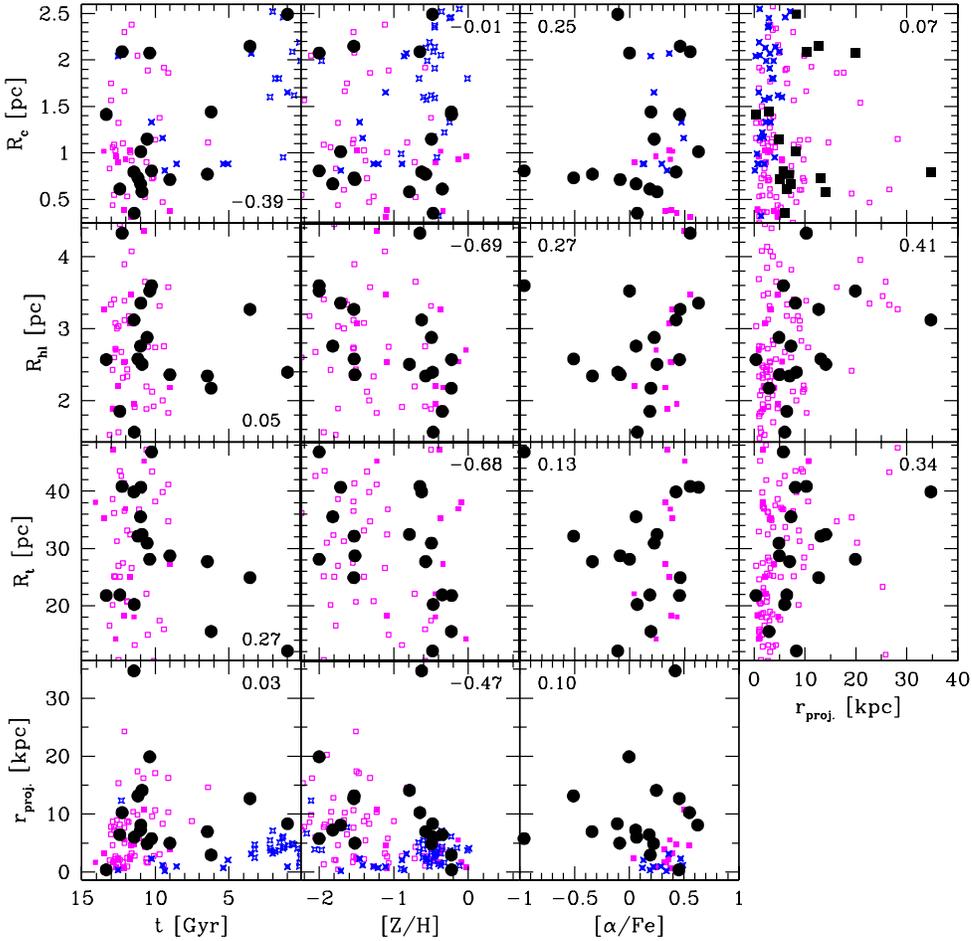}
\caption{Structural parameters, $R_{\rm c}$ (core radius), $R_{\rm hl}$ (half-light
radius), and $R_{\rm t}$ (tidal radius), as a function of age,
metallicity, and [$\alpha$/Fe] ratio for globular clusters in M31
(circles), Milky Way (squares), and LMC (stars). Data on structural
parameters were taken from \cite{barmby01} for M31 globular clusters, from
the 2003 update of the McMaster catalog \citep{harris96} for Milky Way
globular clusters, and from \cite{mackey03} for LMC clusters. Information
on age and metallicity for Galactic globular clusters is parameterized by
their photometric (open squares, data taken from \citealt{salaris02}) and
spectroscopic origin (solid squares, data taken from \citealt{puzia02}).
The structural parameters are shown as a function of projected
galactocentric radius, $r_{\rm proj}$, in the right-column panels. The
bottom row of panels shows the projected galactocentric radius as a
function of age, metallicity, and [$\alpha$/Fe] ratio. In a corner of each
panel the Spearman rank correlation coefficient is given.}
\label{ps:sizecorr}
\end{center}
\end{figure*}

It is interesting that the velocity dispersion around the
\citeauthor{morrison04} model in the thin/thick-disk sample decreases
towards younger globular clusters; from $|0.47|$ km s$^{-1}$ for objects
with ages $>8$ Gyr to 0.38 km s$^{-1}$ for globular clusters younger than
8 Gyr.

The distribution of [$\alpha$/Fe] ratios for halo and thin/thick-disk
globular clusters is shown in the lower panel of Figure~\ref{ps:amakin}.
We find an average [$\alpha$/Fe]~$=0.17\pm0.07$ dex with a dispersion of
0.33 dex, for the halo sample. The thin/thick-disk sample has a similar
mean [$\alpha$/Fe] at $0.14\pm0.01$ dex and a dispersion of 0.32 dex. 

We find a mild age-[$\alpha$/Fe] correlation in the thin/thick-disk sample
of the form ${\rm [}\alpha{\rm /Fe]}=(0.013\pm0.011)\cdot t +
(0.076\pm0.122)$, with a rms of 0.12 dex for a weighted least-square fit,
where $t$ is given in Gyr. We compute weights for individual globular
clusters from an average of upper and lower 1$\sigma$ uncertainties.
Normally, these errors are asymmetric due to the skewness of diagnostic
grids. No significant correlation between [$\alpha$/Fe] and metallicity is
found for thin/thick-disk globular clusters.

\subsection{Structural Parameters}

In the following section we study the correlations between globular
cluster core radi, $R_{c}$, half-light radii, $R_{hl}$, tidal radii,
$R_{t}$ \citep[see][]{king62}, and the projected galactocentric distance,
$r_{\rm proj}$ with previously derived ages, metallicities, and
[$\alpha$/Fe] ratios. We collect data on structural parameters for M31
globular clusters from \cite{barmby01}, for Milky Way globular clusters
from the 2003 update of the McMaster
catalog\footnote{http://physun.physics.mcmaster.ca/\~{}harris/mwgc.dat}
\citep{harris96}, and for LMC clusters from \cite{mackey03}.

Since for extragalactic globular clusters only projected galactocentric
distances are available, we need to project the 3-D positions of the
Galactic globular cluster system on a 2-D surface on the sky for a fair
comparison between Galactic and extragalactic cluster systems.
Fortunately, the McMaster catalog provides the 3-D positions of individual
Galactic globular clusters in the reference system of the Sun, so that the
conversion to a projected galactocentric distance is easy. We simulate the
view towards the Milky Way globular cluster system similar to that towards
M31, by applying a similar pitch angle of $77^{o}$ for the Milky Way disk
and the entire globular cluster system and randomly choosing a line of
sight tilted with respect to the Earth-Galactic center axis.

In Figure~\ref{ps:sizecorr} we compare structural parameters with globular
cluster ages, metallicities, and [$\alpha$/Fe] ratios. We also plot the
structural parameters as a function of projected galactocentric radius in
the very right column of Figure~\ref{ps:sizecorr}. For each sub-panel we
compute the Spearman rank correlation coefficient for the M31 globular
cluster sample. The number is given in a corner of each panel (1 indicates
perfect correlation, $-1$ anti-correlation).

In general, the parameter space coverage between the three globular
cluster systems is in good agreement. We find correlations in the $R_{hl}$
and $R_{t}$ vs. [Z/H] panels, where the Spearman correlation coefficient
is around $-0.7$ with a statistical significance for a correlation of
$>99.7$\% in both cases. Since both the half-light radius $R_{hl}$ and the
tidal radius $R_{t}$ appear to be weakly correlated with galactocentric
radius $r_{\rm proj}$, at least part of their correlation with metallicity
might be due to a correlation of metallicity with galactocentric radius
(reflecting the galactocentric distance dichotomy between metal-rich disk
and metal-poor halo globular clusters; see \citealt{chernoff89, vdB91,
cote99}). This is consistent with the correlation in the $r_{\rm proj}$
vs. [Z/H] plot in Figure~\ref{ps:sizecorr}. However, the sample selection
might produce spurious correlations and it is important to check these
relations with larger samples, in particular because the Galactic sample
does not show such a strong correlation of $R_{hl}$ and $R_{t}$ with
metallicity.

We point out that similar size differences between metal-poor and
metal-rich globular clusters were found in other galaxies \citep{kundu98,
puzia99, larsen01}. A possible explanation for this difference might be
the combined effects of mass segregation and the dependence of
main-sequence lifetimes on metallicity, under the assumption of similar
distribution of central potentials and half-mass radii \citep{jordan04}.
However, the size difference might be also entirely driven by projection
effects which are due to a systematic difference in dynamical evolution of
globular clusters on different orbits \citep{larsen03}. If true, size
differences should disappear at large galactocentric radii, where
dynamical effects similarly affect metal-poor and metal-rich globular
clusters. Intriguingly, Perrett (2001, PhD thesis) found an increasing
size of metal-rich globular clusters with galactocentric radius,
approaching the mean size of metal-poor globular clusters at large
galactocentric distances. No galactocentric distance-size relation was
found in the metal-poor globular cluster sub-population. Size information
on more metal-rich globular clusters at large galactocentric radii is
needed to establish whether the size difference between metal-poor and
metal-rich globular clusters is due to a dynamical effect or reflects
differences in globular cluster formation.

It is well known that the globular cluster populations of M31, LMC, and the
Milky Way fall on the same ''fundamental plane'', which for globular
clusters is defined by velocity dispersion, surface brightness, and radius
\citep{djorgovski95}. Together with the similar correlations between the
structural parameters of globular clusters in M31, Milky Way and LMC with
their ages, metallicites, and abundance ratios, this suggests that the
formation and evolution must have been similar for the studied clusters.

\begin{table*}[!t]
\centering
\caption{Ages, metallicities, and [$\alpha$/Fe] ratios of the integrated light for
three fields in the bulge of M31 and the Milky Way. The spectra of all
fields were taken at different median galactocentric distances, $r_{\rm
median}$, given in the second column. All measurements were performed in
the same way as for the globular cluster data and are given with their
statistical 1$\sigma$ uncertainties.}
\label{tab:amabulges}
\begin{tabular}{lcccc}
\hline\hline
\noalign{\smallskip}
galaxy & $r_{\rm median}$ [kpc] & age [Gyr] & [Z/H] & [$\alpha$/Fe] \\
\noalign{\smallskip}
\hline
\noalign{\smallskip}
MW          & 0.55 & $  9.4\pm1.4$ & $-0.22\pm0.06$ & $0.35\pm0.07$ \\
M31 (z0)  & 0.60 & $14.5\pm1.3$ & $-0.06\pm0.04$ & $0.38\pm0.07$ \\
M31 (z1)  & 2.48 & $14.9\pm2.2$ & $-0.22\pm0.12$ & $0.65\pm0.22$ \\
\noalign{\smallskip}
\hline
\end{tabular}
\end{table*}

\subsection{Comparison with the Integrated Bulge Light of M31 and the Milky Way}

Early speculations by \cite{baade44} claimed that the stellar populations
in Galactic globular clusters would resemble those in the Milky Way bulge
and the nucleus of M31. This simplistic picture was revised soon after,
when a more complete census of stellar populations in Galactic globular
clusters was available. Only the most metal-rich globular clusters appear
to host stellar populations similar to those in spiral bulges and
early-type galaxies \citep[e.g.][]{maraston03}.

In the following section we compare the integrated-light spectra of the M31
nucleus, obtained in this study (see Sect.~\ref{ln:data}), the Milky Way
bulge \citep[taken from][]{puzia02}, and metal-rich globular cluster
populations of both galaxies. Figure~\ref{ps:nmgfehga} shows the location
of the two M31 bulge spectra (as crossed open circles) in age/metallicity
and [$\alpha$/Fe] diagnostic grids for the two different galactocentric
radii (see Fig.~\ref{ps:m31big} and Tab.~\ref{tab:bkgrad}). Using the same
technique as for the globular cluster spectra we derive
luminosity-weighted ages, metallicities, and [$\alpha$/Fe] ratios. The
values are summarized in Table~\ref{tab:amabulges}. Since the inner M31
field and the Milky Way field were taken at similar mean galactocentric
radii ($\sim0.6$ kpc), their stellar populations can be directly compared,
assuming that no significant differences in population gradients are
present between the two galaxies.

Given the statistical and systematic errors, the derived mean ages,
metallicities, and [$\alpha$/Fe] ratios suggest that the stellar
populations in both spiral bulges are similar. Both bulges harbor old
stellar populations, with metallicities close to solar values. There is a
tendency for the M31 bulge to have a slightly higher metallicity. Both
stellar populations have clearly super-solar $\alpha$/Fe ratios, although
the outer M31 field seems to have a higher [$\alpha$/Fe] value.

There is no systematic difference in CN index between metal-rich old M31
globular clusters, which are enhanced in nitrogen (see Sect.~\ref{ln:cn}),
and the corresponding M31 bulge fields (see Fig.~\ref{ps:nmgfecn2}).
However, compared to metal-rich Galactic globular clusters and the
Galactic bulge, the CN index of the M31 bulge is offset by $\sim0.04$ and
$\sim0.1$ mag, respectively, and suggest an enhacement in carbon and/or
nitrogen abundance. The C$_{2}$4668 index for the central M31 field (z0)
is significantly higher with respect to the mean M31 and Galactic globular
clusters at similar [MgFe]\arcmin. This is not the case for the outer
field (z1). However, this C$_{2}$4668 offset is less pronounced when the
central M31 field is compared to the Milky Way bulge. In combination with
the similar G4300 indices for metal-rich globular clusters and the bulges
of M31 (both sectors) and the Milky Way, this suggests that the central
M31 field (z0) is enhanced in carbon {\it and} nitrogen, compared to
Galactic bulge and metal-rich Galactic globular clusters. The outer M31
field is likely to be enhanced in nitrogen only. The reduced Ca4227
indices of both M31 fields with respect to metal-rich Milky Way globular
clusters and Galactic bulge underline these results. Furthermore, we find
an increase in CN index with increasing galactocentric radius for the M31
bulge. This suggests an increasing nitrogen enhancement, which in the
case of the field stellar population in M31 increases with radius. Such
differences were already reported by \cite{morgan57} who discovered that
the integrated spectrum of M31's nucleus is dominated by cyanogen
absorption features. Recently, radial NH gradients in the M31 bulge were
found by \cite{davidge90c}.

In summary, the bulge of M31 appears to be metal-rich, $\alpha$-element
enhanced and old, similar to the Milky Way bulge. However, compared to the
Galactic stellar populations there are significant differences in chemical
composition. We find indications for an enhancement in carbon {\it and}
nitrogen in the central region of M31, while the integrated light of the
outer stellar populations is consistent with a nitrogen enhancement only.
It is well known that stellar populations in early-type galaxies show
super-solar metallicities and $\alpha$-element abundance ratios
\citep{worthey92}. In this respect, our comparison shows that metal-rich
M31 globular clusters and early-type galaxies are more alike than stellar
populations in early-type galaxies and metal-rich Galactic globular
clusters.

\section{Discussion}
\label{ln:discussion}

\subsection{Stellar Streams and Merging}
The existence of the recently discovered thin-disk population of M31
globular clusters \citep{morrison04} is a strong argument for an
undisturbed disk evolution. Depending on the exact kinematical definition
of the thin-disk population, we find that most, but not all, members of
the thin-disk globular cluster population have ages younger than $\sim10$
Gyr and metallicities $-2.0\la$~[Z/H]~$\la0.0$ (see Sect.~\ref{ln:kin}).
The presence of old thin-disk globular clusters suggests an early
formation and smooth kinematical evolution of the thin-disk
globular-cluster population. In particular, it implies {\it in-situ}
formed star clusters. It would be therefore of great interest to determine
the maximum age in a larger dataset of thin-disk globular clusters and to
test how early M31's thin disk formed.

Recent very deep HST observations, pointed towards the halo of M31, made
the age and metallicity distributions of the targeted stellar populations
directly accessible. These CMD studies revealed a dominant metal-rich
stellar population with a surprisingly high metallicity, [Z/H]~$\approx
-0.6$ \citep[][ see also Fig.~\ref{ps:m31big}]{holland96, sarajedini01,
bellazzini03, rich04} and intermediate ages $\sim6-8$ Gyr \citep{brown03}.
There are indications for an even more metal-rich stellar population
[Z/H]~$\ga -0.2$) which is irregularly distributed throughout M31's halo
\citep{sarajedini01, bellazzini03}. The case of a metal-rich
intermediate-age stellar population in the halo of a spiral galaxy is
generally attributed to an accretion or merging event that led to the
incorporation of external stellar populations and/or to infall of fresh
material which fueled new star formation. Depending on the exact merging
kinematics, tidally disrupted satellites remain in the halo for several
Gyr, until their central regions eventually spiral inwards as a result of
dynamical friction. Support for this picture comes from observations of
significant large-scale sub-structure in the form of stellar streams in
the halo of M31 \citep{ibata01, ibata04, ferguson02, mcconnachie03,
mcconnachie04}. Other disk-dominated galaxies show similar indications of
halo sub-structure, pointing to a hierarchical assembly of these systems
\citep[e.g.][]{zibetti04}.

Since accretion events tend in general to heat stellar disks, a good
indicator for a puffed-up disk population is the ratio of disk scale
length, $d$, to scale height $z_{0}$. A minor merger with a mass fraction
of 0.05--0.2 is expected to agitate the stellar disk and increase its
vertical scale height by a factor 1.5--2, compared to an undisturbed
evolving disk \citep{velazquez99}. This is confirmed by observations of
edge-on spirals with and without signs of interaction
\citep{schwarzkopf00, schwarzkopf01}, where interacting spirals show
$d/z_{0}\leq7$. Large-area star counts in M31 find that the scale height
of the disk varies between 50 and 400 pc, while the scale length is
$\sim5-7$ kpc \citep{hopper95}. The resulting $d/z_{0}\approx12.5-140$ is
placing the M31 disk well within the regime of undisturbed stellar disks,
implying a very small past merging rate. If, however, the mass accretion
is smooth enough and/or occurs on preferentially tangential orbits so that
the thin-disk integrity is unaffected, thin-disk evolution and mass
accretion might be two spatially uncorrelated, but temporally coexisting
processes \citep[e.g.][]{font01}.

It is not clear if some M31 globular clusters are associated with the two
stellar streams discovered by \cite{ibata01} and \cite{mcconnachie04},
because of the still too sparse kinematic information. However,
\cite{perrett03} find kinematically coupled groups in the M31 globular
cluster system. Although the location of some globular clusters coincides
with the general locus of the streams (see the northern stream in
Fig.~\ref{ps:m31big}), it is unlikely that accretion contributed
significantly to the assembly of the M31 globular cluster system, in
particular the thin-disk sub-population.

\cite{morrison04} find a velocity dispersion in the thin-disk globular
cluster system of $\sim20-40$ km/s. Globular cluster systems in dwarf
galaxies have typical velocity dispersions $30-80$ km/s
\citep[e.g.][]{olsen04}. In general, the thin-disk globular
clusters in M31 have on average too small a velocity dispersion to be
accreted from merging satellite galaxies, although the very unlikely case
of a merging trajectory which is perfectly aligned with the disk plane
could be a way to preserve a small velocity dispersion of the infalling
cluster system during the accretion event.

Using the numbers from \cite{ibata01}, the most massive stream is
estimated to carry $10^{8}$ to $10^{9}$ M$_{\odot}$ with a total
$M_{V}\!=\!-14$ mag. Assuming an average specific frequency
$S_{N}\!\approx\!10$ for dwarf galaxies of this luminosity
\citep{durrell96}, the estimated number of accreted globular clusters is
$\sim4$. The other stream \citep{mcconnachie04} carries less mass and will
therefore contribute even less globular clusters. However, if both streams
are associated with the two satellites M32 and NGC~205 (both with a total
luminosity $M_{V}\approx\!-\!16.4$ mag), the contributed number of
clusters might be as large as 20--40 objects per galaxy, given typical
specific frequencies $S_{N}\approx2-8$ for these systems. M32 hosts no
globular clusters \citep{ashman98}, although at least $\sim20$ are
expected.

If the two massive stellar streams added some globular clusters to M31, it
is unlikely that they are now part of M31's thin-disk globular cluster
population, based on their trajectories. Both stellar streams have very
radial orbits and therefore do not interact significantly with the
thin-disk globular-cluster population \citep{ibata04}. Despite their
potential past interaction(s), the evolution of a kinematically cool
thin-disk globular-cluster population could have remained undisturbed.

\subsection{Spatial Distribution}
We find spatially clumped globular cluster sub-popu\-lations in M31.
Inter\-mediate-age globular clusters preferentially reside inside $\sim5$
kpc galactocentric distance, which is suggestive of a slightly younger
inner-disk/bulge cluster population. Clusters with ages below 5 Gyr are
predominantly found at large radii along the major axis. Most of these
globular clusters show thin-disk kinematics and have metallicities below
$-0.8$ dex (see also Fig.~\ref{ps:amakin}). Their projected positions (see
Fig.~\ref{ps:m31big}) coincide with luminous star-forming regions in M31's
outer-disk, that were recently observed by
GALEX\footnote{http://www.galex.caltech.edu/popups/gallery-M31.html}.

This spatial clustering suggests that halo, inner-disk/bulge, and
thin-disk globular cluster sub-populations experienced different formation
and/or evolution histories. While the intermediate-age population could
have formed from previously enriched material several Gyr after the old
halo population was formed, the young cluster population might be the
offspring of star-formation triggered by a recent infall of fresh material
in the outskirts of M31. However, a direct connection between these
globular cluster sub-populations and the two giant stellar streams in the
halo is rather unlikely, although we find two young metal-rich young
globular clusters (B337-G68 and B324-S51) close in projection to NGC 205.
Since there is much more sub-structure visible in M31's halo than the two
reported stellar streams \citep[e.g.][]{ferguson02,mcconnachie04}, direct
connections between other stellar streams and globular cluster sub-samples
are conceivable. Only a careful kinematic analysis of the stellar streams
can clarify how much accretion contributed to the formation of M31's halo
and disk globular cluster system. Such ambitious surveys are on the way
\citep[see e.g.][]{ibata04}.

\subsection{Chemical Enrichment}
In the following we discuss chemical characteristics of M31 globular
clusters that might be informative of their formation and evolution.
While type-II supernovae predominantly eject $\alpha$-elements on very
short timescales ($\la10$ Myr), the ejecta of type-Ia supernovae enrich
the interstellar medium with Fe-peak elements with a delay of typically
$\sim1$ Gyr (\citealt{greggio97,matteucci01}, but see also
\citealt{tsujimoto04} for a short-living type-Ia progenitor). These
supernova types mark the two extremes of injection timescales for
processed material back into the interstellar medium. Other elements such
as carbon and nitrogen are predominantly produced by massive stars with a
strong dependence on metallicity, on timescales $\sim10\!-\!100$ Myr
\citep[e.g.][]{maeder92}. In general, the relative delay in progenitor
lifetimes between type-II and Ia supernovae implies that abundance ratios
can be used as tools to clock star-formation timescales.

A detailed model of the chemo-dynamical evolution of globular cluster
systems is not available yet and a comparison of our data with theoretical
predictions is not possible \citep[but see][]{li04}. However, our data on
[$\alpha$/Fe] ratios and nitrogen enhancement of M31 globular clusters
allow us to make some qualitative statements on their formation
timescales.

\subsubsection{$\alpha$-Elements}
Super-solar [$\alpha$/Fe] ratios of M31 globular clusters can be
interpreted as the result of relatively short formation periods, which
ended before a significant number of type-Ia supernovae could contribute
to chemical enrichment of their parent gas clouds. As chemical enrichment
increases with time it requires successively higher star-formation rates
to increase the $\alpha$-element enhancement to a certain level and
''override'' the chemical signature of previous enrichment episodes.
Hence, the average trend of [$\alpha$/Fe] is expected to be a declining
function of metallicity, as enough time will be given to type-Ia
supernovae to lower this ratio, unless previous star-formation events were
truncated before type-Ia supernovae started to eject large amounts of
iron-peak elements. In the same line, a decreasing mean [$\alpha$/Fe] is
expected at younger stellar-population ages.

We find no general correlation between [$\alpha$/Fe] and metallicity for
our M31 globular cluster sample, which indicates that the stellar
populations in most of these objects were predominantly enriched by
core-collapse (type-II) supernovae. In combination with the rather weak
anti-correlation of age vs. [$\alpha$/Fe] and age vs. metallicity, this
suggests that old metal-poor globular clusters formed from material which
experienced a truncated enrichment, dominated by type-II supernova ejecta.
Metal-rich, intermediate-age and young M31 globular clusters, on the other
hand, formed from gas that received its metals from type-II {\it and} Ia
supernovae, perhaps during the simmering star-formation processes in M31's
disk. This indicates that stellar populations in the M31 disk experienced
significant enrichment by type-Ia supernovae. The dominant fraction of M31
globular clusters, however, was enriched by type-II supernovae.

\subsubsection{Nitrogen Enhancement}
\begin{figure*}[!th]
\begin{center}
\includegraphics[width=5.5cm, bb=10 150 630 700]{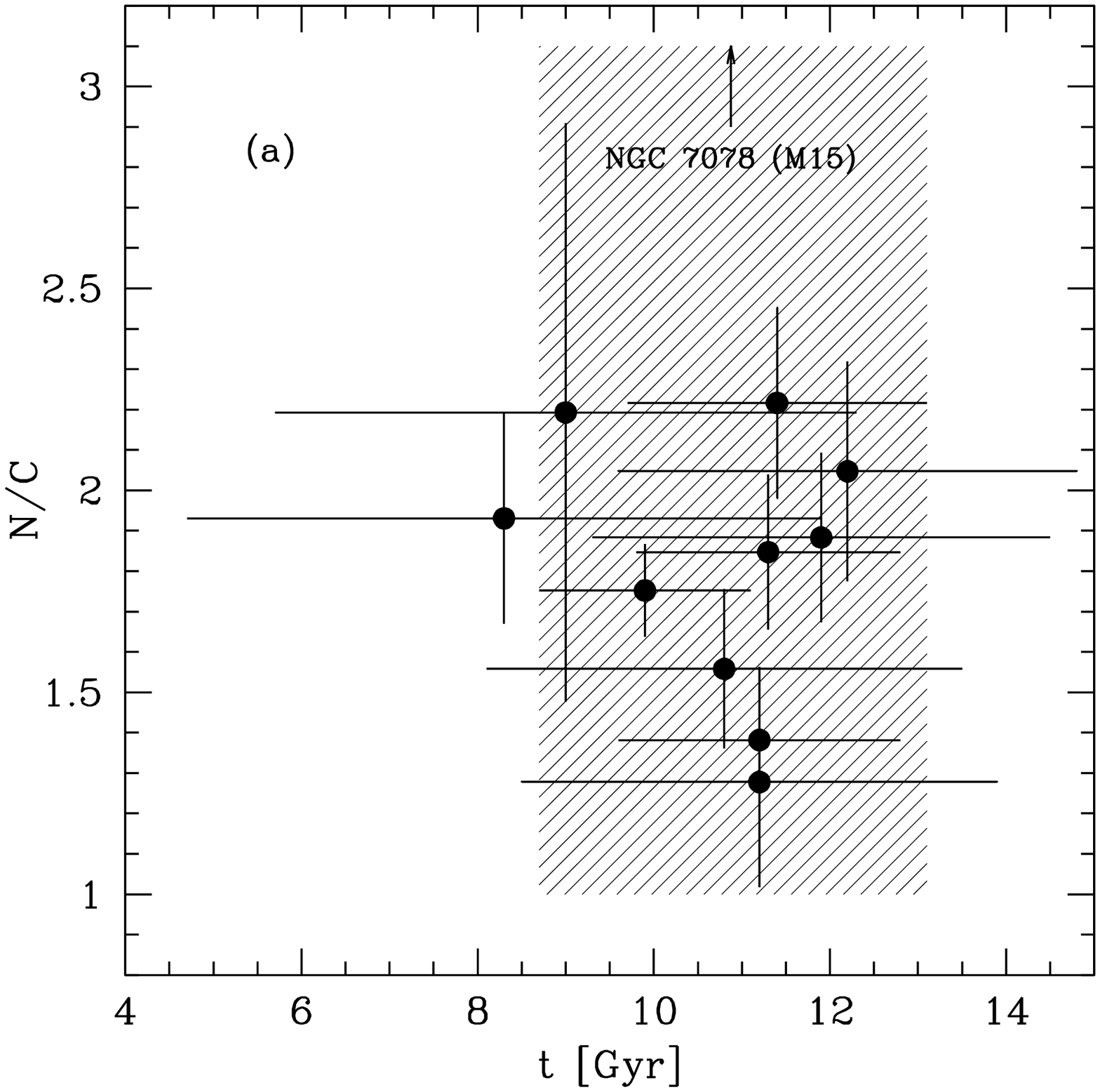}
\includegraphics[width=5.5cm, bb=10 150 630 700]{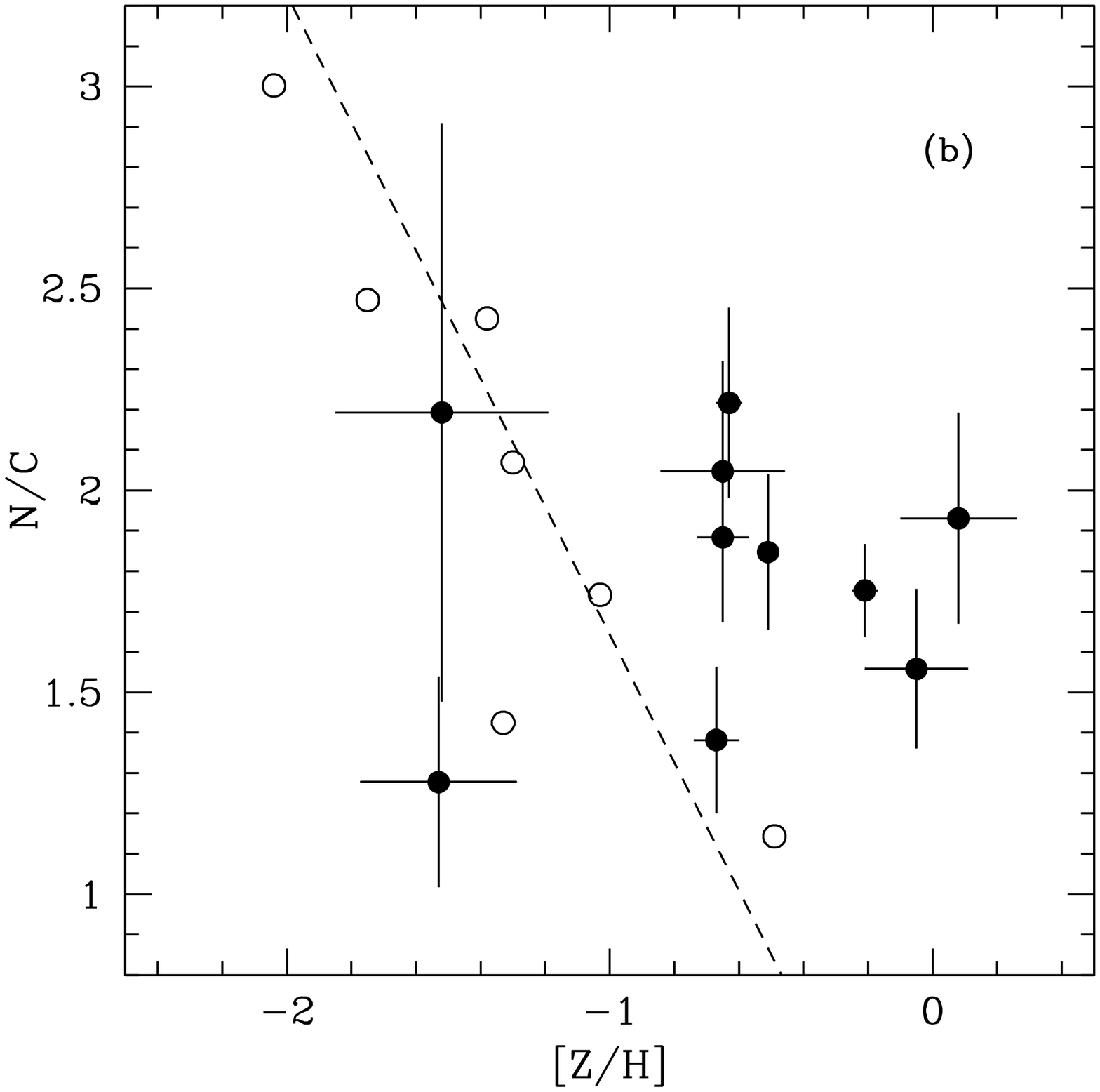}
\includegraphics[width=5.5cm, bb=10 150 630 700]{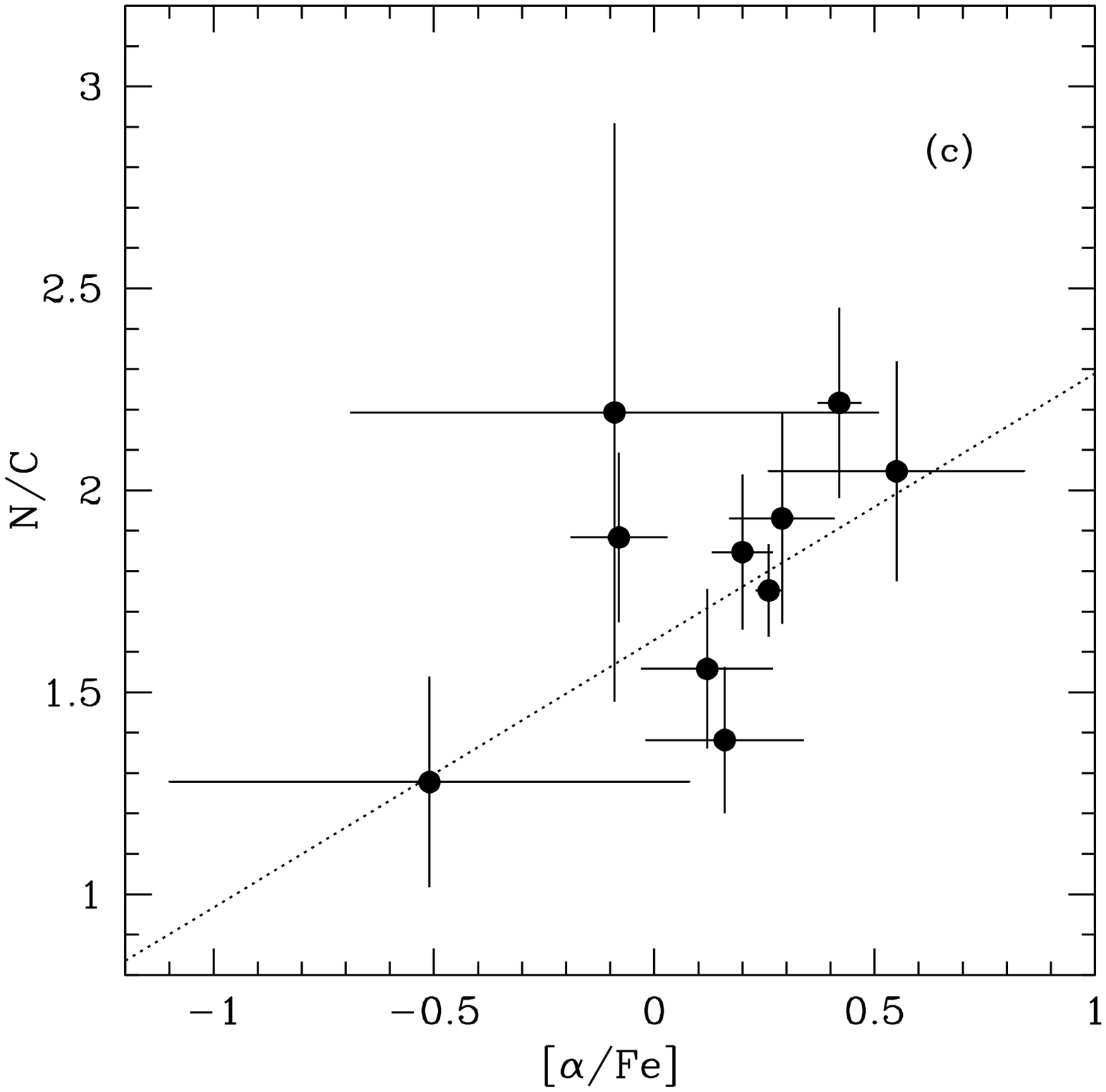}
\caption{Correlation between the N/C ratio, computed as the ratio of the NH index and the CH(G4300) index vs. age, metallicity, and [$\alpha$/Fe] ratio. M31 globular clusters are indicated by solid dots. Open circles mark Milky Way globular clusters. The hatched region in panel (a) shows the typical locus where Galactic globular clusters reside. A linear least-square fit in panel (b) to Galactic globular-cluster data is illustrated by a dashed line. The dashed line in panel (c) shows a linear fit to M31 globular clusters.}
\label{ps:chnhcorr}
\end{center}
\end{figure*}

Nitrogen is produced in two modes. One speaks of a primary and secondary
nitrogen component, depending on whether the seed elements carbon and
oxygen are synthesized from helium by the same star (primary) or were
already in place when the star formed (secondary). Sources of primary
nitrogen are massive, zero-metallicity fast rotating stars that suffer a
hypernova explosion at the end of their very short lifetimes
\citep[typically a few Myr:][]{woosley82, fryer01, heger03}. More massive
and faster rotating hypernova progenitors generally inject more nitrogen
into the interstellar medium \citep{heger00}. However, it is noted by
\citeauthor{heger00} that even a non-rotating 60 M$_{\odot}$ star with a
shallow entropy gradient can produce strong over-abundances of nitrogen.
Secondary nitrogen is mainly ejected by strong stellar winds of
intermediate-mass stars ($\sim4\!-\!7$ M$_{\odot}$) during their AGB phase
\citep[e.g.][]{vdH97}. In general, stars with successively higher mass and
higher angular momentum produce larger amounts of both $\alpha$-elements
and nitrogen \citep[e.g.][]{woosley95}.

This implies that hypernova ejecta have both super-solar [$\alpha$/Fe]
ratios {\it and} enhanced N/C and N/O ratios, as the production of primary
nitrogen is based on the consumption of the seed elements carbon and
oxygen. Moreover, [$\alpha$/Fe], and N/C and N/O ratios should be
positively correlated. This is different for AGB-star yields. \cite{vdH97}
show that the N/C ratio is expected to increase slightly as a function of
stellar mass and metallicity, while N/O is predicted to have a much
steeper correlation. Hence, we expect asymptotically decreasing N/C and
N/O ratios as a function of increasing metallicity. We also expect to see
a correlation of increasing [$\alpha$/Fe] with increasing N/C and N/O
ratios for globular clusters that were mainly enriched by hypernova
ejecta.

The importance of primary and secondary nitrogen enrichment is currently
the subject of a lively debate in the literature. The need for a primary
plateau-enrichment by zero-metallicity massive stars was raised by the
detection of highly super-solar [N/Fe] ratios in extremely metal-poor
[Fe/H]~$\la-4$) Galactic halo stars \citep[e.g.][]{norris01, norris02,
christlieb02, christlieb04} and metal-poor damped Lyman-$\alpha$ systems
\citep[e.g.][]{pettini02}.
         
To study N/C as a function of age, metallicity, and [$\alpha$/Fe], we
match our observations with those of \cite{burstein04} and find ten matched
M31 globular clusters for which NH measurements are available. None of the
Milky Way globular clusters for which NH measurements were performed by
\citeauthor{burstein04} have matched age, metallicity, and [$\alpha$/Fe]
ratios derived from Lick indices in this work. We therefore substitute the
latter with ages and metallicities derived from color-magnitude diagrams
taken from \cite{salaris02} and De Angeli et al. (2004, private
communication). CMD metallicities are on the Zinn-West metallicity scale
and were corrected for an assumed average $\alpha$-element enhancement of
[$\alpha$/Fe]~$=+0.3$ dex, using [Z/H]$=$[Fe/H]$-0.8\cdot$[$\alpha$/Fe]
\citep{trager00a}. No [$\alpha$/Fe] determinations for the matched Galactic
globular clusters are available to us at the moment, so that we have to
postpone a comparison of NH and [$\alpha$/Fe] ratios between Milky Way and
M31 globular clusters.

The Galactic and M31 globular cluster samples of \citealt{burstein04}
cover very different metallicity ranges (see panel b in
Fig.~\ref{ps:chnhcorr}). While most M31 globular clusters fall in the
range $-1\la$~[Z/H]~$\la0$, most of the Milky Way globular clusters have
metallicities below $-1.0$ dex. Hence, a direct comparison of NH
abundances between M31 and Milky Way globular clusters becomes less
meaningful. To bypass this drawback we focus on abundance ratios instead.
Because of their virtually identical dissociation energies
\citep{tomkin84}, the strength of the near-UV NH band and the optical CH
band around 4300 \AA\ can be used to derive the nitrogen to carbon (N/C)
abundance ratio fairly accurately. The strength of the optical CH-band is
measured by the G4300 Lick index which is available for all matched M31
globular clusters as well as Milky Way globular clusters. We therefore
refer to the NH/CH(G4300) index ratio in the following as the N/C ratio.

Figure~\ref{ps:chnhcorr} shows the N/C ratio as a function of
globular cluster age, metallicity, and [$\alpha$/Fe]. We find no clear
correlation of N/C vs. age. Since the age resolution in the covered age
range ($\sim8-13$ Gyr) is poor, it would be very interesting to obtain
more NH data for intermediate-age and young M31 globular clusters. We note
that a correlation of increasing nitrogen-enhancement with older age is
observed in starburst galaxies \citep{esteban95}. If nitrogen-enhancement
is mainly produced by hypernovae, older globular clusters are expected to
show on average higher N/C ratios\footnote{For completeness, we note that
the Galactic globular cluster NGC~7078 shows a relatively high N/C ratio
($\sim4$). This is in contrast to the results of \cite{rauch02} who
measure a carbon-enhancement and a nitrogen depletion (implying a small
N/C ratio) in a sdO cluster star. Further high-resolution measurements for
more stars in NGC~7078 are necessary to resolve this discrepancy.}.

We find enhanced N/C ratios in metal-rich M31 globular clusters compared
to their Galactic counterparts at similar metallicities (see panel b in
Fig.~\ref{ps:chnhcorr}). Almost all M31 clusters with [Z/H]~$\ga-0.8$ dex
appear to have enhanced N/C abundance ratios, while more metal-poor
clusters fall on a similar relation as Milky Way globular clusters (see
dashed line in the middle panel of Fig.~\ref{ps:chnhcorr}). This is clear
evidence against an enrichment scenario dominated by hypernovae, which is
expected to imprint higher N/C ratios at lower metallicities. Instead, the
opposite is observed and lends support to the notion that metal-rich
intermediate-mass AGB stars were given enough time ($\sim250$ Myr) to
eject {\it and} mix secondary nitrogen into the interstellar medium,
before the N-enhanced M31 globular clusters formed.

In panel (c) of Figure~\ref{ps:chnhcorr} we find evidence for a
correlation of the N/C ratio with [$\alpha$/Fe], in the sense that larger
$\alpha$-enhancements are accompanied by higher N/C ratios. A dotted line
shows a weighted linear least-square fit to the data. By itself this
result does not favor either of the nitrogen-enrichment mechanisms
(hypernovae or intermediate-mass AGB stars). However, a significant
hypernova enrichment is expected to produce an extreme [$\alpha$/Fe]
signature with values $\ga1$ dex \citep[e.g.][]{woosley95, fryer01}, which
is observed in some very metal-poor Galactic halo stars
\citep[e.g.][]{aoki02}, but not in the integrated light of
nitrogen-enhanced M31 globular clusters.

\subsection{Chemical Coincidence of Globular Cluster Sub-Populations}
It is intriguing that young M31 and LMC globular clusters and some
counterparts in the M81/Sculptor-group sample show very similar CN indices
at a given metallicity. Two scenarios seem plausible to explain the
chemical similarity: 1) the globular cluster populations experienced similar
enrichment histories, although they formed in galaxies of very
different mass and morphology or 2) the young M31 clusters were 
stripped from accreted dwarf satellites.

Most of the young M31 globular clusters reside in the outskirts of the M31
disk and show radial velocities consistent with a kinematically cold
thin-disk cluster population. However, it is relatively unlikely that an
accretion event was so well aligned to leave no kinematical trace of a
heated globular cluster sub-population. This makes the accretion scenario
less likely.

\cite{forbes04} compile data on ages and metallicities of about a dozen
Galactic globular clusters ($\sim10$\% of the entire system), which are
kinematically associated with giant stellar streams and dwarf galaxies in
the process of accretion\footnote{Note that the potentially accreted
satellite, which might be associated with some of the young M31 globular
clusters, does not necessarily have to leave a luminous stellar component,
and hence a traceable stellar stream.}. They find a range of ages and
metallicities for accreted Galactic globular clusters, from $\sim6$ to 13
Gyr, and metallicities from [Fe/H]~$\approx-2$ to solar values, which does
not allow us to distinguish captured from {\it in-situ} formed globular
clusters using only age and metallicity. However, these authors argue that
the age-metallicity relation for dwarf galaxies (stars and clusters) are
different from the relation of the remaining Milky Way globular clusters.
Different age-metallicity relations point to different chemical enrichment
histories. A comparison of the age-metallicity relations of the Milky Way
and M31 globular cluster systems will be the subject of a future paper,
when more accurate data for the Galactic globular cluster system become
available. At this point, we note that a distinct age-metallicity relation
of accreted globular clusters does not exclude either of the two former
scenarios (chemical coincidence and accretion).

\subsection{Formation Scenarios}
There are several formation and assembly scenarios for globular cluster
systems in spiral galaxies, which can be ordered in three different
categories. Pre-galactic scenarios envision globular cluster formation in
dense metal-poor clouds {\it before} the aggregation of their host galaxy
\citep{peebles68}. In proto-galactic scenarios, globular clusters form in
loose gas clumps {\it during} the coalescence of the host galaxy in more
or less chaotic merging events \citep{searle78}. The third scenario
pictures an {\it a posteriori} assembly where a fraction of the cluster
system is added to the {\it in-situ} formed globular cluster system
through accretion of external globular clusters from satellite galaxies
\citep{cote00}. It is clear that all models overlap in many aspects and
all present valid formation mechanisms for at least some globular cluster 
sub-populations (i.e. halo, disk, bulge) in M31.

The current data does not allow a clear-cut answer on the importance of
each mechanism in the specific case of M31, since our data does not sample
a representative fraction of halo globular clusters. However, our study
has revealed a globular-cluster age distribution which is entirely
inconsistent with the predictions of the pre-galactic and somewhat
inconsistent with the proto-galactic scenario. The formation of
intermediate-age globular clusters appears only plausible in an extended
proto-galactic scenario where the collapse of the gas clumps occurs over
an extended period of time, i.e. for several Gyr: a scenario closely
resembling the hierarchical assembly of galaxies. Hence, the pre- and
proto-galactic models cannot solely describe the assembly of the M31
globular cluster system. Because of their spatial concentration towards
the inner $\sim\!5$ kpc, an accretion scenario also appears less likely as
a formation mechanism for the intermediate-age sub-population.

It should be stressed that the formation of individual globular clusters
might be at stark variance with the suggested formation pictures. For
instance, one might picture the formation of intermediate-age globular
clusters from material which was enriched in the disk and subsequently
funneled into the core regions through bar instabilities. Upscaled
versions of this scenario are observed in other spiral galaxies
\citep[e.g.][]{jogee02}. The fact that most of the intermediate-age
globular clusters show radial velocities that are consistent with
thin-disk kinematics lends support to this formation picture.
Alternatively, one could imagine a two-phase collapse, similar to the one
suggested by \cite{forbes97} for globular cluster systems in early-type
galaxies. In this scenario, halo globular clusters form first from
primordial gas during the proto-galactic phase. The first generation of
stars enriches the remaining gas which later settles in a disk and forms
the second generation of younger and more metal-rich globular clusters.

A fraction of young M31 globular clusters can be explained by the
accretion scenario, as we find a few objects in close projected vicinity
of satellite galaxies and giant stellar streams (see
Fig.~\ref{ps:m31big}). The majority of these young relatively metal-rich
globular clusters, however, resides in the outskirts of M31's disk and
forms a kinematically cold sub-population. As discussed earlier it is
rather unlikely, but not impossible, that a well-aligned, pro-grade
merging event stripped these clusters from an LMC-type satellite galaxy.
It is also possible that we see evidence for a massive star-forming ring
(see GALEX press-release
images\footnote{e.g. http://antwrp.gsfc.nasa.gov/apod/ap031222.html}) in which
young star-clusters are formed from recently compressed molecular clouds,
which might be triggered by recent tidal stress.

For completeness we note that the old globular cluster sub-population has
a wide range of metallicities and its spatial distribution is consistent
with a halo population.

Chemical compositions of intermediate-age and young M31 globular clusters
are consistent with a formation on extended timescales. The chemical
signatures of their old counterparts are expected to be the result of
short star-formation events in which the interstellar medium was
significantly polluted by the thermonuclear ashes of type-II supernovae.
In particular, \cite{li03} put forward the idea that the old M31
globular-cluster population experienced significant enrichment by
nitrogen-rich hypernova ejecta, motivated by the assumption that the
nitrogen overabundance in metal-poor Galactic halo stars and old M31
globular clusters has the same origin. Although conceptually plausible, we
do not subscribe to this idea in its details. It is likely that hypernovae
contributed significantly to the enrichment of the interstellar medium
from which the most metal-poor Galactic stars ([Z/H]~$\la\!-4$) have
formed. However, their metallicity is about two orders of magnitude lower
than that of the most metal-poor known Galactic and M31 globular
cluster\footnote{The most metal-poor Galactic globular cluster NGC~5053
has a [Fe/H]~$\approx-2.3$ dex \citep{harris96}.}. It is hard to believe
that no moderate-mass ($\sim10$ M$_{\odot}$) type-II supernovae
contributed to the chemical signatures of the most metal-poor globular
clusters since the early epochs of chemical enrichment.

Furthermore, the hypernova scenario advocated by \cite{li03}, has
difficulties in explaining the formation of the old metal-poor, but {\it
not} nitrogen-enhanced M31 globular clusters (such as B311, see
Fig.~\ref{ps:chnhcorr}; this globular cluster has the smallest N/C ratio
in our sample). It is conceivable that this might be a result of chemical
variance in the very early epochs of chemical enrichment. Clearly, the
hypernova scenario requires some detailed modeling and comparison with
observations.

It would also be interesting to check if there is a range in nitrogen
enhancement for Milky Way and other globular cluster systems at low
metallicities, resembling the chemical variance during the very early
star-formation epochs. We have embarked on a photometric and spectroscopic
multi-wavelength study of globular clusters in the Milky Way and Local
Group galaxies with ground and space-based telescopes and will address
these issues in the future in more detail.

\section{Summary and Conclusions}
\label{ln:summary}
The presence of a populous M31 globular-cluster subsystem significantly
younger than found in the Galactic globular cluster system is certainly
the most surprising result of this study. Evidence for the counterpart
globular cluster population in the Milky Way is missing, but we have found
for the presence of similar globular cluster sub-populations in other
spiral galaxies. Previous studies have shown that such intermediate-age
and young globular clusters also exist in giant early-type \citep{puzia02,
puzia03proc, puzia03phd, larsen03b, bridges03, puzia04b} and dwarf
galaxies \citep[e.g.][]{beasley02}.

In general, the M31 globular cluster system hosts a wider variety of ages
and chemical compositions than the Galactic globular cluster system, and
is therefore more representative for extragalactic stellar
populations, especially for the comparison with stellar populations in
early-type galaxies. It seems that globular cluster system in spiral
galaxies might have experienced conceptually similar, but quantitatively
different, formation and/or assembly histories. Some extreme cases might
even resemble globular-cluster formation/assembly histories of early-type
galaxies.

Below we summarize the main results of this work.
 
\begin{itemize}
\item[$\bullet$] In addition to a dominant population of globular clusters 
with ages $>10$ Gyr and metallicities from [Z/H]~$\approx-2.0$ to solar
values, we find a population ($\sim20\pm 7$\% of the observed sample) of
intermediate-age globular clusters with ages between $\sim5$ and 9 Gyr and
a mean metallicity around $-0.6$ dex. A few of these objects have
metallicities around $-1.6$ dex.

\item[$\bullet$] We confirm the presence of young M31 globular clusters
recently identified by \cite{beasley04}, which have ages $\la1$
Gyr and relatively high metallicities around $-0.4$ dex; however, two
metal-poor young objects have also been found.

\item[$\bullet$] The M31 globular cluster system has a clearly super-solar
mean $\alpha$/Fe abundance ratio of $+0.14\pm0.04$ dex with a dispersion
$\sigma=0.32$ dex. However, the distribution appears bimodal with peaks
around $+0.2$ and $-0.03$ dex, which have different dispersions, 0.19 and
0.50 dex, respectively.

\item[$\bullet$] Intermediate-age and young globular clusters show roughly
solar abundance ratios. 

\item[$\bullet$] We find evidence for an age-[$\alpha$/Fe] relation in the
sense that younger clusters have smaller mean [$\alpha$/Fe] ratios. This
is driven by the metal-rich intermediate-age and young cluster
population with metallicities higher than [Z/H]~$\approx-1.0$ dex.

\item[$\bullet$] The indices of globular clusters older than $\sim5$ Gyr
are consistent with a factor of three or higher in nitrogen-abundance
enhancement compared to their younger counterparts.

\item[$\bullet$] Using kinematical data for a thin-disk model
\citep{morrison04} we find that the halo population of globular clusters
is old ($\ga9$ Gyr), has a bimodal metallicity distribution and super
solar $\alpha$/Fe-enhancement.

\item[$\bullet$] Thin/thick-disk globular clusters have a wide range of
ages and are predominantly metal-rich with a slightly smaller mean
[$\alpha$/Fe] ratio.

\item[$\bullet$] With the structural parameters of M31 globular clusters
measured by \cite{barmby02}, we confirm a previously found correlation
between half-light/tidal radius and globular clusters metallicity, which
is most likely due to the correlation of half-light/tidal radius and
galactocentric distance.

\item[$\bullet$] Comparing the integrated light of the M31 and Milky Way
bulge, we find that both host metal-rich, $\alpha$-element enhanced and
old stellar populations.

\item[$\bullet$] We find indications for an enhancement in carbon {\it
and} nitrogen in the central region of M31, while the integrated light of
outer M31 stellar populations is consistent with a nitrogen enhancement
only.

\end{itemize}

\begin{acknowledgements}
THP gratefully acknowledges support in form of an ESA Research Fellowship.
KMP acknowledges support of the Natural Sciences and Research Council of
Canada in form of an NSERC Fellowship. TJB thanks David Hanes for
financial assistance during his stay at Queen's University. We are
indebted to Knut Olsen for re-measuring and calibrating Lick indices of
globular clusters in Sculptor-group spiral galaxies and to Francesca
De Angeli for providing CMD ages and metallicities prior to publication.
We thank Nicole Homeier and Maurizio Paolillo for reading early versions
of the draft. It is a pleasure to acknowledge fruitful discussions with
Rupali Chandar, Paul Goudfrooij, and Tom Brown on the results of this
work. We thank the anonymous referee for useful comments and
suggestions that improved the paper. This research has made use of the
SIMBAD database, operated at CDS, Strasbourg, France. The Digitized Sky
Surveys were produced at the Space Telescope Science Institute under U.S.
Government grant NAG W-2166. The images of these surveys are based on
photographic data obtained using the Oschin Schmidt Telescope on Palomar
Mountain and the UK Schmidt Telescope. The plates were processed into the
present compressed digital form with the permission of these institutions.
\end{acknowledgements}

\bibliographystyle{apj}

\appendix
\section{Lick Index Measurements, Ages, Metallicities, and [$\alpha$/Fe] Ratios}
\begin{table*}[h!]
  \caption{Lick indices for our sample M31 globular clusters. Indices were
computed with the passband definitions of \cite{worthey94} and for the
higher-order Balmer lines the definitions of \cite{worthey97}.
We adopted the nomenclature of \cite{galleti04} for all our objects.
Note that B166 is suspected by some studies to be a foreground star. 
  Since the case is not clear, we mark this object as
potentially misclassified but keep it in our globular cluster sample.
However, we point out that all other objects were classified by several
independent studies as genuine globular clusters, based on their
colors and radial velocities \citep[see][]{galleti04}.}

 \label{tab:m31indices}
        {\tiny
        \begin{center}
        \begin{tabular}{lrrrrrrrrrrrrrrrrrrrrrr}
         \hline\hline
         \noalign{\smallskip}

cluster&CN$_1$ &CN$_2$ & Ca4227& G4300 & Fe4383& Ca4455& Fe4531& C$_{2}$4668&H$\beta$&Fe5015&  Mg$_1$ & Mg$_2$ \\
                           &  mag  &  mag  &  \AA  &  \AA  &  \AA  &  \AA  &  \AA  &  \AA  &  \AA  &  \AA  &  mag   &mag\\
\noalign{\smallskip}
\hline
\noalign{\smallskip}
     B006 &   0.088&  0.131&  0.19&  4.14&  4.68&  1.31&  3.26&  1.98&  2.03&  2.35&  0.049&  0.174\\
     B012 &   0.016& $-$0.029& $-$1.25&  2.79& $-$0.11&  0.41&  2.41&  0.17&  2.54&  1.84& $-$0.013&  0.007\\
     B025 &  $-$0.038&  0.011& $-$0.81&  1.95&  3.44&  0.46&  1.66&  0.94&  3.27&  2.17&  0.048&  0.097\\
     B034 &   0.085&  0.172& $-$1.47&  2.25&  2.02& $-$0.51&  3.33&  2.73&  2.21&  1.64&  0.044&  0.143\\
     B045 &  $-$0.033&  0.009& $-$0.00&  3.55&  1.74&  0.95&  2.93&  1.89&  2.37&  1.97&  0.057&  0.166\\
     B048 &   0.084&  0.060&  0.91&  2.91& $-$2.57& $-$0.57&  1.99&  1.85&  2.93&  4.04&  0.060&  0.183\\
     B051 &  $-$0.005&  0.044& $-$0.12&  3.64&  0.62&  0.28&  3.47&  1.72&  1.83&  3.11&  0.080&  0.209\\
     B110 &   0.039&  0.070& $-$0.01&  2.42& $-$2.92&  0.41&  2.32&  1.28&  1.98&  2.91&  0.062&  0.186\\
     B127 &   0.064&  0.133&  0.39&  4.43&  3.61&  0.91&  3.22&  4.30&  1.71&  2.46&  0.100&  0.235\\
     B144 &   0.045&  0.170& $-$0.13&  4.59&  2.19& $-$0.66&  4.07&  4.08&  1.79&  2.21&  0.113&  0.233\\
     B148 &  $-$0.003&  0.056& $-$0.11&  1.83&  2.22&  1.00&  3.25&  3.96&  2.04&  2.71&  0.065&  0.175\\
     B166:&  $-$0.106& $-$0.031&  0.23&  3.72&  2.57&  0.09&  3.98&  1.71&  3.00&  1.88&  0.026&  0.134\\
     B171 &   0.040&  0.113&  0.09&  4.42&  2.68&  0.20&  3.22&  0.82&  1.76&  4.47&  0.088&  0.227\\
     B179 &   0.042&  0.057&  0.01&  3.00&  2.85&  0.43&  2.72&  2.53&  2.12&  2.92&  0.021&  0.103\\
     B182 &  $-$0.011&  0.029&  0.06&  2.31&  4.39&  0.34&  2.55&  1.71&  2.42&  2.51&  0.010&  0.081\\
     B185 &   0.041&  0.100&  0.16&  3.46&  4.16&  0.22&  2.76&  0.55&  1.97&  2.87&  0.071&  0.196\\
     B193 &   0.107&  0.179&  0.38&  3.57&  4.33& $-$0.31&  3.10&  3.63&  2.02&  3.21&  0.098&  0.259\\
     B204 &   0.000&  0.058&  0.93&  3.56&  2.77&  0.64&  3.48&  0.65&  2.01&  2.27&  0.032&  0.142\\
     B218 &  $-$0.004&  0.066&  0.11&  2.22&  1.37&  0.65&  3.71&  2.48&  1.90&  2.64&  0.035&  0.120\\
     B232 &  $-$0.032& $-$0.055& $-$0.05&  1.79&  0.76&  1.10&  1.89&  0.08&  3.11&  0.53&  0.014&  0.058\\
     B235 &   0.009&  0.086&  0.08&  3.46&  3.91&  0.48&  3.16&  1.29&  2.43&  2.42&  0.009&  0.103\\
     B311 &  $-$0.025&  0.003&  0.49&  0.67&  1.14&  0.47&  1.60& $-$1.97&  2.83&  0.89& $-$0.015&  0.044\\
     B312 &   0.047&  0.073& $-$0.36&  1.90&  2.70&  0.25&  2.70&  1.07&  2.97&  1.66&  0.064&  0.138\\
     B315 &  $-$0.168& $-$0.080& $-$0.38& $-$9.53& $-$0.29&  1.00&  4.19& $-$0.60&  4.78&  1.97&  0.059&  0.098\\
     B338 &   0.011&  0.050& $-$0.25&  2.57&  0.38&  0.17&  2.30& $-$1.62&  2.24&  2.84&  0.043&  0.136\\
     B366 &  $-$0.060& $-$0.026& $-$0.17&  0.86&  1.54&  0.98&  1.76&  1.85&  2.11&  1.12& $-$0.014& $-$0.001\\
     B370 &   0.004&  0.008& $-$0.34&  1.25& $-$1.41&  0.24&  1.88&  0.01&  2.74&  0.85&  0.000&  0.045\\
     B372 &   0.122&  0.145& $-$0.24&  2.73&  3.52&  0.21&  2.34&  1.94&  2.39&  1.92&  0.061&  0.137\\
     B386 &  $-$0.005&  0.070& $-$0.03&  1.82&  2.23&  0.92&  1.86& $-$0.82&  2.35&  2.62&  0.032&  0.121\\
\noalign{\smallskip}
\hline
\end{tabular}
\begin{tabular}{lrrrrrrrrrrrrrrrrrrr}
\noalign{\smallskip}
\noalign{\smallskip}
\noalign{\smallskip}
\noalign{\smallskip}
\noalign{\smallskip}
\noalign{\smallskip}
\hline
\hline
\noalign{\smallskip}
cluster& Mg$b$ & Fe5270& Fe5335& Fe5406& Fe5709&H$\delta_A$&H$\gamma_A$&H$\delta_F$&H$\gamma_F$ \\
                   &  \AA  &  \AA  &  \AA  &  \AA  &  \AA  & \AA  &  \AA  &  \AA  &  \AA   \\
\noalign{\smallskip}
\hline
\noalign{\smallskip}
     B006 &   2.65&  2.06&  2.01&  1.05&  0.30&  $-$2.00& $-$4.57& $-$0.09&$-$0.63\\
     B012 &   0.34& $-$0.00&  1.18&  0.22& $-$0.06&   2.21&  1.85&  2.05& 2.35\\
     B025 &   0.75&  1.48&  0.61&  0.14&  0.52&   2.16& $-$1.09&  2.21& 1.23\\
     B034 &   1.84&  1.49&  1.68&  0.75&  0.63&   3.24& $-$0.66&  2.30& 0.47\\
     B045 &   1.30&  2.29&  1.53&  1.21&  0.70&  $-$0.21& $-$3.58& $-$0.09&$-$0.24\\
     B048 &   2.35&  2.37&  2.38&  0.38&  0.75&   1.34& $-$5.22& $-$0.15&$-$2.74\\
     B051 &   2.04&  2.53&  1.63&  1.28&  1.17&   1.08& $-$4.20&  1.26&$-$0.34\\
     B110 &   2.61&  2.49&  1.79&  1.04&  0.80&  $-$0.27& $-$1.94&  0.31& 0.88\\
     B127 &   3.00&  3.02&  0.95&  1.09&  1.02&  $-$0.99& $-$4.19&  0.41&$-$1.29\\
     B144 &   2.91&  2.43&  1.57&  1.03&  0.60&  $-$5.11& $-$5.08& $-$0.07&$-$0.96\\
     B148 &   2.26&  1.96&  2.04&  1.15&  0.61&  $-$0.02& $-$2.76&  1.04&$-$0.00\\
     B166:&   1.60&  1.99&  0.42&  0.37& \dots&   1.45& $-$2.51&  1.50& 0.26\\
     B171 &   3.42&  2.53&  2.57&  1.44&  0.99&   0.02& $-$5.97&  0.38&$-$1.46\\
     B179 &   1.81&  1.40&  1.27&  0.65&  0.52&   0.53& $-$1.91&  0.82& 0.52\\
     B182 &   1.67&  1.74&  1.66&  0.69&  0.28&   0.66& $-$0.57&  1.52& 1.31\\
     B185 &   2.59&  3.00&  2.11&  1.01&  0.91&  $-$0.69& $-$3.94&  0.74&$-$0.14\\
     B193 &   3.98&  2.87&  2.03&  1.19&  0.62&  $-$1.17& $-$5.28&  0.68&$-$0.96\\
     B204 &   2.04&  1.65&  2.25&  1.00&  0.44&  $-$0.86& $-$4.94&  0.26&$-$0.74\\
     B218 &   1.85&  1.61&  2.00&  0.86&  1.07&   0.46& $-$1.77&  0.93& 0.55\\
     B232 &   0.47&  1.67&  0.26&  0.12& \dots&   3.64&  0.66&  2.00& 1.95\\
     B235 &   1.65&  1.82&  1.78&  0.73&  0.63&   0.96& $-$2.56&  0.80& 0.37\\
     B311 &   0.49&  1.10&  0.61&  0.29& $-$0.00&   0.84&  0.83&  3.21& 2.46\\
     B312 &   1.34&  2.15&  0.69&  0.33&  0.54&  $-$0.73& $-$2.29&  1.54& 1.01\\
     B315 &   0.33&  1.93&  0.11&  0.88&  0.96&   6.42&  5.34&  4.59& 4.62\\
     B338 &   1.25&  1.51&  0.91&  0.56&  0.86&  $-$1.59& $-$1.53&  0.82& 0.37\\
     B366 &   0.20&  0.25&  1.07&  0.61& $-$0.37&   2.57&  1.88&  2.31& 2.27\\
     B370 &   0.64& $-$0.08&  0.73&  0.43& $-$0.02&   3.42&  1.08&  2.67& 1.74\\
     B372 &   1.25&  1.93&  1.61&  0.61&  0.63&  $-$0.03& $-$0.92&  1.50& 0.64\\
     B386 &   1.38&  2.21&  0.66&  0.72&  1.16&   2.26& $-$1.01&  1.95& 0.27\\
\noalign{\smallskip}
\hline
\end{tabular}
\end{center}
}

\end{table*}

\newpage
\begin{table*}[h!]
  \caption{Statistical errors of Lick indices for our sample globular clusters.}

 \label{tab:m31indexerr}
        {\tiny
        \begin{center}
        \begin{tabular}{lrrrrrrrrrrrrrrrrrrrrrr}
         \hline\hline
         \noalign{\smallskip}

cluster&CN$_1$ &CN$_2$ & Ca4227& G4300 & Fe4383& Ca4455& Fe4531& C$_{2}$4668&H$\beta$&Fe5015&  Mg$_1$ & Mg$_2$ \\
                     &  mag  &  mag  &  \AA  &  \AA  &  \AA  &  \AA  &  \AA  &  \AA  &  \AA  &  \AA  &  mag   &mag\\
\noalign{\smallskip}
\hline
\noalign{\smallskip}
     B006 &    0.013&  0.019&  0.58&  0.64&  0.31&  0.35&  0.45&  0.54&  0.56&  0.59&  0.016&  0.016\\
     B012 &    0.012&  0.018&  0.65&  0.87&  0.25&  0.28&  0.33&  0.50&  0.51&  0.53&  0.011&  0.011\\
     B025 &    0.020&  0.040&  1.41&  1.51&  0.53&  0.56&  0.62&  0.77&  0.78&  0.84&  0.019&  0.019\\
     B034 &    0.019&  0.027&  0.85&  0.90&  0.26&  0.29&  0.32&  0.46&  0.47&  0.47&  0.011&  0.011\\
     B045 &    0.005&  0.007&  0.25&  0.35&  0.33&  0.37&  0.41&  0.56&  0.56&  0.60&  0.014&  0.014\\
     B048 &    0.014&  0.022&  0.69&  0.85&  0.56&  0.60&  0.67&  0.76&  0.77&  0.84&  0.019&  0.019\\
     B051 &    0.015&  0.017&  0.56&  0.72&  0.41&  0.45&  0.50&  0.66&  0.67&  0.69&  0.016&  0.016\\
     B110 &    0.010&  0.014&  0.49&  0.53&  0.44&  0.45&  0.47&  0.50&  0.51&  0.52&  0.013&  0.013\\
     B127 &    0.002&  0.004&  0.12&  0.13&  0.16&  0.16&  0.17&  0.20&  0.20&  0.21&  0.004&  0.004\\
     B144 &    0.004&  0.009&  0.29&  0.41&  0.26&  0.30&  0.32&  0.45&  0.46&  0.46&  0.010&  0.011\\
     B148 &    0.010&  0.011&  0.36&  0.39&  0.27&  0.30&  0.35&  0.37&  0.37&  0.38&  0.010&  0.010\\
     B166:&    0.007&  0.011&  0.38&  0.43&  0.35&  0.41&  0.49&  0.66&  0.67&  0.69&  0.016&  0.016\\
     B171 &    0.007&  0.011&  0.33&  0.39&  0.44&  0.45&  0.49&  0.61&  0.62&  0.63&  0.014&  0.014\\
     B179 &    0.005&  0.007&  0.22&  0.25&  0.23&  0.25&  0.30&  0.33&  0.34&  0.36&  0.009&  0.010\\
     B182 &    0.005&  0.008&  0.26&  0.29&  0.21&  0.25&  0.32&  0.41&  0.42&  0.44&  0.011&  0.011\\
     B185 &    0.004&  0.008&  0.25&  0.27&  0.44&  0.57&  0.61&  0.75&  0.75&  0.76&  0.016&  0.016\\
     B193 &    0.005&  0.015&  0.46&  0.52&  0.27&  0.31&  0.35&  0.53&  0.53&  0.54&  0.013&  0.013\\
     B204 &    0.004&  0.007&  0.24&  0.28&  0.20&  0.23&  0.26&  0.31&  0.32&  0.34&  0.009&  0.009\\
     B218 &    0.004&  0.006&  0.19&  0.23&  0.17&  0.19&  0.21&  0.29&  0.29&  0.30&  0.007&  0.007\\
     B232 &    0.010&  0.016&  0.53&  0.60&  0.41&  0.43&  0.54&  0.66&  0.66&  0.69&  0.015&  0.015\\
     B235 &    0.013&  0.017&  0.56&  0.80&  0.37&  0.46&  0.54&  0.61&  0.62&  0.65&  0.018&  0.018\\
     B311 &    0.005&  0.011&  0.38&  0.40&  0.42&  0.43&  0.44&  0.52&  0.52&  0.53&  0.012&  0.012\\
     B312 &    0.010&  0.011&  0.36&  0.47&  0.38&  0.39&  0.41&  0.43&  0.43&  0.45&  0.010&  0.010\\
     B315 &    0.010&  0.027&  0.97&  0.99&  0.27&  0.30&  0.35&  0.40&  0.40&  0.46&  0.011&  0.011\\
     B338 &    0.006&  0.012&  0.38&  0.52&  0.58&  0.70&  0.72&  0.75&  0.75&  0.79&  0.018&  0.018\\
     B366 &    0.024&  0.025&  0.84&  0.94&  0.28&  0.33&  0.44&  0.53&  0.55&  0.58&  0.015&  0.015\\
     B370 &    0.006&  0.016&  0.54&  0.61&  0.57&  0.58&  0.66&  0.71&  0.71&  0.72&  0.015&  0.015\\
     B372 &    0.013&  0.023&  0.73&  0.77&  0.30&  0.36&  0.48&  0.63&  0.63&  0.65&  0.016&  0.016\\
     B386 &    0.003&  0.012&  0.41&  0.46&  0.32&  0.35&  0.39&  0.51&  0.51&  0.52&  0.012&  0.012\\
\noalign{\smallskip}
\hline
\end{tabular}
\begin{tabular}{lcrrrrrrrrrrrrrrrrrrr}
\noalign{\smallskip}
\noalign{\smallskip}
\noalign{\smallskip}
\noalign{\smallskip}
\noalign{\smallskip}
\noalign{\smallskip}
\hline
\hline
\noalign{\smallskip}
cluster&Mg$b$ & Fe5270& Fe5335& Fe5406& Fe5709&H$\delta_A$&H$\gamma_A$&H$\delta_F$&H$\gamma_F$ \\
                   &  \AA  &  \AA  &  \AA  &  \AA  &  \AA  & \AA  &  \AA  &  \AA  &  \AA   \\
\noalign{\smallskip}
\hline
\noalign{\smallskip}
     B006 &  0.62&  0.63&  0.63&  0.63&  0.64& 0.98&  1.00&  1.09& 1.10\\
     B012 &  0.53&  0.54&  0.55&  0.56&  0.56& 1.22&  1.43&  1.56& 1.59\\
     B025 &  0.87&  0.88&  0.88&  0.88&  0.89& 1.73&  1.75&  1.79& 1.80\\
     B034 &  0.49&  0.50&  0.50&  0.50&  0.50& 1.13&  1.22&  1.28& 1.29\\
     B045 &  0.64&  0.64&  0.64&  0.65&  0.65& 0.60&  0.69&  0.76& 0.77\\
     B048 &  0.86&  0.86&  0.88&  0.88&  0.89& 1.10&  1.20&  1.24& 1.27\\
     B051 &  0.73&  0.74&  0.74&  0.74&  0.74& 1.05&  1.07&  1.09& 1.09\\
     B110 &  0.52&  0.53&  0.53&  0.53&  0.53& 0.85&  1.22&  1.31& 1.34\\
     B127 &  0.21&  0.21&  0.21&  0.21&  0.21& 0.23&  0.30&  0.30& 0.31\\
     B144 &  0.47&  0.47&  0.48&  0.48&  0.48& 0.81&  0.82&  0.83& 0.86\\
     B148 &  0.39&  0.39&  0.39&  0.39&  0.39& 0.77&  0.78&  0.78& 0.79\\
     B166:&  0.70&  0.71&  0.72&  0.72& \dots& 0.59&  0.61&  0.63& 0.63\\
     B171 &  0.65&  0.65&  0.65&  0.66&  0.66& 0.53&  0.57&  0.67& 0.69\\
     B179 &  0.37&  0.37&  0.37&  0.37&  0.38& 0.30&  0.32&  0.33& 0.35\\
     B182 &  0.46&  0.47&  0.47&  0.47&  0.47& 0.46&  0.56&  0.57& 0.58\\
     B185 &  0.76&  0.76&  0.76&  0.76&  0.76& 0.66&  0.76&  0.76& 0.77\\
     B193 &  0.57&  0.57&  0.57&  0.58&  0.58& 0.88&  0.93&  0.94& 0.95\\
     B204 &  0.35&  0.35&  0.36&  0.36&  0.36& 0.42&  0.44&  0.46& 0.47\\
     B218 &  0.30&  0.30&  0.31&  0.31&  0.31& 0.28&  0.30&  0.35& 0.37\\
     B232 &  0.70&  0.70&  0.70&  0.71& \dots& 0.76&  0.77&  0.84& 0.84\\
     B235 &  0.66&  0.68&  0.70&  0.70&  0.70& 1.16&  0.73&  1.35& 0.74\\
     B311 &  0.56&  0.56&  0.56&  0.56&  0.57& 0.75&  0.76&  0.79& 0.80\\
     B312 &  0.45&  0.45&  0.45&  0.45&  0.45& 0.83&  0.88&  0.89& 0.90\\
     B315 &  0.47&  0.48&  0.50&  0.51&  0.51& 1.03&  1.05&  1.06& 1.06\\
     B338 &  0.82&  0.83&  0.83&  0.83&  0.83& 0.61&  0.63&  0.65& 0.73\\
     B366 &  0.59&  0.59&  0.60&  0.60&  0.60& 1.17&  1.19&  1.21& 1.22\\
     B370 &  0.73&  0.73&  0.74&  0.74&  0.74& 0.94&  1.03&  1.04& 1.06\\
     B372 &  0.66&  0.67&  0.67&  0.67&  0.67& 0.91&  0.93&  0.94& 0.95\\
     B386 &  0.54&  0.54&  0.54&  0.54&  0.54& 0.51&  0.63&  0.64& 0.66\\
\noalign{\smallskip}
\hline
\end{tabular}
\end{center}
}
\end{table*}

\newpage
\begin{table*}[h!]
  \caption{Ages, Metallicities, and [$\alpha$/Fe] ratios for M31 globular clusters.
We point out that the quoted uncertainties are the propagated
statistical uncertainties of the data in combination with the SSP models
used in this work. Systematic uncertainties of the presented parameters
are of the order (for [Z/H] and [$\alpha$/Fe]) or larger (for age) than the
statistical errors and are discussed in Section 3.3.}

 \label{tab:ama}
        {\tiny
        \begin{center}
        \begin{tabular}{lcrrr}
         \hline\hline
         \noalign{\smallskip}

cluster&D$^{\mathrm{a}}$& Age [Gyr] & [Z/H] & [$\alpha$/Fe] \\
\noalign{\smallskip}
\hline
\noalign{\smallskip}
      B006&1& $ 12.4\pm2.8$&$ -0.35\pm0.14$&$ 0.18\pm0.18$\\
      B012&1& $ 10.2\pm2.9$&$ -2.00\pm0.30$&$-0.96\pm1.20$\\
      B025&1& $  9.6\pm3.3$&$ -1.17\pm0.38$&$ 0.43\pm0.34$\\
      B034&1& $  8.1\pm3.9$&$ -0.62\pm0.15$&$ 0.31\pm0.17$\\
      B045&1& $ 10.6\pm2.7$&$ -0.50\pm0.11$&$ 0.22\pm0.19$\\
      B048&1& $  7.5\pm3.6$&$ -0.20\pm0.30$&$ 0.04\pm0.22$\\
      B051&1& $ 11.3\pm3.2$&$ -0.40\pm0.17$&$ 0.32\pm0.20$\\
      B110&1& $  6.2\pm2.7$&$ -0.23\pm0.18$&$ 0.19\pm0.13$\\
      B127&1& $ 13.3\pm1.7$&$ -0.22\pm0.05$&$ 0.45\pm0.05$\\
      B144&1& $ 14.4\pm3.0$&$ -0.25\pm0.10$&$ 0.45\pm0.12$\\
      B148&1& $ 11.1\pm3.1$&$ -0.47\pm0.12$&$ 0.22\pm0.11$\\
     B166:&1& $  9.3\pm2.9$&$ -0.52\pm0.13$&$ 0.56\pm0.27$\\
      B171&1& $ 10.8\pm2.7$&$ -0.05\pm0.21$&$ 0.12\pm0.15$\\
      B179&1& $ 11.2\pm1.6$&$ -0.67\pm0.07$&$ 0.16\pm0.18$\\
      B182&1& $  7.1\pm4.3$&$ -0.61\pm0.19$&$-0.48\pm0.23$\\
      B185&1& $  5.8\pm4.0$&$ -0.12\pm0.28$&$ 0.01\pm0.18$\\
      B193&1& $  8.3\pm3.6$&$  0.08\pm0.21$&$ 0.29\pm0.12$\\
      B204&1& $ 12.3\pm1.8$&$ -0.44\pm0.05$&$ 0.00\pm0.12$\\
      B218&1& $ 11.9\pm2.6$&$ -0.65\pm0.12$&$-0.08\pm0.12$\\
      B232&1& $  9.0\pm3.3$&$ -1.52\pm0.52$&$-0.09\pm0.66$\\
      B235&1& $ 10.2\pm3.8$&$ -0.63\pm0.23$&$-0.27\pm0.31$\\
      B311&1& $ 11.2\pm2.7$&$ -1.53\pm0.38$&$-0.51\pm0.85$\\
      B312&1& $  9.4\pm1.7$&$ -0.58\pm0.09$&$ 0.40\pm0.17$\\
      B315&1& $  3.5\pm2.2$&$ -1.54\pm0.61$&$ 0.46\pm0.23$\\
      B338&1& $ 12.2\pm2.6$&$ -0.65\pm0.19$&$ 0.55\pm0.29$\\
      B366&1& $ 11.0\pm3.0$&$ -1.84\pm0.32$&$-0.91\pm1.18$\\
      B370&1& $  9.1\pm2.5$&$ -1.73\pm0.21$&$ 0.99\pm0.95$\\
      B372&1& $ 11.9\pm3.5$&$ -0.77\pm0.26$&$ 0.09\pm0.24$\\
      B386&1& $ 10.9\pm3.2$&$ -0.79\pm0.22$&$ 0.25\pm0.22$\\
      B126&2& $  7.2\pm3.1$&$ -1.25\pm0.21$&$ 0.15\pm0.19$\\
      B134&2& $ 11.3\pm1.8$&$ -0.54\pm0.06$&$ 0.09\pm0.10$\\
      B158&2& $ 11.3\pm1.5$&$ -0.51\pm0.05$&$ 0.20\pm0.07$\\
      B163&2& $  7.7\pm1.0$&$  0.09\pm0.05$&$ 0.20\pm0.03$\\
      B222&2& $  1.1\pm0.4$&$ -0.46\pm0.13$&$ 0.20\pm0.13$\\
      B225&2& $  9.9\pm1.2$&$ -0.21\pm0.04$&$ 0.26\pm0.03$\\
      B234&2& $ 11.4\pm2.0$&$ -0.47\pm0.05$&$ 0.07\pm0.09$\\
      B292&2& $  9.2\pm3.3$&$ -1.53\pm0.39$&$ 0.11\pm0.31$\\
      B301&2& $  4.1\pm3.7$&$ -0.61\pm0.13$&$-0.12\pm0.24$\\
      B304&2& $ 13.2\pm3.1$&$ -1.27\pm0.37$&$-0.12\pm0.29$\\
      B310&2& $ 10.8\pm3.1$&$ -1.55\pm0.37$&$-0.33\pm0.39$\\
      B313&2& $ 11.2\pm1.2$&$ -0.53\pm0.03$&$ 0.28\pm0.11$\\
      B314&2& $  1.0\pm0.1$&$ -0.28\pm0.06$&$ 0.07\pm0.23$\\
      B321&2& $  1.0\pm0.1$&$ -1.79\pm0.08$&$ 0.20\pm0.31$\\
      B322&2& $  2.3\pm0.7$&$ -1.73\pm0.33$&$ 0.28\pm0.30$\\
      B324&2& $  1.0\pm0.1$&$ -0.48\pm0.02$&$-0.11\pm0.11$\\
      B327&2& $  5.4\pm1.4$&$ -1.69\pm0.31$&$ 0.30\pm0.16$\\
      B328&2& $ 11.0\pm2.4$&$ -1.71\pm0.30$&$ 0.62\pm0.32$\\
      B337&2& $  4.9\pm2.9$&$ -0.59\pm0.05$&$ 0.00\pm0.11$\\
      B347&2& $ 10.2\pm2.4$&$ -1.97\pm0.14$&$ 0.68\pm0.36$\\
      B350&2& $  9.3\pm2.3$&$ -1.45\pm0.26$&$ 0.38\pm0.28$\\
      B365&2& $  6.6\pm3.1$&$ -0.95\pm0.13$&$ 0.08\pm0.16$\\
      B380&2& $  1.0\pm0.1$&$ -1.51\pm0.03$&$-0.19\pm0.13$\\
      B383&2& $ 11.3\pm2.4$&$ -0.33\pm0.06$&$ 0.28\pm0.05$\\
      B393&2& $ 11.2\pm1.4$&$ -0.63\pm0.05$&$-0.03\pm0.14$\\
      B398&2& $ 15.0\pm2.4$&$ -0.32\pm0.06$&$ 0.23\pm0.09$\\
      B401&2& $ 10.2\pm2.4$&$ -1.94\pm0.14$&$ 0.47\pm0.31$\\
      NB16&2& $  7.2\pm3.5$&$ -0.86\pm0.11$&$ 0.27\pm0.11$\\
      NB67&2& $ 10.1\pm3.2$&$ -0.98\pm0.18$&$ 0.38\pm0.13$\\
      NB68&2& $  7.6\pm2.3$&$ -0.28\pm0.06$&$ 0.19\pm0.07$\\
      NB74&2& $ 15.0\pm1.6$&$ -0.04\pm0.06$&$ 0.56\pm0.13$\\
      NB81&2& $  4.5\pm1.1$&$ -0.19\pm0.06$&$ 0.08\pm0.06$\\
      NB83&2& $  7.3\pm3.6$&$ -0.68\pm0.09$&$ 0.05\pm0.13$\\
      NB87&2& $  8.8\pm2.4$&$  0.48\pm0.08$&$ 0.09\pm0.04$\\
      NB89&2& $  6.8\pm1.7$&$ -0.41\pm0.04$&$ 0.18\pm0.07$\\
      NB91&2& $  7.8\pm2.4$&$ -0.28\pm0.06$&$ 0.13\pm0.07$\\
      B240&3& $ 11.0\pm1.1$&$ -1.82\pm0.11$&$ 0.06\pm0.24$\\
      B058&3& $  6.4\pm4.1$&$ -0.57\pm0.10$&$-0.34\pm0.15$\\
      B178&3& $  6.3\pm1.8$&$ -0.61\pm0.06$&$ 0.38\pm0.09$\\
      G001&3& $ 11.4\pm1.7$&$ -0.63\pm0.06$&$ 0.42\pm0.05$\\
      B358&3& $ 10.4\pm1.9$&$ -2.00\pm0.11$&$-0.00\pm0.27$\\
\noalign{\smallskip}
\hline
\end{tabular}
\end{center}
}
\begin{list}{}{}
\item[$^{\mathrm{a}}$] Dataset label: 1 -- this work, 2 -- \cite{beasley04}, 3 -- \cite{worthey94} and \cite{kuntschner02}.
\end{list}
\end{table*}


\end{document}